\documentclass[12pt,preprint]{aastex}
\usepackage{amsmath}
\usepackage{natbib}
\usepackage{hyperref}
\usepackage{rotating}
\usepackage{color}

\begin{document}

\bibliographystyle{plainnat}

\title{\Large{Gamma-ray burst prompt correlations}}

\author{Dainotti M. G.\altaffilmark{1,2,3}, Del Vecchio R.\altaffilmark{3}, Tarnopolski M.\altaffilmark{3}}

\altaffiltext{1}{Physics Department, Stanford University, Via Pueblo Mall 382, Stanford, CA, USA, E-mail: mdainott@stanford.edu}
\altaffiltext{2}{INAF-Istituto di Astrofisica Spaziale e Fisica cosmica, Via Gobetti 101, 40129, Bologna, Italy}
\altaffiltext{3}{Astronomical Observatory, Jagiellonian University, Orla 171, 30-244 Krak{\'o}w, Poland, E-mails: delvecchioroberta@hotmail.it, mariagiovannadainotti@yahoo.it, mariusz.tarnopolski@uj.edu.pl}

\begin{abstract}
The mechanism responsible for the prompt emission of gamma-ray bursts (GRBs) is still a debated issue. The prompt phase-related GRB correlations can allow to discriminate among the most plausible theoretical models explaining this emission. We present an overview of the observational two-parameter correlations, their physical interpretations, their use as redshift estimators and possibly as cosmological tools. The nowadays challenge is to make GRBs, the farthest stellar-scaled objects observed (up to redshift $z=9.4$), standard candles through well established and robust correlations. However, GRBs spanning several orders of magnitude in their energetics are far from being standard candles. We describe the advances in the prompt correlation research in the past decades, with particular focus paid to the discoveries in the last 20 years.
\end{abstract}

\maketitle
\pagebreak
\tableofcontents

\pagebreak

\section{Introduction}

Gamma-ray bursts (GRBs) are highly energetic events with the total isotropic energy released of the order of $10^{48}-10^{55}\,{\rm erg}$ (for recent reviews, see \citealt{nakar2007,zhang11,gehrels2013,berger2014,kumar15,meszaros2015}). GRBs were discovered by military satellites {\it Vela} in late 1960's and were recognized early to be of extrasolar origin \citep{klebesadel73}. A bimodal structure (reported first by \citealt{mazets}) in the duration distribution of GRBs detected by the Burst and Transient Source Experiment (BATSE) onboard the {\it Compton Gamma-Ray Observatory} ({\it CGRO}) \citep{186}, based on which GRBs are nowadays commonly classified into short (with durations $T_{90}<2\,{\rm s}$, SGRBs) and long (with $T_{90}>2\,{\rm s}$, LGRBs), was found \citep{kouveliotou93}. BATSE observations allowed also to confirm the hypothesis of \citet{klebesadel73} that GRBs are of extragalatic origin due to isotropic angular distribution in the sky combined with the fact that they exhibited an intensity distribution that deviated strongly from the $-3/2$ power law \citep{paczynski91a,paczynski91b,186,fishman1995,briggs1996}. This was later corroborated by establishing the first redshift measurement, taken for GRB970508, which with $0.835<z\lesssim 2.3$ was placed at a cosmological distance of at least $2.9\,{\rm Gpc}$ \citep{metzger97}. Despite initial isotropy, SGRBs were shown to be distributed anisotropically on the sky, while LGRBs are distributed isotropically \citep{balazs1998,meszaros2000a,meszaros2000b,meszaros2003,maglio2003,bernui,vavrek,tarnopolski2015a}. Cosmological consequences of the anisotropic celestial distribution of SGRBs were discussed lately by \citet{meszaros2009} and \citet{meszaros2015}. Finally, the progenitors of LGRBs are associated with supernovae (SNe) \citep{hjorth03,malesani04,woosley06,sparre11,schulze14} related with collapse of massive stars. Progenitors of SGRBs are thought to be neutron star--black hole (NS--BH) or NS--NS mergers \citep{eichler,paczynski91b,narayan1992,nakar2004}, and no connection between SGRBs and SNe has been proven \citep{zhang09}.

While the recent first direct detection of gravitational waves (GW), termed GW150914, by the Laser Interferometer Gravitational Wave Observatory (LIGO) \citep{abbott}, interpreted as a merger of two stellar-mass BHs with masses $36^{+5}_{-4}M_\odot$ and $29^{+4}_{-4}M_\odot$, is by itself a discovery of prime importance, it becomes especially interesting in light of the finding of \citet{connaughton} who reported a weak transient source lasting $1\,{\rm s}$ and detected by {\it Fermi}/GBM \citep{bhat2016} only $0.4\,{\rm s}$ after the GW150914, termed GW150914-GBM. Its false alarm probability is estimated to be 0.0022. The fluence in the energy band $1\,{\rm keV}-10\,{\rm MeV}$ is computed to be $1.8^{+1.5}_{-1.0}\times10^{49}\,{\rm erg}\,{\rm s}^{-1}$. While these GW and GRB events are consistent in direction, its connection is tentative due to relatively large uncertainties in their localization. This association is unexpected as SGRBs have been thought to originate from NS--NS or NS--BH mergers. Moreover, neither {\it INTEGRAL} \citep{savchenko}, nor {\it Swift} \citep{evans} detected any signals that could be ascribed to a GRB. Even if it turns out that it is only a chance coincidence \citep{lyutnikov}, it has already triggered scenarios explaining how a BH--BH merger can become a GRB, e.g. the nascent BH could generate a GRB via accretion of a mass $\simeq 10^{-5}M_\odot$ \citep{xli}, indicating its location in a dense medium (see also \citealt{loeb}), or two high-mass, low-metallicity stars could undergo an SN explosion, and the matter ejected from the last exploding star can form---after some time---an accretion disk producing an SGRB \citep{perna}. Also the possible detection of an afterglow that can be visible many months after the event \citep{morsony} could shed light on the nature of the GW and SGRB association.

From a phenomenological point of view, a GRB is composed of the prompt emission, which consists of high-energy photons such as $\gamma$-rays and hard X-rays, and the afterglow emission, i.e. a long lasting multi-wavelength emission (X-ray, optical, and sometimes also radio), which follows the prompt phase. The first afterglow observation (for GRB970228) was due to the {\it BeppoSAX} satellite \citep{costa97,vanparadijs1997}. Another class, besides SGRBs and LGRBs, i.e. intermediate in duration, was proposed to be present in univariate duration distributions \citep{horvath98,horvath02,horvath08,horvath09,huja09,ripa09}, as well as in higher dimensional parameter spaces \citep{mukherjee,horvath06,ripa09,horvath10,veres10,koen12}. On the other hand, this elusive intermediate class might be a statistical feature that can be explained by modelling the duration distribution with skewed distributions, instead of the commonly applied standard Gaussians \citep{zitouni,tarnopolski2015b,tarnopolski2016a,tarnopolski2016b,tarnopolski2016c}. Additionally, GRB classification was shown to be detector dependent \citep{nakar2007,bromberg,tarnopolski2015c}. Moreover, a subclass classification of LGRBs was proposed \citep{gao10}, and \cite{norris2006} discovered the existence of an intermediate class or SGRBs with extended emission, that show mixed properties between SGRBs and LGRBs. GRBs with very long durations (ultra-long GRBs, with $T_{90}>1000\,{\rm s}$) are statistically different than regular (i.e., with $T_{90}<500\,{\rm s}$) LGRBs \citep{boer2015}, and hence might form a different class (see also \citealt{virgili2013,zhang2014,levan2014,levan2015}). Another relevant classification appears related to the spectral features distinguishing normal GRBs from X-ray flashes (XRFs). The XRFs \citep{Heise2001,Kippen2001} are extragalactic transient X-ray sources with spatial distribution, spectral and temporal characteristics similar to LGRBs. The remarkable property that distinguishes XRFs from GRBs is that their $\nu F_{\nu}$ prompt emission spectrum peaks at energies which are observed to be typically an order of magnitude lower than the observed peak energies of GRBs. XRFs are empirically defined by a greater fluence (time integrated flux) in the X-ray band ($2-30\,{\rm keV}$) than in the $\gamma$-ray band ($30-400\,{\rm keV}$). This classification is also relevant for the investigation of GRB correlations since some of them become stronger or weaker by introducing different GRB categories. \citet{grupe2013}, using 754 {\it Swift} GRBs, performed an exhaustive analysis of several correlations as well as the GRB redshift distribution, discovering that the bright bursts are more common in the high-$z$ (i.e., $z\gtrsim 3$) than in the local universe.

This classification has further enhanced the knowledge of the progenitor system from which GRBs originate. It was soon after their discovery that LGRBs were thought to originate from distant star-forming galaxies. Since then, LGRBs have been firmly associated with powerful core-collapse SNe and the association seems solid. Nevertheless, there have been puzzling cases of LGRBs that were not associated with bright SNe \citep{fynbo06,dellavalle06}. This implies that it is possible to observe GRBs without an associated bright SNe or there are other progenitors for LGRBs than core-collapse of massive stars. Another relevant uncertainty concerning the progenitor systems for LGRBs is the role of metallicity $Z$. In the collapsar model \citep{woosley06}, LGRBs are only formed by massive stars with $Z/Z_\odot$ below $\simeq 0.1-0.3$. However, several GRBs have been located in very metal-rich systems \citep{perley15} and it is an important goal to understand whether there are other ways to form LGRBs than through the collapsar scenario \citep{greiner15}. One of the models used to explain the GRB phenomenon is the ``fireball'' model \citep{wijers97,meszaros1998,meszaros2006} in which a compact central engine (either the collapsed core of a massive star or the merger product of an NS--NS binary) launches a highly relativistic, jetted electron-positron-baryon plasma. Interactions of blobs within the jet are believed to produce the prompt emission. Instead, the interaction of the jet with the ambient material causes the afterglow phase. However, problems in explaining the light curves within this model have been shown by \citet{Willingale2007}. Specifically, for $\simeq 50\%$ of GRBs the observed afterglows are in agreement with the model, but for the rest the temporal and spectral indices do not conform and suggest a continued late energy injection. \citet{melandri2008} performed a multiwavelength analysis and found that the forward shock (FS) model does not explain almost 50\% of the examined GRBs, even after taking into account energy injection. \citet{rykoff2009} showed that the fireball model does not model correctly early afterglows. \citep{oates2011} analyzed the prompt and afterglow light curves, and pointed out that some GRBs required energy injection to explain the outflows. The crisis of the standard fireball models appeared when {\it Swift} \citep{gehrels2004} observations revealed a more complex behaviour of the light curves than observed in the past \citep{Obrien06,sakamoto07,zhang07c} and pointed out that GRBs often follow ``canonical'' light curves \citep{Nousek2006}. Therefore, the discovery of correlations among relevant physical parameters in the prompt phase is very important in this context in order to use them as possible model discriminators. In fact, many theoretical models have been presented in the literature in order to explain the wide variety of observations, but each model has some advantages as well as drawbacks, and the use of the phenomenological correlations can boost the understanding of the mechanism responsible for the prompt emission. Moreover, given the much larger (compared to SNe) redshift range over which GRBs can be observed, it is tempting to include them as cosmological probes, extending the redshift range by almost an order of magnitude further than the available SNe Ia. GRBs are observed up to redshift $z=9.4$ \citep{cucchiara11}, much more distant than SNe Ia, observed up to $z=2.26$ \citep{Rodney2015}, and therefore they can help to understand the nature of dark energy and determine the evolution of its equation of state at very high $z$. However, contrary to SNe Ia, which originate from white dwarves reaching the Chandrasekhar limit and always releasing the same amount of energy, GRBs cannot yet be considered standard candles with their (isotropic-equivalent) energies spanning 8 orders of magnitude (see also \citealt{lin2015} and references therein). Therefore, finding universal relations among observable properties can help to standardize their energetics and/or luminosities. They 
can serve as a tracer of the history of the cosmic star formation rate \citep{totani1997,porciani2001,bromm2006,kistler2009,souza2011} and provide invaluable information on the physics in the intergalactic medium \citep{barkana2004,ioka2005,inoue2007}. This is the reason why the study of GRB correlations is so relevant for understanding the GRB emission mechanism, for finding a good distance indicator, and for exploring the high-redshift universe \citep{salvaterra2015}.

This paper is organized in the following manner. In Sect.~\ref{notations} we explain the nomenclature and definitions adopted in this work, and in Sect.~\ref{Prompt correlations} we analyze the correlations between various prompt parameters. We summarize in Sect.~\ref{summary}.

\section{Notations and nomenclature}\label{notations}
For clarity we report a summary of the nomenclature adopted in the review. $L$, $F$, $E$, $S$ and $T$ indicate the luminosity, the energy flux, the energy, the fluence and the timescale, respectively, which can be measured in several wavelengths. More specifically:
\begin{itemize}
\item $T_{90}$ is the time interval in which 90\% of the GRB's fluence is accumulated, starting from the time at which 5\% of the total fluence was detected \citep{kouveliotou93}.

\item $T_{50}$ is defined, similarly to $T_{90}$, as the time interval from 25\% to 75\% of the total detected fluence.

\item $T_{45}$ is the time spanned by the brightest 45\% of the total counts detected above background \citep{reichart2001}.

\item$T_{\rm peak}$ is the time at which the pulse (i.e., a sharp rise and a slower, smooth decay \citep{fishman94,92,stern96,ryde02}) in the prompt light curve peaks (see Fig.~\ref{fig1}).
\begin{figure}[htbp]
\centering
\includegraphics[width=12.2cm,height=7cm,angle=0]{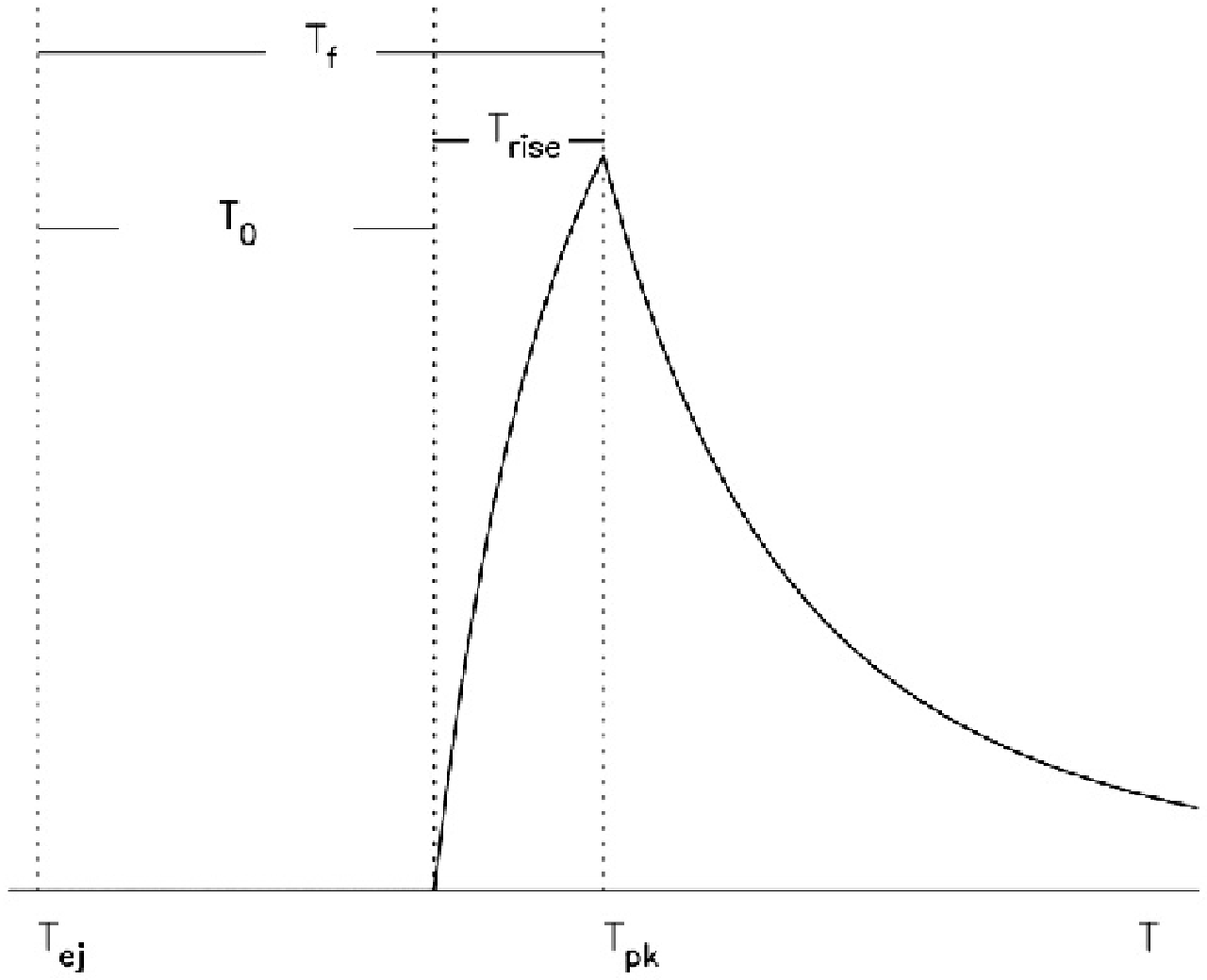}
\caption{A sketch of the pulse displaying $T_{\rm ej}$ and $T_{\rm peak}$ (denoted by $T_{\rm pk}$ here) and 
the quantities $T_{f}$ and $T_{0}=T_{f}-T_{\rm rise}$. (Figure after \cite{willingale2010}; see Fig. 1 therein.)}
\label{fig1}
\end{figure}

\item $T_{\rm break}$ is the time of a power law break in the afterglow light curve \citep{SPH99,willingale2010}, i.e. the time when the afterglow brightness has a power law decline that suddenly steepens due to the slowing down of the jet until the relativistic beaming angle roughly equals the jet opening angle $\theta_{\rm jet}$ \citep{Rhoads97}

\item $\tau_{\rm lag}$ and $\tau_{\rm RT}$ are the difference of arrival times to the observer of the high energy photons and low energy photons defined between $25-50\,{\rm keV}$ and $100-300\,{\rm keV}$ energy band, and the shortest time over which the light curve increases by $50\%$ of the peak flux of the pulse.

\item $T_p$ is the end time prompt phase at which the exponential decay switches to a power law, which is usually followed by a shallow decay called the plateau phase, and $T_a$ is the time at the end of this plateau phase \citep{Willingale2007}.

\item $T_f$ is the pulse width since the burst trigger at the time $T_{\rm ej}$ of the ejecta.

\item $E_{\rm peak}$, $E_{\rm iso}$, $E_{\gamma}$ and $E_{\rm prompt}$ are the peak energy, i.e. the energy at the peak of the $\nu F_{\nu}$ spectrum \citep{mallozzi95}, the total isotropic energy emitted during the whole burst (e.g., \citealt{AmatiEtal02}), the total energy corrected for the beaming factor [the latter two are connected via $E_\gamma=(1-\cos\theta_{\rm jet})E_{\rm iso}$], and the isotropic energy emitted in the prompt phase, respectively.

\item $F_{\rm peak}$, $F_{\rm tot}$ are the peak and the total fluxes respectively \citep{lee96}.

\item $L_a$, $L_{X,p}$ and $L_f$ are the luminosities respective to $T_a$, $T_p$ (specified in the X-ray band) and $T_f$.

\item $L$ is the observed luminosity, and specifically $L_{\rm peak}$ and $L_{\rm iso}$ are the peak luminosity (i.e., the luminosity at the pulse peak, \citealt{norris2000}) and the total isotropic luminosity, both in a given energy band. More precisely, $L_{\rm peak}$ is defined as follows:
\begin{equation}
L_{\rm peak} = 4\pi D_L(z,\Omega_M, \Omega_{\Lambda})^2 F_{\rm{peak}}\,,
\label{eq1}
\end{equation}
with $D_L(z,\Omega_M, \Omega_{\Lambda})$ the luminosity distance given by
\begin{equation}
D_L(z,\Omega_M, \Omega_{\Lambda}) = \frac{c(1+z)}{H_0}\int^{z}_0 \frac{dz'}{\sqrt{\Omega_M (1+z')^3+\Omega_{\Lambda}}},
\label{dl}
\end{equation}
where $\Omega_M$ and $\Omega_{\Lambda}$ are the matter and dark energy density parameters, $H_0$ is the present-day Hubble constant, and $z$ is the redshift. Similarly, $L_{\rm iso}$ is given by
\begin{equation}
L_{\rm iso} = 4\pi D_L(z,\Omega_M, \Omega_{\Lambda})^2 F_{\rm tot}.
\end{equation}

\item $S_{\gamma}$, $S_{\rm obs}$, $S_{\rm tot}$ indicate the prompt fluence in the whole gamma band (i.e., from a few hundred keV to a few MeV), the observed fluence in the range $50-300\,{\rm keV}$ and the total fluence in the $20\,{\rm keV}-1.5\,{\rm MeV}$ energy band.

\item $V$ is the variability of the GRB's light curve. It is computed by taking the difference between the observed light curve and its smoothed version, squaring this difference, summing these squared differences over time intervals, and appropriately normalizing the resulting sum \citep{reichart2001}. Different smoothing filters may be applied (see also \citealt{li2006} for a different approach). $V_f$ denotes the variability for a certain fraction of the smoothing timescale in the light curve.
\end{itemize}

Most of the quantities described above are given in the observer frame, except for $E_{\rm iso}$, $E_{\rm prompt}$, $L_{\rm peak}$ and $L_{\rm iso}$, which are already defined in the rest frame. With the upper index ``$*$'' we explicitly denote the observables in the GRB rest frame. The rest frame times are the observed times divided by the cosmic time expansion, for example the rest frame time in the prompt phase is denoted with $T^*_p=T_p/\left(1+z\right)$. The energetics are transformed differently, e.g. $E^*_{\rm peak}=E_{\rm peak}(1+z)$.

The Band function \citep{Band1993} is a commonly applied phenomenological spectral profile, such that
\begin{equation}
N_E(E)=A_{\rm norm}\times \left\{
\begin{array}{ll}
\left(\frac{E}{100\,{\rm keV}}\right)^\alpha \exp\left(-\frac{E}{E_0}\right), & E \leq (\alpha-\beta)E_0  \\
\left[\frac{(\alpha-\beta)E_0}{100\,{\rm keV}}\right]^{\alpha-\beta} \left(\frac{E}{100\,{\rm keV}}\right)^\beta \exp (\alpha-\beta), & E \geq (\alpha-\beta)E_0 \\
\end{array}
\right.
\end{equation}
where $A_{\rm norm}$ is the normalization. Here, $\alpha$ and $\beta$ are the low- and high-energy indices of the Band function, respectively. $N_E(E)$ is in units of ${\rm photons}\,{\rm cm}^{-2}\,{\rm s}^{-1}\,{\rm keV}^{-1}$. For the cases $\beta<-2$ and $\alpha>-2$, the $E_{\rm peak}$ can be derived as $E_{\rm peak}=(2+\alpha)E_0$, which corresponds to the energy at the maximum flux in the $\nu F_\nu$ spectra \citep{Band1993,yonetoku04}.

The Pearson correlation coefficient \citep{Kendall,Pearson} is denoted with $r$, the Spearman correlation coefficient \citep{Spearman} with $\rho$, and the $p$-value (a probability that a correlation is drawn by chance) is denoted with $P$.

Finally, we mostly deal with correlations of the form $y=ax+b$. However, when the intercept $b$ is neglected in the text, but its value is non-negligible (or not known due to lacking in the original paper), we use the notation $y\sim ax$ to emphasize the slope.

\section{The Prompt Correlations}\label{Prompt correlations}
Several physical relations between relevant quantities in GRBs were found since the 1990's. In each paragraph below we follow the discovery of the correlation with the definition of the quantities, the discussions presented in literature and their physical interpretation.

\subsection{The $L_{\rm peak}-\tau_{\rm{lag}}$ correlation}\label{L-lag time}

\subsubsection{Literature overview}

\citet{liang96}, using 34 bright GRBs detected by BATSE, found that $E_{\rm peak}$ depends linearly on the previous flux emitted by the pulse, i.e. that the rate of change of $E_{\rm peak}$ is proportional to the instantaneous luminosity. Quantitatively,
\begin{equation}
\frac{L_{\rm peak}}{N}=-\frac{{\rm d}E_{\rm peak}}{{\rm d}t},
\label{eqLK}
\end{equation}
where $N$ is a normalization constant expressing the luminosity for each pulse within a burst, and $L_{\rm peak}$ was calculated from the observed flux via Eq.~(\ref{eq1}).

\begin{figure}[htbp]
\centering
\includegraphics[width=7.9cm,height=7cm,angle=0]{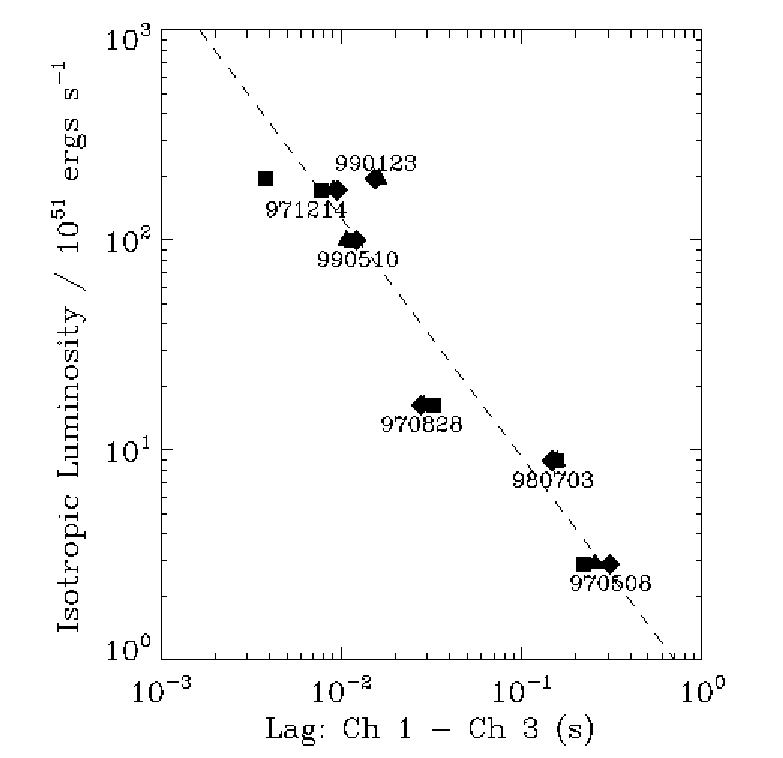}
\includegraphics[width=8cm,height=7cm,angle=0]{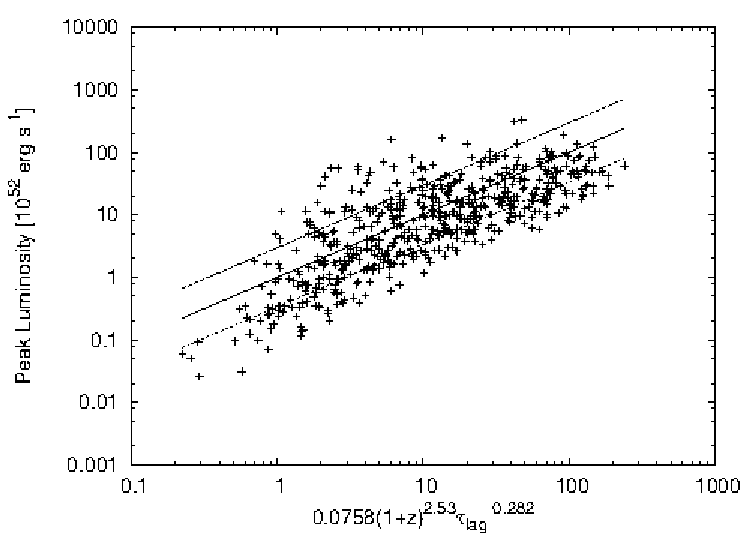}
\caption{\footnotesize {\bf Left panel:} $L_{\rm peak}$ vs. $\tau^*_{\rm lag}$  distribution for six GRBs with measured 
redshifts. The dashed line represents the power law fit to the lag times for ranges consisting 
of count rates larger than 0.1 $\times$ peak intensity (squares), 
yielding $\log L_{\rm peak} \sim -1.14\log(\tau^*_{\rm lag}/0.01\,{\rm s})$. The lag time is computed using channel 
1 ($25-50\,{\rm keV}$) and channel 3 ($100-300\,{\rm keV}$) of the BATSE instrument. (Figure after \cite{norris2000}; 
see Fig. 6 therein. @ AAS. Reproduced with permission.) 
{\bf Right panel:} The $L_{\rm peak}-\tau_{\rm lag}$ distribution in the $\log L_{\rm peak}$ vs. 
$\sim 2.53 \log (1 + z)-0.282 \log \tau_{\rm lag}$ plane. The correlation coefficient is 
$\rho=0.77$, $P=7.9 \times 10^{-75}$. The solid line represents the best fit line and two dashed lines delineate 
$1\sigma$ deviation. (Figure after \cite{tsutsui2008}; see Fig. 4 therein. Copyright @ 2008 AIP Publishing.)}
\label{fig:1}
\end{figure}

The $L_{\rm peak}-\tau_{\rm{lag}}$ correlation was introduced for the first time by \citet{norris2000} who examined a sample of 174 GRBs detected by BATSE, among which 6 GRBs had an established redshift and those were used to find an anticorrelation between $L_{\rm peak}$ and $\tau_{\rm{lag}}$ in the form of (see the left panel of Fig.~\ref{fig:1})
\begin{equation}
\log L_{\rm peak} = 55.11-1.14 \log\tau^*_{\rm lag},
\end{equation}
with $L_{\rm peak}$, in units of $10^{53}\,{\rm erg}\,{\rm s}^{-1}$, computed in the $50-300\,{\rm keV}$ range, and $\tau^*_{\rm lag}$ is measured in seconds. A remarkably consistent relation was found by \citet{schaefer01}, who used a sample of 112 BATSE GRBs and reported that
\begin{equation}
\log L_{\rm peak}=52.46-(1.14 \pm 0.20)\log \tau_{\rm lag},
\end{equation}
being in perfect agreement with the result of \citet{norris2000}. Here, $L_{\rm peak}$ is in units of $10^{51}\,{\rm erg}\,{\rm s}^{-1}$, and $\tau_{\rm lag}$ in seconds. This relation has been confirmed by several studies (e.g. \citealt{salmonson2000,daigne03,zhang06}).

\citet{schaefer2004} showed that the $L_{\rm peak}-\tau_{\rm{lag}}$ relation is a consequence of the \citet{liang96} empirical relation from Eq.~(\ref{eqLK}), and he derived this dependence to be of the form $\log L_{\rm peak} \sim -\log \tau_{\rm lag}$. This correlation was useful in the investigation of \citet{kocevski03}, who used a sample of 19 BATSE GRBs and the $L_{\rm peak}-\tau_{\rm lag}$ relation from \citep{schaefer01} to infer their pseudo-redshifts. Their approach was to vary the guessed $z$ until it allowed to match the luminosity distance $D_L$ measured with the GRB's energy flux and the $D_L$ that can be calculated from the guessed redshift within a flat $\Lambda$CDM model, until the agreement among the two converged to within $10^{-3}$. Next, the rate of $E_{\rm peak}$ decay, as in \citep{liang96}, was measured. Finally, \citet{kocevski03} showed that the $L_{\rm peak}$ is directly related to the GRB's spectral evolution. However, \citet{hakkila08} found a different slope, $-0.62\pm 0.04$, and argued that the $L_{\rm peak}-\tau_{\rm{lag}}$ relation is a pulse rather than a burst property, i.e. each pulse is characterized by its own $\tau_{\rm{lag}}$, distinct for various pulses within a GRB.

\citet{tsutsui2008}, using pseudo-redshifts estimated via the Yonetoku relation (see Sect.~\ref{Yonetoku}) for 565 BATSE GRBs, found that the $L_{\rm peak}-\tau_{\rm lag}$ relation has a $\rho$ of only $0.38$ (see the right panel of Fig. \ref{fig:1}). However, assuming that the luminosity is a function of both the redshift and the lag, a new redshift-dependent $L_{\rm peak}-\tau_{\rm{lag}}$ relation was found as
\begin{equation}
\log L_{\rm peak} = 50.88 + 2.53 \log (1 + z)-0.282 \log\tau_{\rm lag},
\end{equation}
with $L_{\rm peak}$ in units of $10^{50}\,{\rm erg}\,{\rm s}^{-1}$, $\tau_{\rm lag}$ in seconds, $\rho=0.77$ and $P=7.9 \times 10^{-75}$. Although the spectral lag is computed from two channels of BATSE, this new $L_{\rm peak}-\tau_{\rm{lag}}$ relation suggests that a future lag-luminosity relation defined within the {\it Swift} data should also depend on the redshift.

Afterwards, \citet{Sultana2012} presented a relation between the $z$- and $k$-corrected $\tau_{\rm{lag}}$ for the {\it Swift} energy bands $50-100$ keV and $100-200$ keV, and $L_{\rm{peak}}$, for a subset of 12 {\it Swift} long GRBs. The $z$-correction takes into account the time dilatation effect by multiplying the observed lag by $(1+z)^{-1}$ to translate it into the rest frame. The $k$-correction takes into account a similar effect caused by energy bands being different in the observer and rest frames via multiplication by $(1+z)^{0.33}$ \citep{gehrels06}. The net corrected $\tau^*_{\rm lag}$ is thence $(1+z)^{-0.67}\tau_{\rm lag}$. In addition, \cite{Sultana2012} demonstrated that this correlation in the prompt phase can be extrapolated into the $L_a-T^*_a$ relation \citep{Dainotti2008,Dainotti2010,dainotti11a,Dainotti2013a}. \citet{Sultana2012} found\footnote{Note that \citet{Sultana2012} used $L_{\rm iso}$ to denote the peak isotropic luminosity.}:
\begin{equation}
\log L_{\rm peak}=(54.87 \pm 0.29)-(1.19 \pm 0.17) \log\left[(1+z)^{-0.67} \tau_{\rm lag}\right],
\end{equation}
and
\begin{equation}
\log L_a=(51.57 \pm 0.10)-(1.10 \pm 0.03) \log T^*_a,
\end{equation}
with $\tau_{\rm lag}$ in ms, $T^*_a$ in seconds, and $L$ in ${\rm erg}\,{\rm s}^{-1}$. The correlation coefficient is significant for these two relations ($\rho = -0.65$ for the $L_{\rm{peak}}-\tau_{\rm{lag}}$ and $\rho = -0.88$ for the $L_a-T^{*}_a$ relations) and have surprisingly similar best-fit power law indices ($-1.19 \pm 0.17$ and $-1.10 \pm 0.03$, respectively). Although $\tau_{\rm{lag}}$ and $T^{*}_a$ represent different GRB time variables, it appears distinctly that the $L_{\rm{peak}}-\tau_{\rm{lag}}$ relation extrapolates into $L_a-T^{*}_a$ for timescales $\tau_{\rm{lag}} \simeq T^{*}_a$. A discussion and comparison of this extrapolation with the $L_f-T_f$ relation is extensively presented in \citep{Dainotti2015b}.

\citet{ukwatta10} confirmed that there is a correlation between $L^*_{\rm peak}$ and the $z$- and $k$-corrected $\tau_{\rm{lag}}$ among 31 GRBs observed by {\it Swift}, with $r=-0.68$, $P=7\times 10^{-2}$ and the slope equal to $-1.4\pm 0.3$, hence confirming the $L_{\rm peak}-\tau_{\rm{lag}}$ relation, although with a large scatter. This was followed by another confirmation of this correlation \citep{ukwatta12} with the use of 43 {\it Swift} GRBs with known redshift, which yielded $r=-0.82$, $P=5.5\times 10^{-5}$, and a slope of $-1.2\pm 0.2$, being consistent with the previous results.

Finally, \citet{margutti2010} established that the X-ray flares obey the same $L_{\rm{peak}}-\tau^*_{\rm{lag}}$ relation (in the rest-frame energy band $0.3-10\,{\rm keV}$) as GRBs, and proposed that their underlying mechanism is similar.

\subsubsection{Physical interpretation of the $L_{\rm peak}-\tau_{\rm lag}$ relation}\label{PhysicalofLpeak-taulag}

The physical assumption on which the work by \cite{norris2000} was based is that the initial mechanism for the energy formation affects the development of the pulse much more than dissipation. From the study of several pulses in bright, long BATSE GRBs, it was claimed that for pulses with precisely defined shape, the rise-to-decay ratio is $\leq 1$. In addition, when the ratio diminishes, pulses show a tendency to be broader and weaker.

\citet{salmonson2000} proposed that the $L_{\rm peak}-\tau_{\rm{lag}}$ relation arises from an entirely kinematic effect. In this scenario, an emitting region with constant (among the bursts) luminosity is the source of the GRB's radiation. He also claimed that variations in the line-of-sight velocity should affect the observed luminosity proportionally to the Lorentz factor of the jet's expansion, $\Gamma=\left[1-(v/c)^2\right]^{-1/2}$ (where $v$ is the relative velocity between the inertial reference frames and $c$ is the speed of light), while the apparent $\tau_{\rm lag}$ is proportional to $1/\Gamma$. The variations in the velocity among the line-of-sight is a result of the jet's expansion velocity combined with the cosmological expansion. The differences of luminosity and lag between different bursts are due to the different velocities of the individual emitting regions. In this case, the luminosity is expected to be proportional to $1/\tau_{\rm lag}$, which is consistent with the observed relation. This explanation, however, requires the comoving luminosity to be nearly constant among the bursts, which is a very strong condition to be fulfilled. Moreover, this scenario has several other problems (as pointed out by \citealt{schaefer2004}):
\begin{enumerate}
\item it requires the Lorentz factor and luminosity to have the same range of variation. However, the observed $L_{\rm peak}$ span more than three orders of magnitude (e.g., \citealt{schaefer01}), while the Lorentz factors span less than one order of magnitude (i.e., a factor of 5) \citep{panaitescu02};
\item it follows that the observed luminosity should be linearly dependent on the jet's Lorentz factor, yet this claim is not justified. In fact, a number of corrections is to be taken into account, leading to a significantly nonlinear dependence. The forward motion of the jet introduces by itself an additional quadratic dependence \citep{fenimore96}.
\end{enumerate}

\citet{ioka01} proposed another interpretation for the $L_{\rm peak}-\tau_{{\rm lag}}$ correlation. 
From their analysis, a model in which the peak luminosity depends on the viewing angle is elaborated: the viewing angle is the off-axis angular position from which the observer examines the emission. Indeed, it is found that a high-luminosity peak in GRBs with brief spectral lag is due to an emitted jet with a smaller viewing angle than a fainter peak with extended lag. It is also claimed that the viewing angle can have implications on other correlations, such as the luminosity-variability relation presented in Sect.~\ref{var}. As an additional result from this study, it was pointed out that XRFs can be seen as GRBs detected from large angles with high spectral lag and small variability.

On the other hand, regarding the jet angle distributions, \citet{liang2008} found an anticorrelation between the jet opening angle and the isotropic kinetic energy among 179 X-ray GRB light curves and the afterglow data of 57 GRBs. Assuming that the GRB rate follows the star formation rate, and after a careful consideration of selection effects, \citet{lu12jet} found in a sample of 77 GRBs an anticorrelation between the jet opening angle $\theta_{\rm jet}$ and the redshift in the form
\begin{equation}
\log \theta_{\rm jet}=(-0.90\pm 0.09)-(0.94\pm0.19)\log (1+z),
\end{equation}
with $\rho=0.55$ and $P<10^{-4}$. Using a mock sample and bootstrap technique, they showed that the observed $\theta_{\rm jet}-z$ relation is most likely due to instrumental selection effects. Moreover, they argued that other types of relation, e.g. $\tau_{\rm lag}-z$ \citep{yi08} or the redshift dependence of the shallow decays in X-ray afterglows \cite{stratta09}, while might have connections with the jet geometry, are also likely to stem from observational biases or sample selection effects. Also, \citet{ryan2015} investigated the jet opening angle properties using a sample of 226 {\it Swift}/XRT GRBs with known redshift. They found that most of the observed afterglows were observed off-axis, hence the expected behaviour of the afterglow light curves can be significantly affected by the viewing angle.

\citet{zhang09} argued, on the basis of the kinematic model, that the origin of the $L_{\rm peak}-\tau_{\rm lag}$ relation is due to a more intrinsic $L_{\rm peak}-V$ relation (see Sect.~\ref{var}). They also gave an interpretation of the latter relation within the internal shock model (see Sect.~\ref{intvar}). Recently, \citet{uhm16} constructed a model based on the synchrotron radiation mechanism that explains the physical origin of the spectral lags and is consistent with observations.

Another explanation for the origin of the $L_{\rm peak}-\tau_{\rm{\rm lag}}$ relation, given by \cite{Sultana2012}, involves only kinematic effects. In this case, $L_{\rm peak}$ and $\tau_{\rm lag}$ depend on the quantity:
\begin{equation}
D=\frac{1}{\Gamma(1-\beta_0 \cos\theta)(1+z)},
\label{doppler}
\end{equation}
depicting the Doppler factor of a jet at a viewing angle $\theta$ and with velocity $\beta_0\equiv v/c$ at redshift $z$. In this study there is no reference to the masses and forces involved and, as a consequence of the Doppler effect, the factor $D$ associates the GRB rest frame timescale $\tau$ with the observed time $t$ in the following way:
\begin{equation}
t = \frac{\tau}{D}.
\label{tfrac}
\end{equation}
Therefore, considering a decay timescale $\Delta \tau$ in the GRB rest frame, Eq.~(\ref{tfrac}) in the observer frame will give $\Delta t = \Delta \tau/D$. Furthermore, taking into account a spectrum given by $\Phi(E) \propto E^{-\alpha}$, the peak luminosity (as already pointed out by \citealt{salmonson2000}) can be computed as
\begin{equation}
L_{\rm peak} \propto D^{\alpha}, 
\label{lpeak}
\end{equation}
with $\alpha \approx 1$. In such a way, Eqs.~(\ref{tfrac}) and (\ref{lpeak}) allow to retrieve the observed $L_{\rm peak}-\tau_{\rm{\rm lag}}$ relation. Finally, the analogous correlation coefficients and best-fit slopes of the $L_{\rm peak}-\tau_{\rm lag}$ and $L_a-T^{*}_a$ correlations obtained by \cite{Sultana2012} seem to hint toward a similar origin for these two relations.

\subsection{The $L_{\rm peak}-V$ correlation}\label{var}

The first correlation between $L_{\rm peak}$ and $V$ was discovered by \citet{fenimore2000}, and was given as
\begin{equation}
\log L_{\rm peak} = 56.49 +3.35\log V,
\end{equation}
with $L_{\rm peak}$ measured in ${\rm erg}\,{\rm s}^{-1}$. Here, the luminosity is per steradian in a specified (rest frame) energy bandpass ($50-300$ keV), averaged over $256\,{\rm ms}$. First, seven BATSE GRBs with a measured redshift were used to calibrate the $L_{\rm peak}-V$ relation. Next, the obtained relationship was applied to 220 bright BATSE GRBs in order to obtain the luminosities and distances, and to infer that the GRB formation rate scales as $(1+z)^{3.3\pm 0.3}$. Finally, the authors emphasized the need of confirmation of the proposed $L_{\rm peak}-V$ relation.

\subsubsection{Literature overview}

\begin{figure}[htbp]
\centering
\includegraphics[width=5.2cm,height=4.3cm,angle=0]{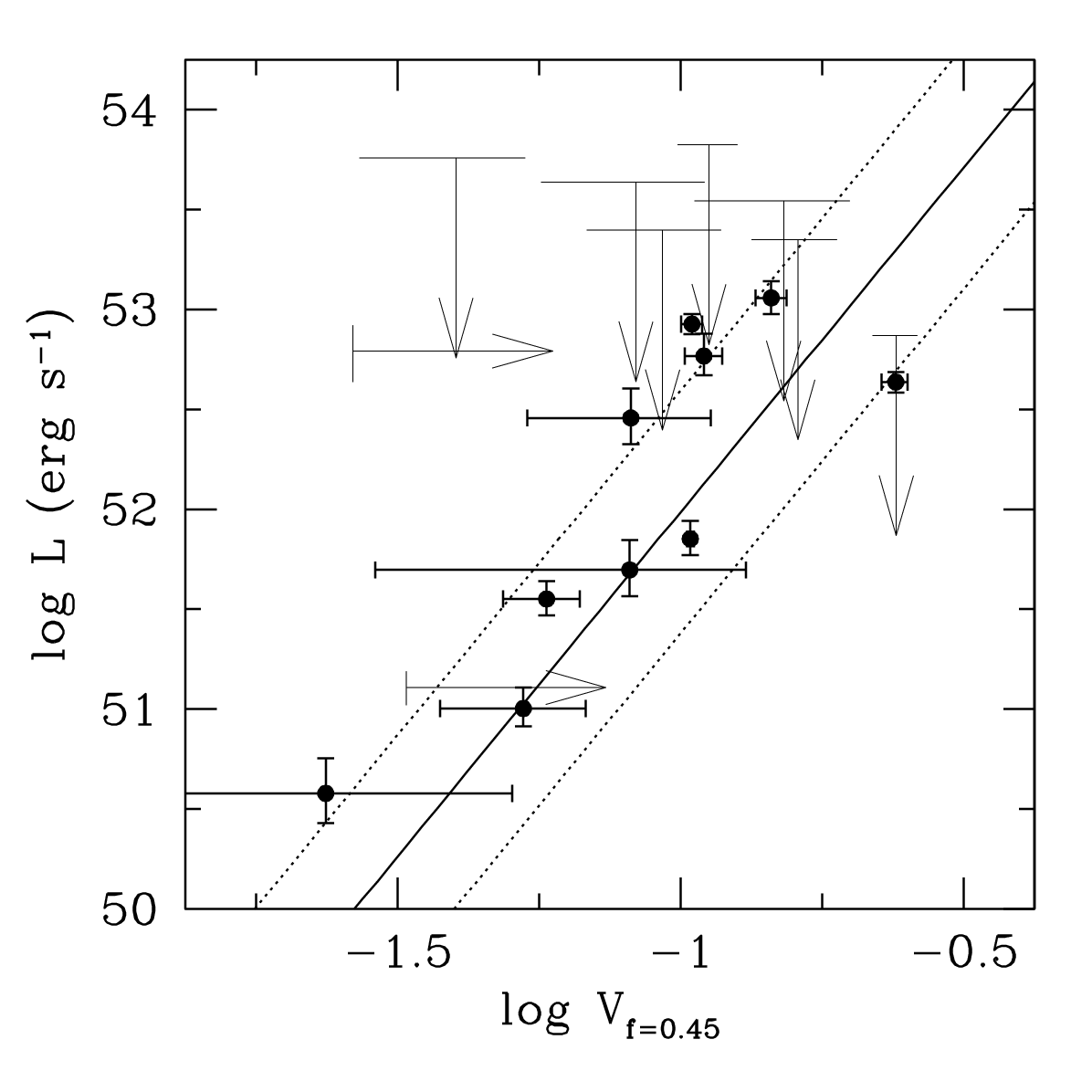}  
\hspace{0.2cm}
\includegraphics[width=5.2cm,height=4cm,angle=0]{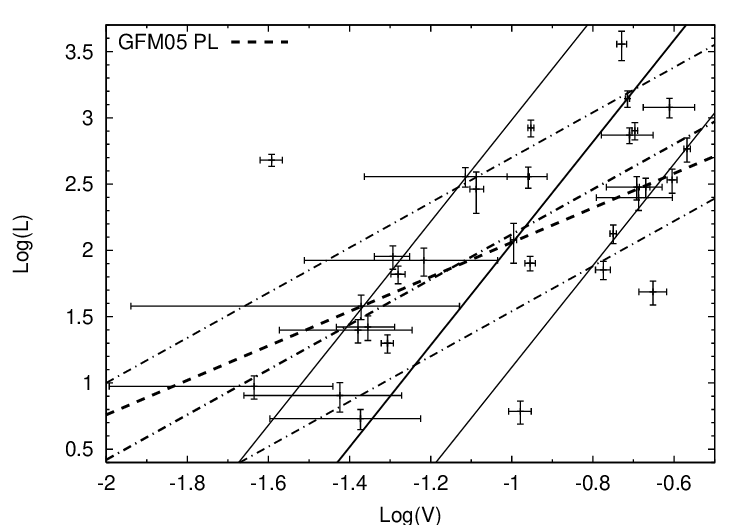}   
\includegraphics[width=5.2cm,height=4cm,angle=0]{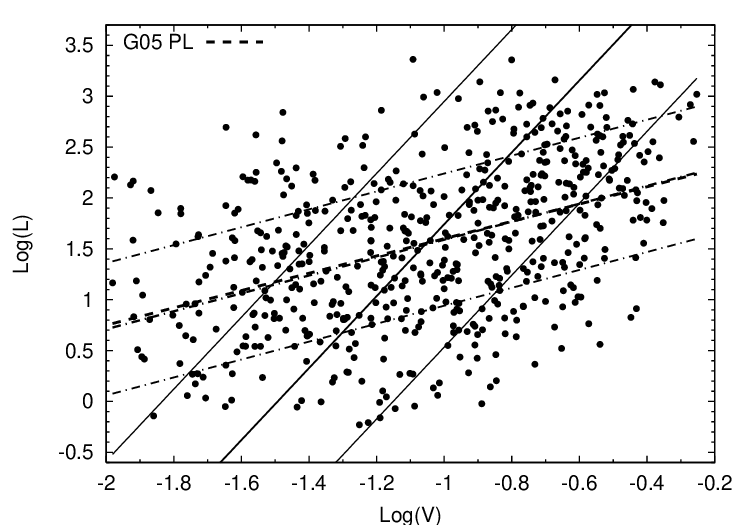} 
\vspace{0.8cm}
\caption{\footnotesize {\bf Left panel:} the variabilities $V_{f=0.45}$ and peak luminosities $L_{\rm peak}$ of the 
data set, excluding GRB980425. In this case, $V_{f=0.45}$ indicates the variabilities for the 
45\% smoothing timescale of the light curve. The solid and dotted lines are the best fit line and $1\sigma$ deviation 
respectively in the $\log L_{\rm peak}-\log V_{f=0.45}$ plane. (Figure after \cite{reichart2001}; see Fig. 9 therein. @ AAS. Reproduced with permission.)
{\bf Middle panel:} The $\log L_{\rm peak}-\log V$ plane for the sample of 32 GRBs with measured redshift. 
The best fit lines and $1\sigma$ deviations are also displayed: solid lines are computed with 
the \citet{reichart2001} method, dashed-dotted lines with the \citet{Dagostini2005} method and the dashed lines are 
recovered by \citet{guidorzi2005}. (Figure after \cite{guidorzi2006}; see Fig. 1 therein.)
{\bf Right panel:} $\log L_{\rm peak}-\log V$ relation for the set of 551 BATSE GRBs. 
The best-fit lines and $1\sigma$ regions are also shown, the solid lines are fitted with the \citet{reichart2001} 
method, the dashed-dotted lines with the  \citet{Dagostini2005} method and the dashed lines are recovered by 
\cite{guidorzi2005}. (Figure after \cite{guidorzi2006}; see Fig. 2 therein.)}
\label{fig:3}
\end{figure}

\citet{reichart2001} used a total of 20 GRBs observed by {\it CGRO}/BATSE (13 bursts), the {\it KONUS}/Wind (5 bursts), the {\it Ulysses}/GRB (1 burst), and the {\it NEAR}/XGRS (1 burst), finding:
\begin{equation}
\log L_{\rm peak} \sim (3.3^{+1.1}_{-0.9}) \log V.
\end{equation} 
with $\rho=0.8$ and $P=1.4 \times 10^{-4}$ (see the left panel of Fig.~\ref{fig:3}); $L_{\rm peak}$ was computed in the $50-300\,{\rm keV}$ observer-frame energy band, which corresponds roughly to the range $100-1000\,{\rm keV}$ in the rest frame for $z\simeq 1-2$, typical for GRBs in the sample examined. The distribution of the sample's bursts in the $\log L_{\rm peak}-\log V_f$ plane appears to be well modeled by the following parameterization:
\begin{equation}
\log V_f (L) = \log\bar{V}_f + b + m (\log L_{\rm peak}-\log \bar{L}_{\rm peak}),
\label{var2}
\end{equation}
where $b=0.013^{+0.075}_{-0.092}$ is the intercept of the line, $m=0.302^{+0.112}_{-0.075}$ is its slope, and $\bar{V}_f$ and  $\bar{L}_{\rm peak}$ are the median values of $V_f$ and $L_{\rm peak}$ for the bursts in the sample for which spectroscopic redshifts, peak fluxes, and 64-ms or better resolution light curves are available.

Later, \cite{guidorzi2005} updated the sample to 32 GRBs detected by different satellites, i.e. {\it BeppoSAX}, {\it CGRO}/BATSE, {\it HETE-2} and {\it KONUS} (see the middle panel of Fig.~\ref{fig:3}). The existence of a correlation was confirmed, but they found a dramatically different relationship with respect to the original one:
\begin{equation}
\log L_{\rm peak} = 3.36^{+0.89}_{-0.43}+1.30^{+0.84}_{-0.44} \log V,
\end{equation}
with $\rho=0.625$ and $P=10^{-4}$, and $L_{\rm peak}$ in units of $10^{50}\,{\rm erg}\,{\rm s}^{-1}$.

However, \citet{reichart2005} using the same sample claimed that this result was the outcome of an improper statistical methodology, and confirmed the previous work of \citet{reichart2001}. Indeed, they showed that the difference among their results and the ones from \cite{guidorzi2005} was due to the fact that the variance of the sample in the fit in \citep{guidorzi2005} was not taken into account. They used an updated data set, finding that the fit was well described by the slope $m=3.4^{+0.9}_{-0.6}$, with a sample variance $\sigma_V=0.2 \pm 0.04$.

Subsequently, \citet{guidorzi2006} using a sample of 551 BATSE GRBs with pseudo-redshifts derived using the $L_{\rm peak}-\tau_{\rm lag}$ relation \citep{guidorzi05b}, tested the $L_{\rm peak}-V$ correlation (see right panel of Fig.~\ref{fig:3}). They also calculated the slope of the correlation of the samples using the methods implemented by \citet{reichart2001} and \citet{Dagostini2005}. The former method provided a value of the slope in the $L_{\rm peak}-V$ correlation consistent with respect to the previous works:
\begin{equation}
\log L_{\rm peak} \sim 3.5^{+0.6}_{-0.4} \log V.
\end{equation}
Instead, the slope for this sample using the latter method is much lower than the value in \citep{reichart2001}:
\begin{equation}
\log L_{\rm peak} \sim 0.88^{+0.12}_{-0.13} \log V.
\end{equation}
The latter slope $m$ is consistent with the results obtained by \citet{guidorzi2005}, but inconsistent with the results derived by \citet{reichart2005}.

Afterwards, \citet{rizzuto2007} tested this correlation with a sample of 36 LGRBs detected by {\it Swift} in the $15-350$ keV energy range and known redshifts. The sample consisted of bright GRBs with $L_{\rm peak} > 5 \times 10^{50}\,{\rm erg}\,{\rm s}^{-1}$ within a $100-1000\,{\rm keV}$ energy range. In their study, they adopted two definitions of variability, presented by \citet{reichart2001}, called $V_{\rm R}$, and by \citet{li2006}, hereafter $V_{\rm LP}$. $V_{\rm R}$ and $V_{\rm LP}$ differ from each other with a different smoothing filter which, in the second case, selects only high-frequency variability. Finally, \cite{rizzuto2007} confirmed the correlation and its intrinsic dispersion around the best-fitting power law given by
\begin{equation}
\log L_{\rm peak} \sim (2.3 \pm 0.17) \log V_{\rm LP}, 
\end{equation}
with $\rho=0.758$ and $P=0.011$, and

\begin{equation}
\log L_{\rm peak} \sim (1.7 \pm 0.4) \log V_{\rm R}, 
\end{equation}
with $\sigma_{\log L} = 0.58^{+0.15}_{-0.12}$, $\rho=0.115$, and $P=0.506$. Six low-luminosity GRBs (i.e., GRB050223, GRB050416A, GRB050803, GRB051016B, GRB060614 and GRB060729), out of a total of 36 in the sample, are outliers of the correlation, showing values of $V_{\rm R}$ higher than expected. Thus, the correlation is not valid for low-luminosity GRBs.

As is visible from this discussion, the scatter in this relation is not negligible, thus making it less reliable than the previously discussed ones. However, investigating the physical explanation of this correlation is worth to be depicted for further developments.

\subsubsection{Physical interpretation of the $L_{peak}-V$ relation}\label{intvar}

We here briefly describe the internal and external shock model \citep{100,98}, in which the GRB is caused by emission from a relativistic, expanding baryonic shell with a Lorentz bulk factor $\Gamma$. Let there be a spherical section with an opening angle $\theta_{\rm jet}$. In general, $\theta_{\rm jet}$ can be greater than $\Gamma^{-1}$, but the observer can detect radiation coming only from the angular region with size $\simeq\Gamma^{-1}$. An external shock is formed when the expanding shell collides with the external medium. In general, there might be more than one shell, and the internal shock takes place when a faster shell reaches a slower one. In both cases one distinguishes an FS, when the shock propagates into the external shell or the external medium, and a reverse shock (RS), when it propagates into the inner shell.

\citet{fenimore2000} pointed out that the underlying cause of the $L_{\rm peak}-V$ relation is unclear. In the context of the internal shock model, larger initial $\Gamma$ factors tend to produce more efficient collisions. After changing some quantities such as the $\Gamma$ factors, the ambient density, and/or the initial mass of the shells, the observed variability values are not recovered. Therefore, the central engine seems to play a relevant role in the explanation for the observed $L_{\rm peak}-V$ correlation. In fact, this  correlation was also explored within the context of a model in which the GRB variability is due to a change in the jet-opening angles and narrower jets have faster outflows \citep{salmonson2002}. As a result, this model predicts bright luminosities, small pulse lags and large variability as well as an early jet break time for on-axis observed bursts. On the other hand, dimmer luminosities, longer pulse lags, flatter bursts and later jet break times will cause larger viewing angles.

\citet{guidorzi2006} gave an interpretation for the smaller value of the correlation in the context of the jet-emission scenario where a stronger dependence of the $\Gamma$ of the expanding shells on the jet-opening angle is expected. 
However, \citet{schaefer2007} attributed the origin of the $L_{\rm peak}-V$ relation to be based on relativistically shocked jets. Indeed, $V$ and $L_{\rm peak}$ are both functions of $\Gamma$, where $L_{\rm peak}$ is proportional to a high power of $\Gamma$, as was already demonstrated in the context of the  $L_{\rm peak}-\tau_{\rm lag}$ relation (see Sect. \ref{PhysicalofLpeak-taulag}), and hence fast rise times and short pulse durations imply high variability.

\subsection{The $L_{\rm iso}-\tau_{\rm RT}$ correlation and its physical interpretation}\label{LisotauRT}

\citet{schaefer02} predicted that $\tau_{\rm RT}$ should be connected with $L_{\rm iso}$ in a following manner:
\begin{equation}
L_{\rm iso}\propto\tau^{-N/2}_{\rm RT},
\end{equation}
with the exponent $N\simeq 3$ (see also \citealt{schaefer2002,schaefer2007}). Therefore, fast rises indicate high luminosities and slow rises low luminosities. The $\tau_{\rm{RT}}$ can be directly connected to the physics of the shocked jet. Indeed, for a sudden collision of a material within a jet (with the shock creating an individual pulse in the GRB light curve), $\tau_{\rm{RT}}$ will be determined as the maximum delay between the arrival time of photons from the center of the visible region versus their arrival time from its edge.
\begin{figure}[htbp]
\includegraphics[width=8.1cm,height=6.2cm,angle=0]{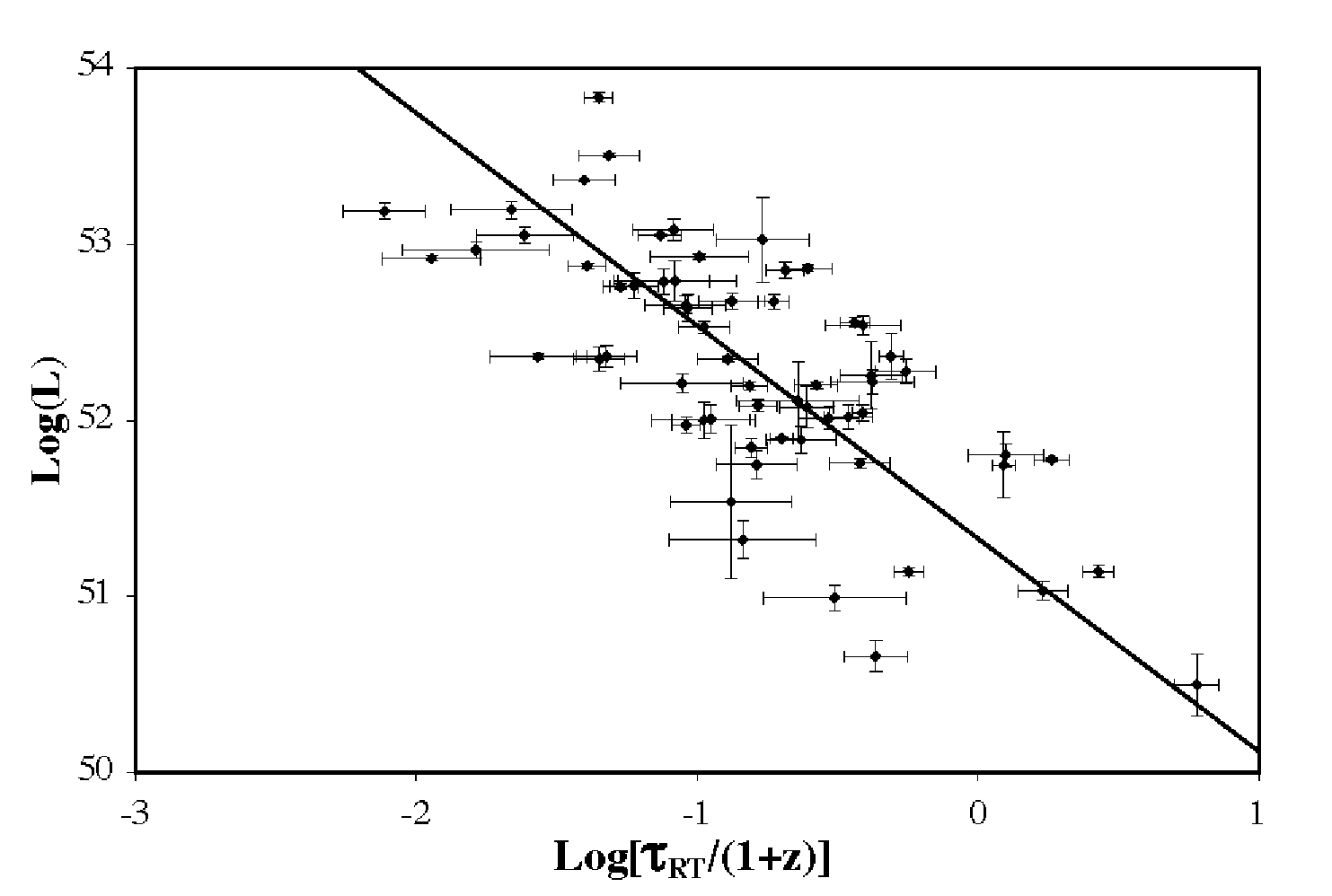}
\includegraphics[width=8.1cm,height=6cm,angle=0]{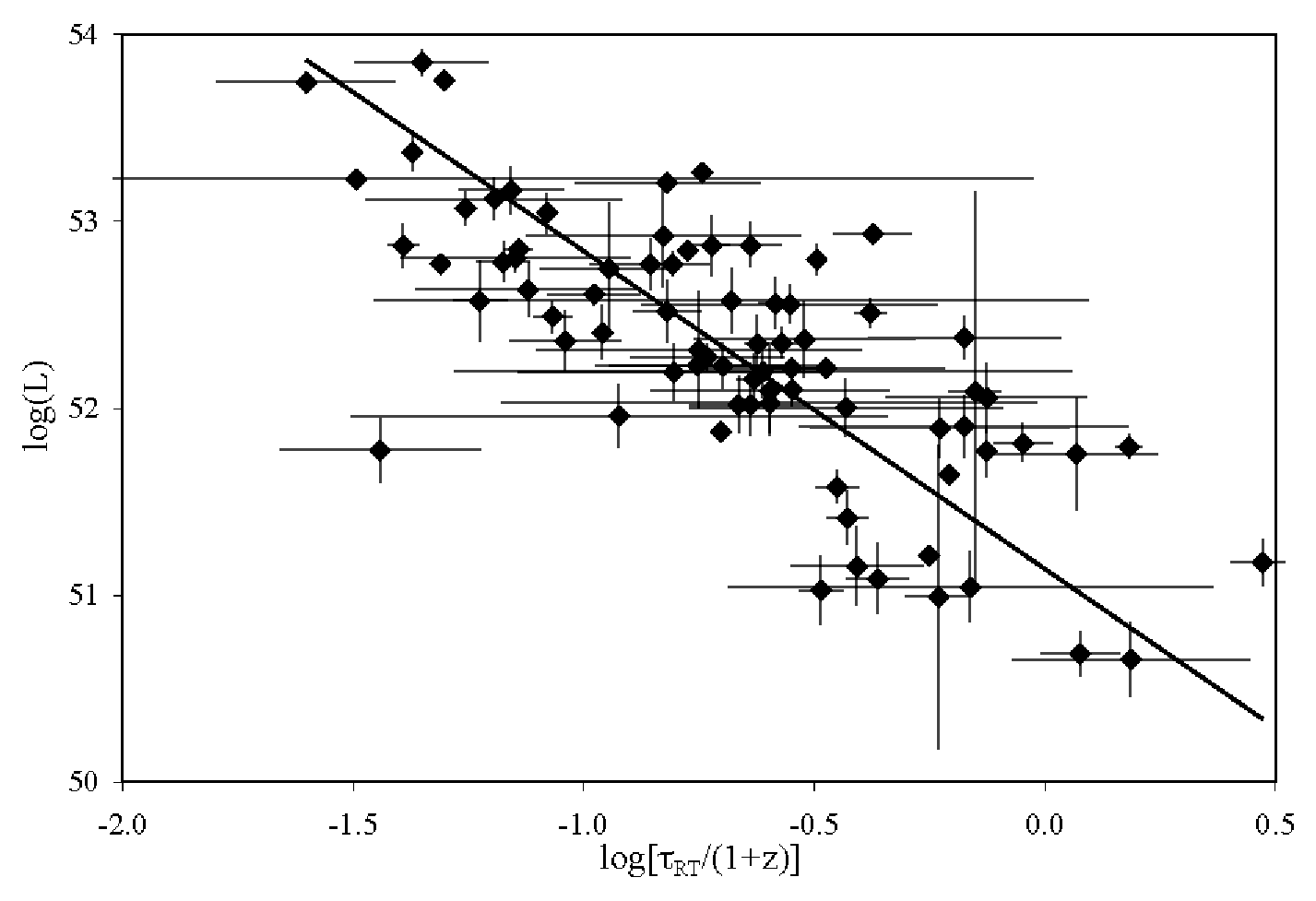}
\caption{\footnotesize {\bf Left panel:} the $\log L_{\rm iso}-\log \tau^*_{\rm RT}$ relation with the best fit line 
displayed. The errors are given by the $1\sigma$ confidence interval. (Figure after \cite{schaefer2007}; see 
Fig. 5 therein. @ AAS. Reproduced with permission.) 
{\bf Right panel:} the $\log L_{\rm iso}-\log \tau^*_{\rm RT}$ correlation with the best fit line. 
(Figure after \cite{xiao09}; see Fig. 3 therein. @ AAS. Reproduced with permission.)}
\label{fig:5}
\end{figure} 
 
The angular opening of the emitted jet, usually associated with $\Gamma$, could cause this delay leading to a relation $\tau_{\rm RT} \propto \Gamma^{-2}$. The radius at which the material is shocked affects the proportionality constant, and the minimum radius under which the material cannot radiate efficiently anymore should be the same for each GRB \citep{panaitescu02}. In addition, a large scatter is expected because it will depend on how near this minimum radius the collisions are observed.

With both $\tau_{\rm RT}$ and $L_{\rm iso}$ being functions of $\Gamma$, \citet{schaefer2007} confirmed that $\log L_{\rm iso}$ should be $\sim -N/2\log\tau_{\rm RT}$. From 69 GRBs detected by BATSE and {\it Swift}, the following relation was obtained:
\begin{equation}
\log L_{\rm iso} = 53.54 -1.21\log\tau^*_{\rm RT},
\label{equ}
\end{equation}
with $L_{\rm iso}$ in ${\rm erg}\,{\rm s}^{-1}$ and $\tau^*_{\rm RT}$ measured in seconds. The $1\sigma$ uncertainties in the intercept and slope are $\sigma_a=0.06$ and $\sigma_b=0.06$ (see the left panel of Fig. \ref{fig:5}). The uncertainty in the log of the burst luminosity is
\begin{equation}
\sigma_{\log L_{\rm iso}}^2 = \sigma_a^2 + \left[\sigma_b \log \frac{\tau^*_{\rm RT}}{0.1\,{\rm s}}\right]^2 + \left(\frac{0.43 b \sigma_{\rm RT}}{\tau_{\rm RT}}\right)^2 + \sigma_{\rm RT,sys}^2,
\end{equation}
where \citet{schaefer2007} takes into account the extra scatter, $\sigma_{\rm sys}$. When $\sigma_{\rm RT,sys}=0.47$, the $\chi^{2}$ of the best fit line is unity.

\citet{xiao09} explained in details the procedure of how they calculated the $\tau_{\rm{RT}}$ using 107 GRBs with known spectroscopic redshift observed by BATSE, {\it HETE}, {\it KONUS} and {\it Swift} (see the right panel of Fig. \ref{fig:5}), taking into account also the Poissonian noise. Their analysis yielded
\begin{equation}
\log L_{\rm iso} = 53.84-1.70\log\tau^*_{\rm RT},
\end{equation}
with the same units as in Eq.~(\ref{equ}). As a consequence, the flattening of the light curve before computing the rise time is an important step. The problem is that the flattening should be done carefully, in fact if the light curve is flattened too much, a rise time comparable with the smoothing-time bin is obtained, while if it is flattened not enough, the Poissonian noise dominates the apparent fastest rise time, giving a too small rise time. Therefore, for some of the dimmest bursts, the Poissonian-noise dominant region and the smoothing-effect dominant region can coincide, thus not yielding $\tau_{\rm RT}$ values for the weakest bursts. Finally, the physical interpretation of this correlation is given by \citet{schaefer2007}. It is shown that the fastest rise in a light curve is related to the Lorentz factor $\Gamma$ simply due to the geometrical rise time for a region subtending an angle of $1/\Gamma$, assuming that the minimum radius for which the optical depth of the jet material is of order of unity remains constant. The luminosity of the burst is also a power law of $\Gamma$, which scales as $\Gamma^{N}$ for $3<N<5$. Therefore, the $\tau_{\rm RT}-\Gamma$ and the $L_{\rm iso}-\Gamma$ relations together yield the observed $L_{\rm iso}-\tau_{\rm RT}$ relation.

\subsection{The $\Gamma_0-E_{\rm prompt}$ and $\Gamma_0-L_{\rm iso}$ correlations and their physical interpretation}

\citet{freedman01} in their analysis of the GRB emission, considering a relativistic velocity for the fireball, showed that the radiation detected by an observer is within an opening angle $\simeq 1/\Gamma(t)$. Hence, the total fireball energy $E$ should be interpreted as the energy that the fireball would have carried if this is assumed spherically symmetric. In particular, it was claimed that the afterglow flux measurements in X-rays gave a strong evaluation for the fireball energy per unit solid angle represented by $\epsilon_e =\xi_eE/4\pi$, within the observable opening angle $1/\Gamma(t)$, where $\xi_e$ is the electron energy fraction. It was found that
\begin{equation}
\Gamma(t) = 10.6\left(\frac{1+z}{2}\right)^{3/8}\left(\frac{E_{\rm prompt}}{n_0}\right)^{1/8}t^{-3/8},
\end{equation}
where $E_{\rm prompt}$ is in units of $10^{53}\,{\rm erg}$, $n_0$ is the uniform ambient density of the expanding fireball 
in units of ${\rm cm}^{-3}$, and $t$ is the time of the fireball expansion in days. Finally, it was pointed out that $\xi_e$ from the afterglow observations should be close to equipartition, namely $\xi_e\simeq\frac{1}{3}$. For example, for GRB970508 it was found that $\xi_e\simeq0.2$ \citep{waxman97,wijers99,granot99}. A similar conclusion, i.e. that it is also close to equipartition, could be drawn for GRB971214, however \citet{wijers99} proposed another interpretation for this GRB's data, demanding $\xi_e\simeq1$.

\begin{figure}[htbp]
\centering
\includegraphics[width=8.1cm,height=6.1cm,angle=0,clip]{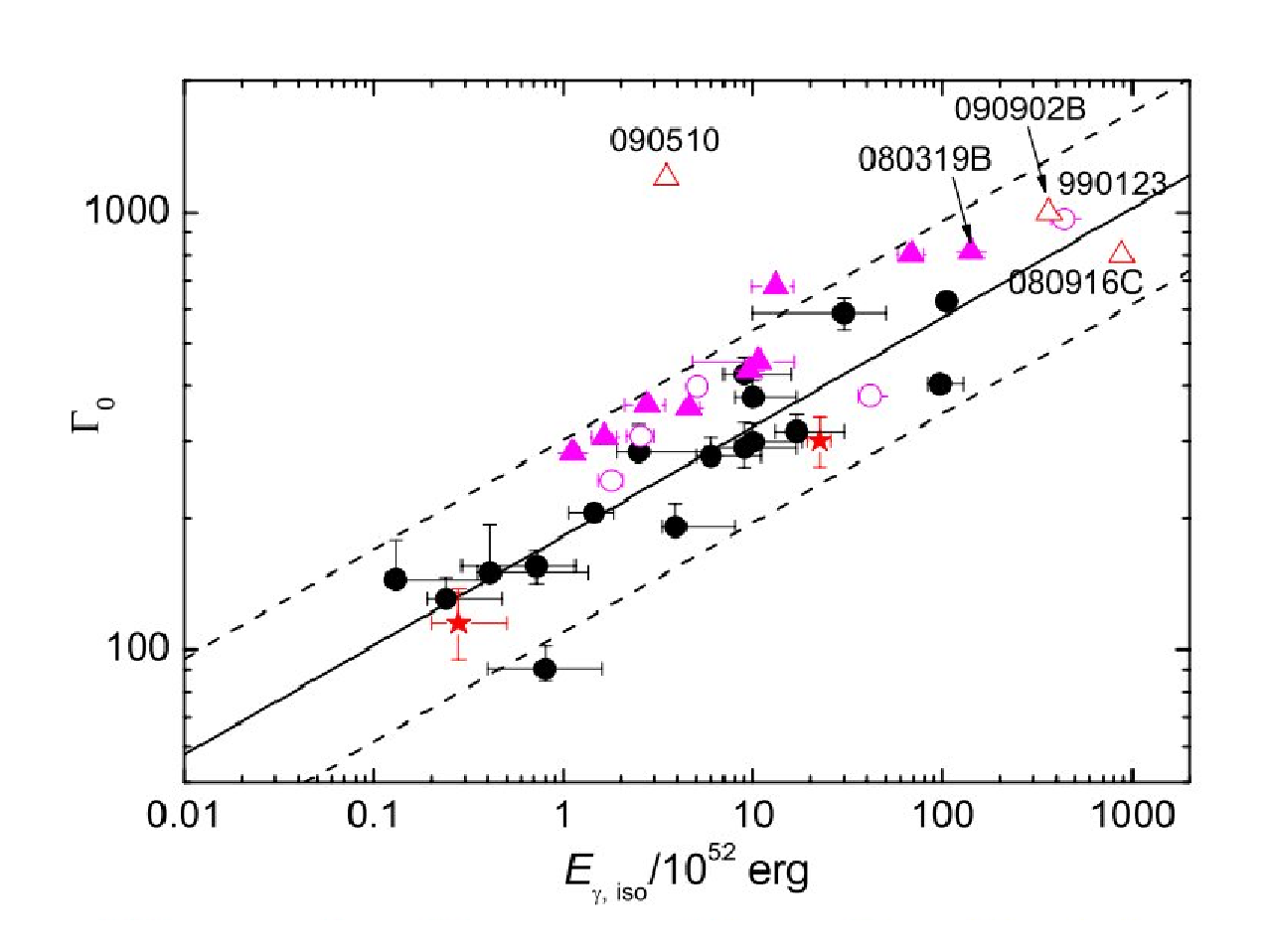}  
\includegraphics[width=8.1cm,height=6.2cm,angle=0,clip]{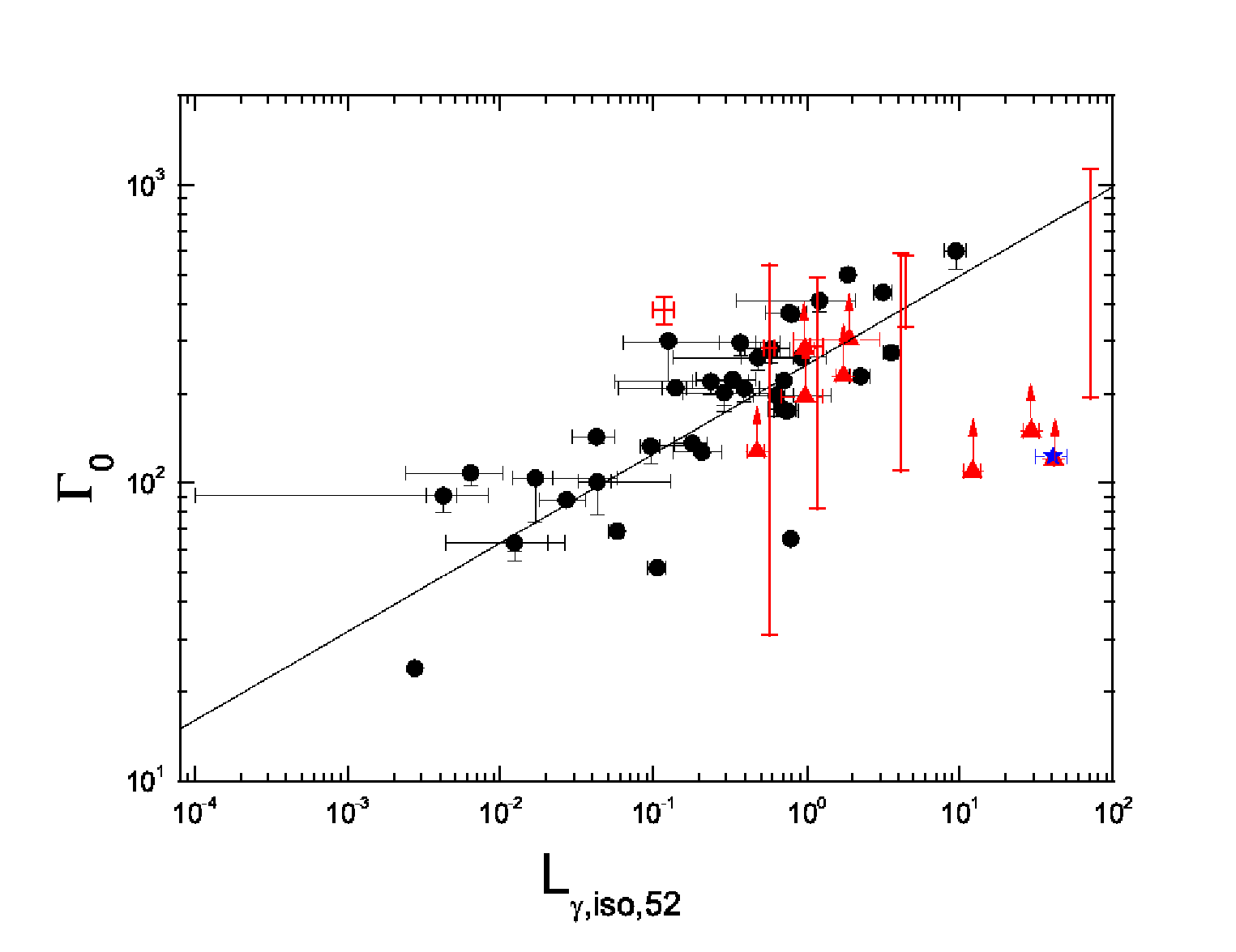} 
\caption{\footnotesize {\bf Left panel:} $\log \Gamma_0-\log E_{\rm prompt}$ relation with the addition of GRBs with an onset trend in the X-ray band (GRBs 070208 and 080319C; red stars), $\Gamma_0$ 
computed by RS peaks or probable afterglow peaks (pink open circles), lower values of $\Gamma_0$ obtained from single 
power law decay light curves (pink solid triangles), and strong lower values of $\Gamma_0$ for {\it Fermi}/LAT GRBs 
080916C, 090902B, and 090510 (red open triangles) calculated from opacity limits with {\it Fermi}/LAT observations. 
The solid line indicates the best fit of the $\Gamma_0-E_{\rm prompt}$ relation, 
$\log \Gamma_0 = 2.26+0.25 \log E_{\rm prompt}$. The two dashed lines represent the $2\sigma$ deviation, where the 
standard deviation of the ratio $\Gamma_0/E_{\rm prompt}^{0.25}$ for the data sample is $\sigma = 0.11$. 
(Figure after \cite{Liang2010}; see Fig. 8 therein. @ AAS. Reproduced with permission.)
{\bf Right panel:} $\log \Gamma_{0}$ vs. $\log L_{\rm iso}$ distribution. The best fit line is 
given by $\log \Gamma_{0} \simeq 2.40+0.30 \log L_{\rm iso}$ with $r=0.79$. The triangles represent the bursts with 
only lower values and the star indicates the only short burst in the sample, GRB090510. (Figure after \cite{Lu2012}; 
see Fig. 2 therein. @ AAS. Reproduced with permission.)}
\label{fig:liang}
\end{figure}

\citet{Liang2010} selected from the {\it Swift} catalogue 20 optical and 12 X-ray GRBs showing the onset of the afterglow 
shaped by the deceleration of the fireball due to the circumburst medium. The optically selected GRBs were used to fit a linear relation in the $\log\Gamma_0-\log E_{\rm prompt}$ plane, where $\Gamma_0$ is the initial Lorentz factor of the fireball and $E_{\rm prompt}$ is in units of $10^{52}\,{\rm erg}$ (see left panel of Fig.~\ref{fig:liang}). The best fit line of the $\Gamma_0-E_{\rm prompt}$ relation is given by
\begin{equation}
\log \Gamma_0 = (2.26 \pm 0.03) + (0.25 \pm 0.03) \log E_{\rm prompt},
\end{equation}
with $\rho=0.89$, $P<10^{-4}$, and $\sigma=0.11$ which can be measured with the deviation of the ratio $\Gamma_0/E^{0.25}_{\rm prompt}$. It was found that most of the GRBs with a lower limit of $\Gamma_0$ are enclosed within the $2\sigma$ region represented by the dashed lines in the left panel of Fig.~\ref{fig:liang}, and it was pointed out that GRBs with a tentative $\Gamma_0$ derived from RS peaks or the afterglow peaks, as well as those which lower limits of $\Gamma_0$ were derived from light curves with a single power law, are systematically above the best fit line. The lower values of $\Gamma_0$, obtained from a set of optical afterglow light curves with a decaying trend since the start of the detection, were compatible with this correlation.

Later, this correlation was verified by \citet{ghirlanda11} and \citet{Lu2012}. \cite{ghirlanda11}, studying the spectral evolution of 13 SGRBs detected by {\it Fermi}/GBM, investigated spectra resolved in the $8\,{\rm keV}-35\,{\rm MeV}$ energy range and confirmed the results of \citet{Liang2010}.

\citet{Lu2012} enlarged this sample reaching a total of 51 GRBs with spectroscopically confirmed redshifts, and engaged three methods to constrain $\Gamma_0$: (1) the afterglow onset method \citep{sari99} which considers $T_{\rm peak}$ of the early afterglow light curve as the deceleration time of the external FS; (2) the pair opacity constraint method \citep{lithwick01} which requires that the observed high energy $\gamma$-rays (i.e., those in the GeV range) are optically thin to electron-positron pair production, thus leading to a lower limit on $\Gamma_0$ of the emitting region; (3) the early external forward emission method \citep{zou10} where an upper limit of $\Gamma_0$ can be derived from the quiescent periods between the prompt emission pulses, in which the signal of external shock has to go down the instrument thresholds. Considering some aspects of the external shock emission, the $\Gamma_0-E_{\rm prompt}$ correlation was statistically re-analysed using 38 GRBs with $\Gamma_0$ calculated using method (1) (as the other two provide only a range of the Lorentz factors, not a definite value), finding
\begin{equation}
\log \Gamma_0 = (1.96 \pm 0.002) + (0.29 \pm 0.002) \log E_{\rm prompt},
\end{equation}
with $r=0.67$, and $E_{\rm prompt}$ in units of $10^{52}\,{\rm erg}$. In addition, applying the beaming correction, a relation between $\Gamma_0$ and $L_{\rm iso}$, using the same sample (see right panel of Fig.~\ref{fig:liang}), was found to be
\begin{equation}
\log \Gamma_0 = (2.40 \pm 0.002) + (0.30 \pm 0.002)\log L_{\rm iso},
\end{equation}
with $r=0.79$, and $L_{\rm iso}$ in units of $10^{52}\,{\rm erg}\,{\rm s}^{-1}$.

Regarding the physical interpretation, \citet{Liang2010} claimed that this correlation clearly shows 
the association of $E_{\rm prompt}$ with $\Gamma_0$ angular structure, and this result yielded another evidence for the fireball deceleration model. Instead, \cite{Lu2012} found that this relation is well explained by a neutrino-annihilation-powered jet during the emission, indicating a high accretion rate and not very fast BH spin. Besides, evidence for a jet dominated by a magnetic field have already been presented \citep{zhang2009,fan10,zhang2011}. From the studies of the BH central engine models it was also indicated that magnetic fields are a fundamental feature \citep{Lei09}. Nevertheless, the baryon loading mechanism in a strongly magnetized jet is more complex, and it has still to be fully investigated.

\subsection{Correlations between the energetics and the peak energy}

\subsubsection{The $\langle E_{\rm peak}\rangle-F_{\rm peak}$ and the $E_{\rm peak}-S_{\rm tot}$ correlations}\label{Mallozzi}

\begin{figure}[htbp]
\centering
\includegraphics[width=13.cm,height=8.5cm,angle=0]{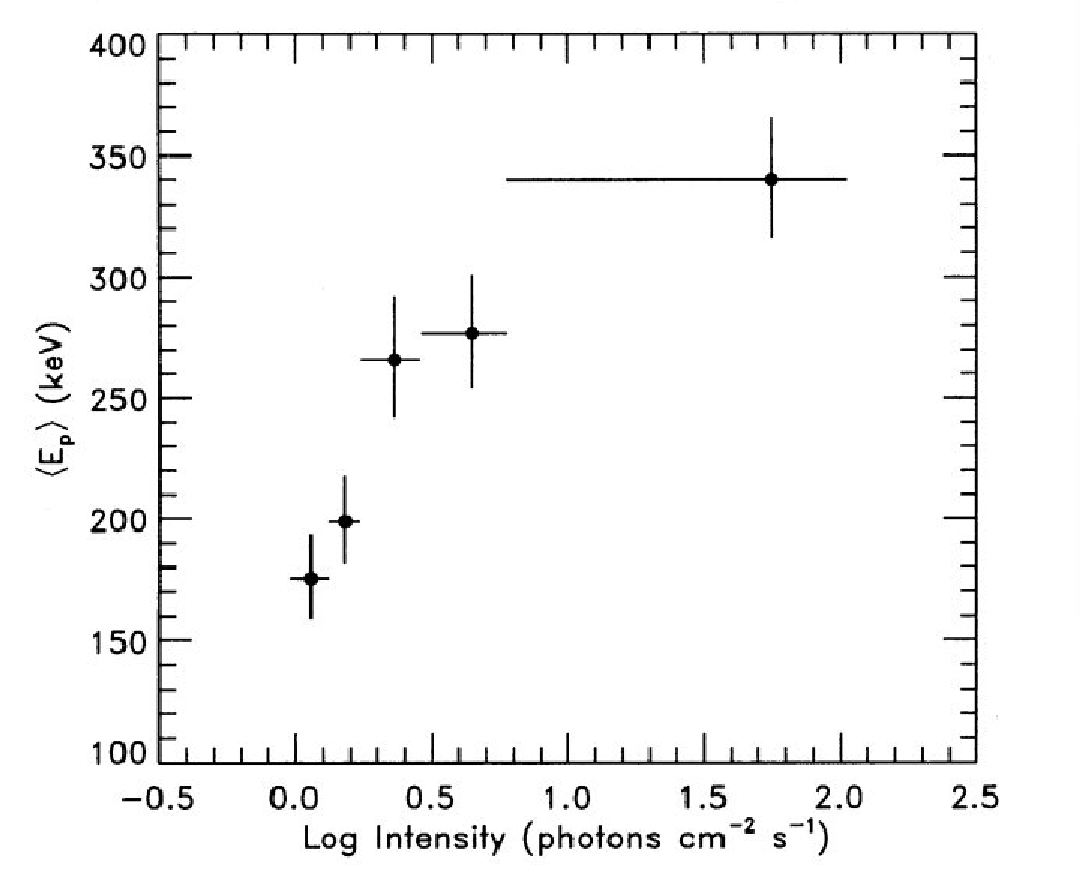}
\caption{\footnotesize The average $\nu F_{\nu}$ peak energies as a function of intensity for five groups of GRB spectra. 
The vertical bars represent a $1\sigma$ estimated error in the mean, where the peak energy 
distributions were assumed to be approximately Gaussian in logarithm of energy. The horizontal bars mark the bin widths. 
(Figure after \cite{mallozzi95}; see Fig. 2 therein. @ AAS. Reproduced with permission.)}
\label{fig:mallozzi95}
\end{figure}

\citet{mallozzi95} analysed 399 GRBs observed by BATSE and discovered a correlation between the logarithmic average peak energies $\langle E_{\rm peak}\rangle$ and $F_{\rm peak}$. Choosing as a selection criterion for the bursts $F_{\rm peak} \ge 1\,{\rm ph}\,{\rm cm}^{-2}\,{\rm s}^{-1}$, they derived $F_{\rm peak}$ from the count rate data in $256\,{\rm ms}$ time bins in the energy band $50-300\,{\rm keV}$ and used the $E_{\rm peak}$ distribution derived from the Comptonized photon model (the differential photon number flux per unit energy):
\begin{equation}
\frac{{\rm d}N}{{\rm d}E}= A{\rm e}^{-E(2+\beta_S)/E_{\rm peak}}\left(\frac{E}{E_{\rm piv}}\right)^{\beta_S},
\end{equation}
with $A$ the normalization, $\beta_S$ the spectral index, and $E_{\rm piv}=100\,{\rm keV}$. Then, they grouped the 
sample into 5 different width $F_{\rm peak}$ bins of about 80 events each 
(see Fig.~\ref{fig:mallozzi95}). 
The bursts were ranked such that group 1 had the lowest peak flux values and group 5 had the highest values. They found a 
correlation with $\rho=0.90$ and $P=0.04$. Lower intensity GRBs exhibited a lower $\langle E_{peak}\rangle$.

\begin{figure}[htbp]
\centering
\hspace{0.44cm}
\includegraphics[width=7.4cm,height=6.8cm,angle=0]{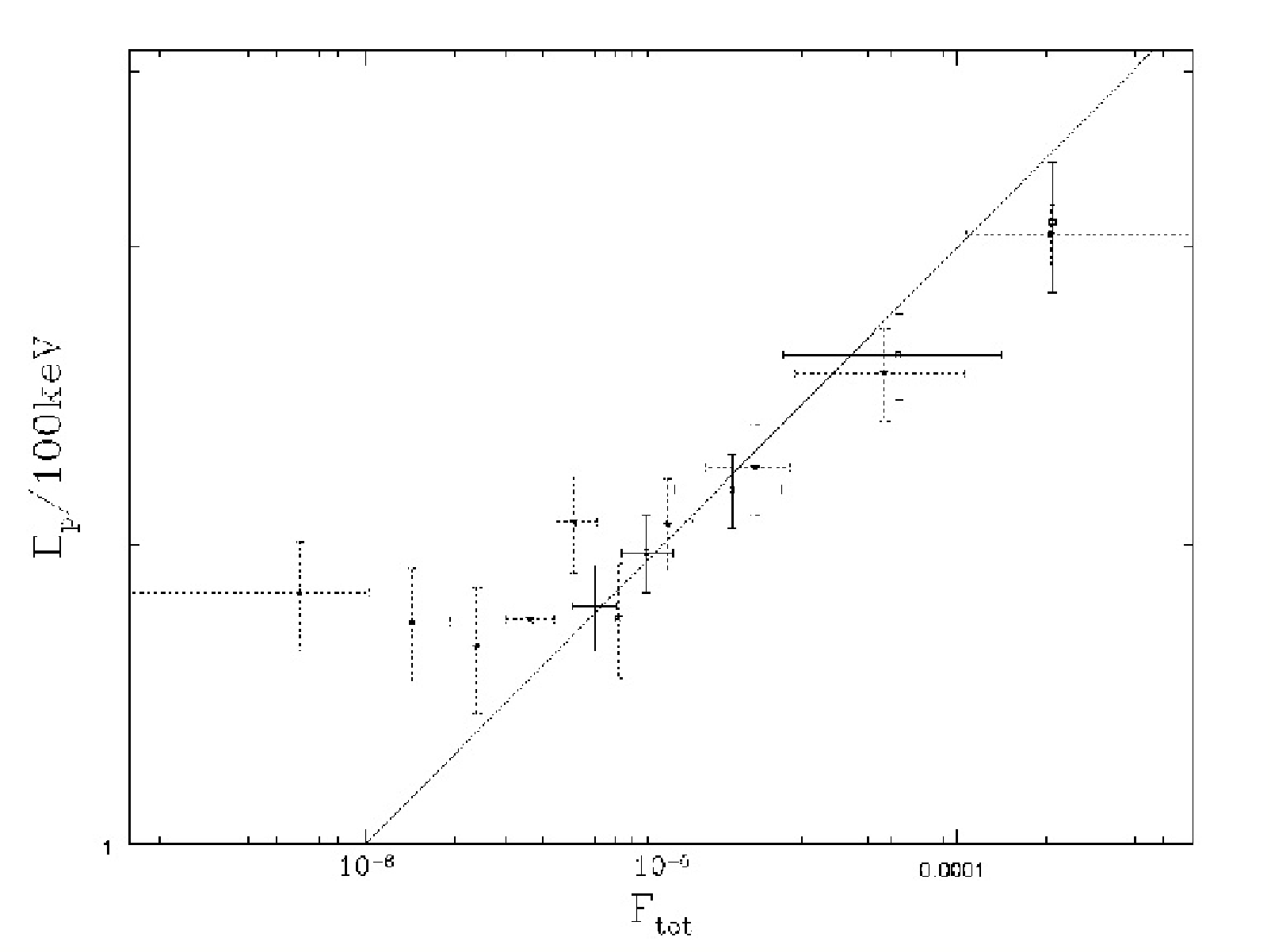}
\includegraphics[width=7.4cm,height=6.8cm,angle=0]{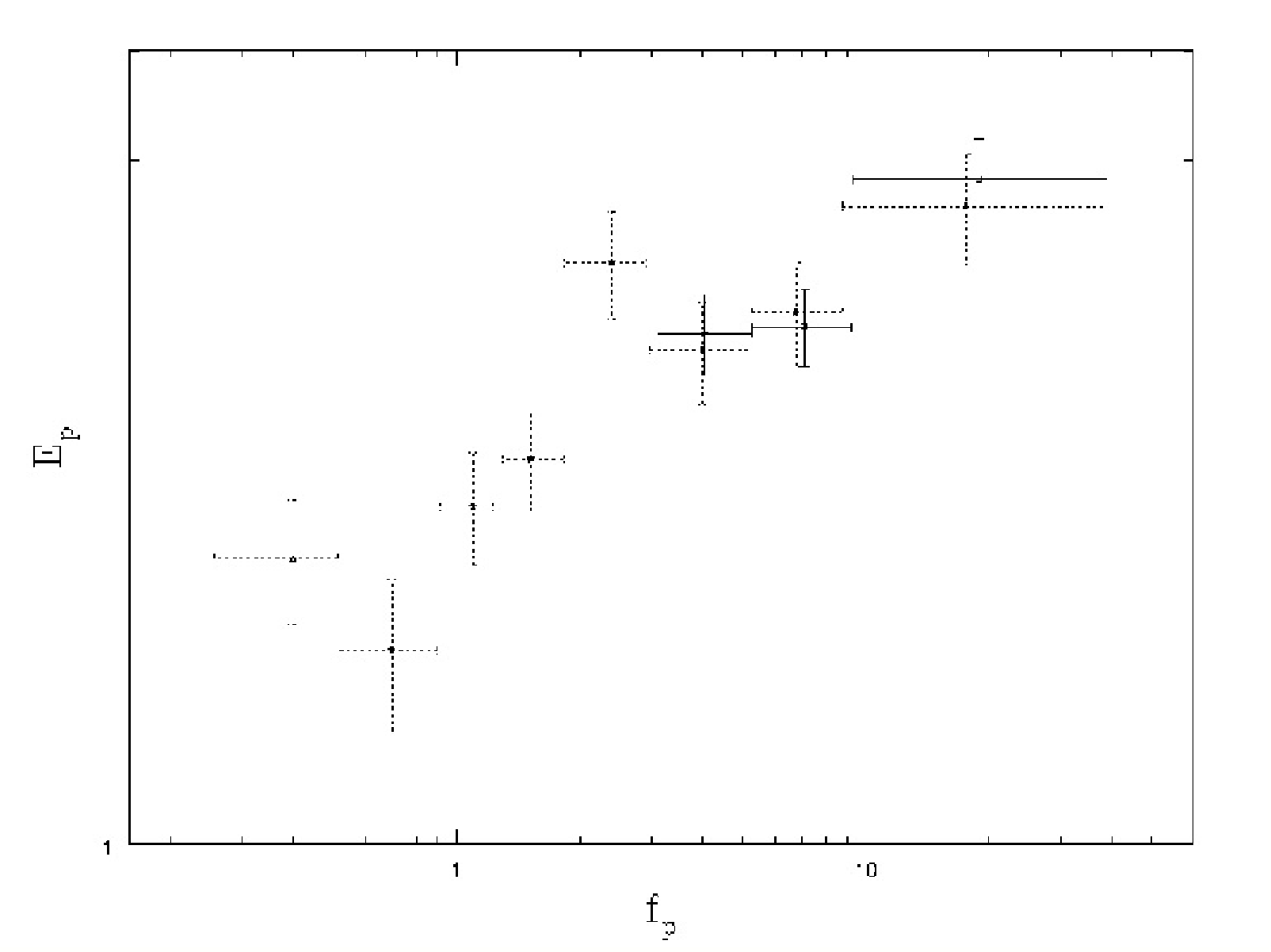}
\caption{\footnotesize $E_{\rm peak}$ vs. ({\bf left panel}) $S_{\rm tot}$ and ({\bf right panel}) $F_{\rm peak}$ 
distributions for the complete (dashed elements) and sub (solid elements) spectral sample. 
The flux suggests a tight correlation at low values, but not for the most luminous GRBs. The solid line represents a
least squares fit compatible with the correlation computed by statistical methods. (Figures after \cite{lloydpetr2000}; 
see Fig. 3 therein. @ AAS. Reproduced with permission.)}
\label{fig:Lloyd2000}
\end{figure}

Later, \citet{lloydpetr2000} examined the $E_{\rm peak}-S_{\rm tot}$ correlation with 1000 simulated bursts in the same energy 
range as \citet{mallozzi95}, and found a strong correlation between $E_{\rm peak}$ and $S_{\rm tot}$ (see the left panel of 
Fig.~\ref{fig:Lloyd2000}). The relation between the two variables was as follows:
\begin{equation}
\log E_{\rm peak} \sim 0.29\log S_{\rm tot},
\end{equation}
with the Kendall correlation coefficient \citep{KendallCorr} $\tau=0.80$ and $P=10^{-13}$. In addition, they compared it to the 
$E_{\rm peak}-F_{\rm peak}$ relation (see right panel of Fig.~\ref{fig:Lloyd2000}). This relation was for the whole spectral sample, and consistent with earlier results \citep{mallozzi95,Mallozzi98}. However, they selected a subsample composed of only the most luminous GRBs, because spectral parameters obtained from bursts near the detector threshold are not robust. Therefore, to better understand the selection effects relevant to $E_{\rm peak}$ and burst strength, they considered the following selection criteria: $F_{\rm peak} \ge 3\,{\rm ph}\,{\rm cm}^{-2}\,{\rm s}^{-1}$, $S_{\rm obs} \ge 10^{-6}\,{\rm erg}\,{\rm cm}^{-2}$, and $S_{\rm tot} \ge 5\times 10^{-6}\,{\rm erg}\,{\rm cm}^{-2}$. Due to the sensitivity over a certain energy band of all the detectors, especially BATSE, and to some restrictions to the trigger, the selection effects are inevitable. However, the subsample of the most luminous GRBs presents a weak $E_{\rm peak}-F_{\rm peak}$ correlation. Instead, a tight $E_{\rm peak}-S_{\rm tot}$ correlation was found for the whole sample as well as the subsample of the brightest GRBs. \citet{lloydpetr2000} paid more attention to the $E_{\rm peak}-S_{\rm tot}$ correlation for the brightest GRBs because it is easier to deal with the truncation effects in this case, and the cosmological interpretation is simpler.

This correlation has been the basis for the investigation of the Amati relation (see Sect.~\ref{Amati}), and the Ghirlanda relation (see Sect. \ref{ghirlanda}). \citet{lloydpetr2000} concluded that ``the observed correlation can be explained by cosmological expansion alone if the total radiated energy (in the $\gamma$-ray range) is constant''. 
In fact, their finding does not depend on the GRB rate density or on the distribution of other parameters. However, the data from GRBs with known redshift are incompatible with a narrow distribution of radiated energy or luminosity.

Following a different approach, \citet{goldstein10} pointed out that the ratio $E_{\rm peak}/S_{\rm tot}$ can serve as an indicator of the ratio of the energy at which most of the $\gamma$-rays are radiated to the total energy, and claimed that the $E_{\rm peak}-S_{\rm tot}$ relation is a significant tool for classifying LGRBs and SGRBs. The fluence indicates the duration of the burst without providing a biased value of $T_{90}$, and $E_{\rm peak}/S_{\rm tot}$ displays, as a spectral hardness ratio, an increased hardness for SGRBs in respect to LGRBs, in agreement with \citep{kouveliotou93}. This correlation is quite interesting, since the energy ratio, being dependent only on the square of the luminosity distance, gets rid of the cosmological dependence for the considered quantities. Therefore, it was evaluated that the energy ratio could be used as a GRB classifier.

Later, \citet{lu12} with the results of time-resolved spectral analysis, computed the $E_{\rm peak}-S_{\rm tot}$
relation for 51 LGRBs and 11 bright SGRBs observed with {\it Fermi}/GBM. For each GRB, they fitted a simple power law function. They measured its scatter with the distance of the data points from the best fit line. The measured scatter of the $E_{\rm peak}-S_{\rm tot}$ relation is $0.17 \pm 0.08$. This result was reported for the first time by 
\citet{gole}, and later confirmed by \citet{borgonovo01,ghirlanda10,guiriec2010,ghirlanda11}.

\subsubsection{The \texorpdfstring{$E_{\rm peak}-E_{\rm iso}$}{Lg} correlation}\label{Amati}

Evidence for a correlation between $E_{\rm peak}$ and $S_{\rm tot}$ was first found by \citet{Lloyd99} and \citet{lloyd2000} based on 46 BATSE events, but this relation was in the observer frame due to the paucity of the data with precise redshift measurement, as was shown in previous paragraphs. Evidence for a stronger correlation between $E_{\rm peak}$ and $E_{\rm iso}$, also called the Amati relation, was reported by \citet{AmatiEtal02} based on a limited sample of 12 GRBs with known redshifts (9 with firm redshift and 3 with plausible values) detected by {\it BeppoSAX}. They found that
\begin{equation}
\log E_{\rm peak} \sim (0.52 \pm 0.06) \log E_{\rm iso},
\end{equation}
with $r=0.949$, $P=0.005$, and $E_{\rm iso}$ calculated as 
\begin{equation}
E_{\rm iso} = 4\pi D_L(z,\Omega_M, \Omega_{\Lambda})^2 S_{\rm tot} (1+z)^{-2}. 
\end{equation}
Regarding the methodology considered, instead of fitting the observed spectra, as done for example by \cite{Bloom2001}, the GRB spectra were blue-shifted to the rest frames to obtain their intrinsic form. Then, the total emitted energy is calculated by integrating the \citet{Band1993} spectral model in $1-10^{4}\,{\rm keV}$ energy band and scaling for the luminosity distance. This was computed employing a flat Friedman-Lema\^{\i}tre-Robertson-Walker cosmological model with $H_0 = 65\,{\rm km}\,{\rm s}^{-1}\,{\rm Mpc}^{-1}$, $\Omega_M=0.3$, $\Omega_\Lambda=0.7$, and taking into account both the cosmological time dilation and spectral redshift.

\citet{amati03} enlarged the set of \cite{AmatiEtal02} by including 20 GRBs from {\it BeppoSAX} with known redshift for which new spectral data ({\it BeppoSAX} events) or published best-fitting spectral parameters (BATSE and {\it HETE-2} events) were accessible. The relation was found to be
\begin{equation}
\log E_{\rm peak} =(2.07 \pm 0.03) + (0.35 \pm 0.06) \log E_{\rm iso},
\label{equ2}
\end{equation}
with $r=0.92$, $P=1.1\times10^{-8}$, $E_{\rm peak}$ in keV and $E_{\rm iso}$ in units of $10^{52}{\rm erg}$. Therefore, its statistical significance increased, providing a correlation coefficient 
comparable to that obtained by \citet{AmatiEtal02}, but based on a larger set.

Based on {\it HETE-2} measurements, \citet{lamb04} and \citet{sakamoto04} verified the previous results and considered also XRFs, finding out that the Amati relation remains valid over three orders of magnitude in $E_{\rm peak}$ and five orders of magnitude in $E_{\rm iso}$. The increasing amount of GRBs with measured redshift allowed to verify this relation and strengthen its validity, as found by \citet{Ghirlanda2004} with 29 events ($r=0.803$ and $P=7.6 \times 10^{-7}$; see left panel in Fig.~\ref{fig:9}).

\begin{figure}[htbp]
\includegraphics[width=8.1cm, height=6cm,angle=0,clip]{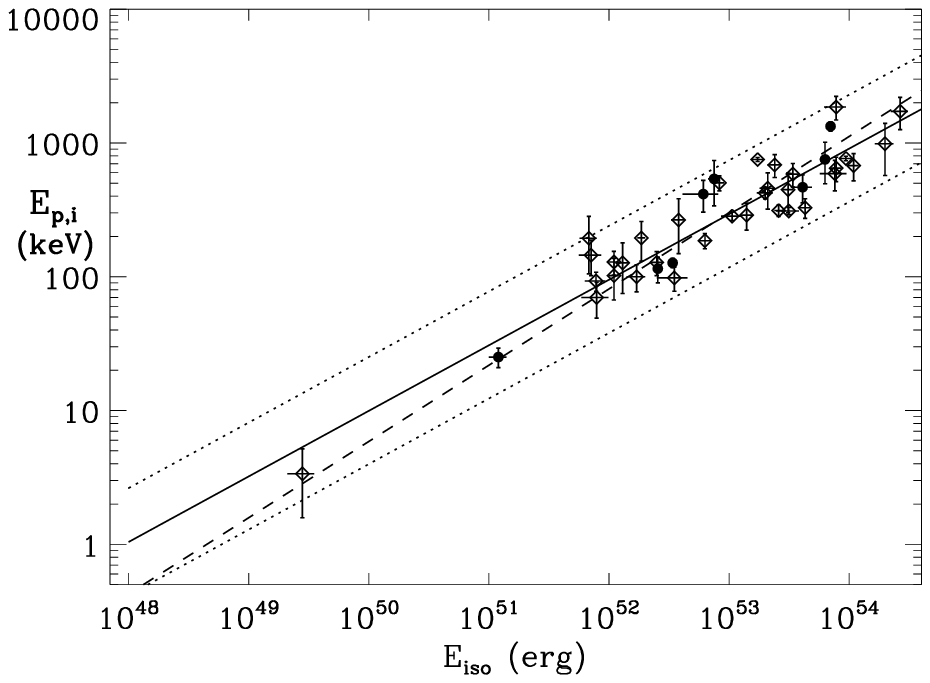} 
\includegraphics[width=8.1cm, height=6.2cm,clip]{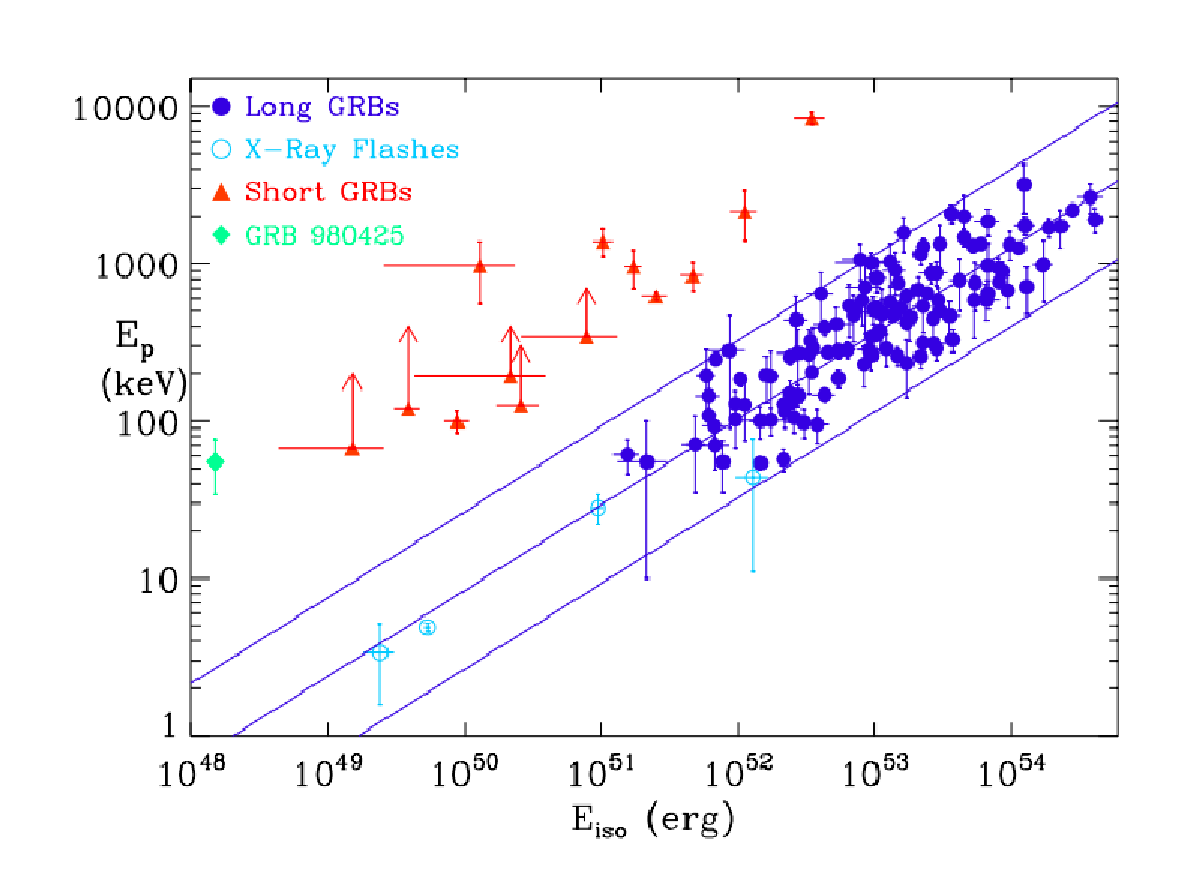}\\
\includegraphics[width=8.1cm, height=6cm,angle=0,clip]{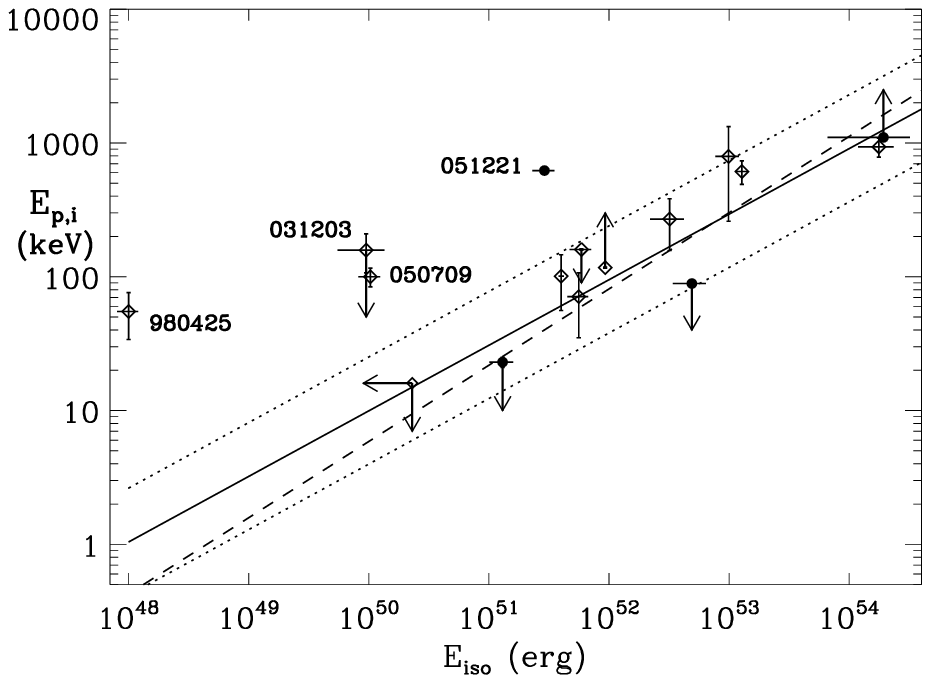} 
\includegraphics[width=8.1cm, height=6cm,angle=0,clip]{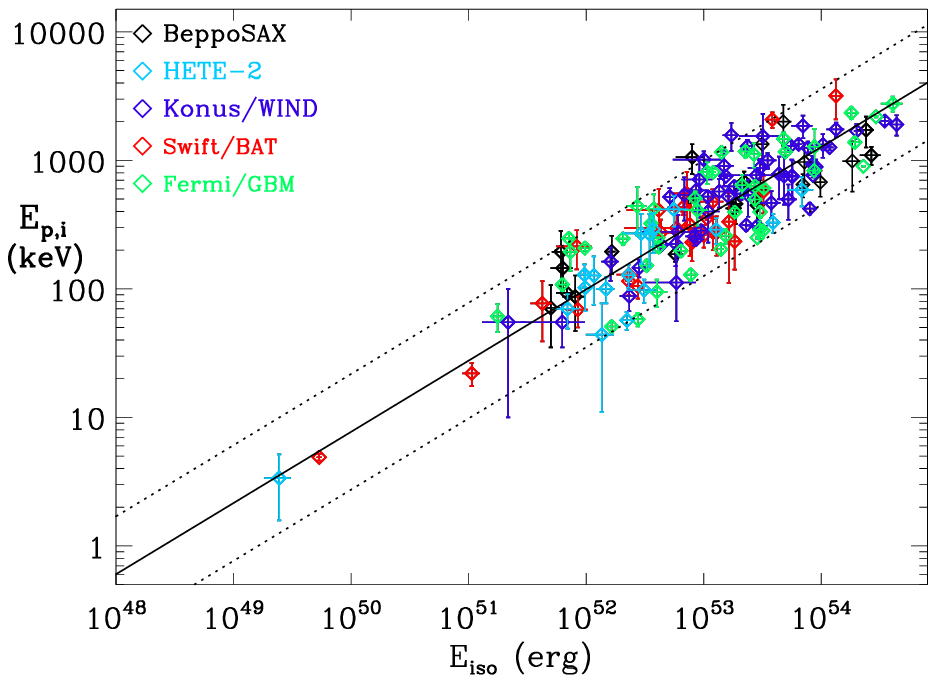} 
\caption{\footnotesize {\bf Upper left panel:} $E_{\rm peak}-E_{\rm iso}$ distribution for 41 GRBs/XRFs with measured 
redshifts and $E_{\rm peak}$ values. Filled circles indicate Swift GRBs. The solid line represents 
the best fit line $\log E_{\rm peak} = 1.98+ 0.49\log E_{\rm iso}$; the dotted lines show the region within a vertical 
logarithmic deviation of 0.4. The dashed line represents the best fit line $\log E_{\rm peak}=1.89+0.57\log E_{\rm iso}$ 
computed without taking into account the sample variance. (Figure after \cite{Amati2006}; see Fig. 2 therein.) 
{\bf Upper right panel:} the distribution of the sample in the $E_{\rm peak}-E_{\rm iso}$ plane. 
The lines indicate the best fit line and the $\pm 2\sigma$ confidence region for LGRBs and XRFs. 
(Figure after \cite{amati12}; see Fig. 4 therein. Copyright @ 2012 World Scientific Publishing Company.) 
{\bf Bottom left panel:} $E_{\rm peak}-E_{\rm iso}$ distribution of 12 GRBs with uncertain values of
$z$ and/or $E_{\rm peak}$, for the sub-energetic event GRB980425 and for the two SGRBs 050709 and 051221. 
Filled circles represent {\it Swift} GRBs. The solid line is the best fit line $\log E_{\rm peak} = 1.98+ 0.49\log E_{\rm iso}$; the dotted lines mark the region 
within a vertical deviation in logarithmic scale of 0.4. The dashed line is the best fit line 
$\log E_{\rm peak}=1.89+0.57\log E_{\rm iso}$ computed without taking into account the sample variance. 
(Figure after \cite{Amati2006}; see Fig. 3 therein.) 
{\bf Bottom right panel:} the $E_{\rm peak}-E_{\rm iso}$ distribution for the LGRBs. 
The black line represents the best fit line and, for each point, the color indicates the instrument which performed 
the spectral measurement. (Figure after \cite{amati13}; see Fig. 4 therein. Copyright @ 2013 World Scientific Publishing Company.)}
\label{fig:6}
\end{figure}

\citet{ghirlanda05} verified the $E_{\rm peak}-E_{\rm iso}$ correlation among LGRBs considering a set of 442 BATSE GRBs with measured $E_{\rm peak}$ and with pseudo-redshifts computed via the $L_{\rm peak}-\tau_{\rm lag}$ correlation. It was shown that the scatter of the sample around the best fitting line is comparable with that of another set composed of 27 GRBs with measured spectroscopic redshifts. This is because the weights of the outliers were marginal. It was noted that the relation for the 442 BATSE GRBs has a slope slightly smaller (0.47) than the one obtained for the 27 GRBs with measured spectroscopic redshifts (0.56).

Afterwards, \citet{Amati2006} (see the upper left and bottom left panels in Fig.~\ref{fig:6}) updated 
the study of the $E_{\rm peak}-E_{\rm iso}$ correlation considering a sample of 41 LGRBs/XRFs with firm values of $z$ and $E_{\rm peak}$, 12 GRBs with uncertain $z$ and/or $E_{\rm peak}$, 2 SGRBs with certain values of $z$ and $E_{\rm peak}$, and the sub-energetic events GRB980425/SN1998bw and GRB031203/SN2003lw. The different sets are displayed in the upper right panel in Fig.~\ref{fig:6}. Taking into account also the sample variance it was found:
\begin{equation}
\log E_{\rm peak} = 1.98^{+0.05}_{-0.04} + \left(0.49^{+0.06}_{-0.05}\right) \log E_{\rm iso},
\label{equ3}
\end{equation}
with $\rho=0.89$, $P= 7\times 10^{-15}$, and units the same as in Eq.~(\ref{equ2}). Moreover, sub-energetic GRBs (980425 and possibly 031203) and SGRBs were incompatible with the $E_{\rm peak}-E_{\rm iso}$ relation, suggesting that it can be an important tool for distinguishing different classes of GRBs. Indeed, the increasing number of GRBs with measured $z$ and $E_{\rm peak}$ can provide the most reliable evidence for the existence of two or more subclasses of outliers for the $E_{\rm peak}-E_{\rm iso}$ relation. Moreover, the relation is valid also for the particular sub-energetic event GRB060218. Finally, the normalization considered by \cite{Amati2006} is consistent with those obtained by other instruments.

\citet{Ghirlanda2008} confirmed the $E_{\rm peak}-E_{\rm iso}$ correlation for softer events (XRFs). The sample consisted of 76 GRBs observed by several satellites, mainly {\it HETE-2}, {\it KONUS}/Wind, {\it Swift} and {\it Fermi}/GBM.
The most important outcome is a tight correlation with no new outliers (with respect to the classical GRB980425 and GRB031203) in the $E_{\rm peak}-E_{\rm iso}$ plane. The obtained relation was
\begin{equation}
\log E_{\rm peak} \sim (0.54\pm 0.01)\log E_{\rm iso}.
\end{equation}

\citet{amati09} studied 95 {\it Fermi} GRBs with measured $z$ and obtained an updated $E_{\rm peak}-E_{\rm iso}$ relation, which read
\begin{equation}
\log E_{\rm peak} \sim 0.57\log E_{\rm iso},
\end{equation}
with $\rho=0.88$ and $P < 10^{-3}$.
In particular, they investigated two GRBs (080916C and 090323) with very energetic prompt emission, and found that they follow the $E_{\rm peak}-E_{\rm iso}$ relation well. On the other hand, an SGRB, 090510, also a very luminous and energetic event, was found not to obey the relation. Hence, \citet{amati09} proposed that the correlation might serve as a discriminating factor among high-energetic GRBs. In addition, they claimed that the physics of the radiation process for really luminous and energetic GRBs is identical to that for average-luminous and soft-dim long events (XRFs), because all these groups follow the Amati relation.

Later, \citet{amati12} provided an update of the analysis by \citet{Amati2008} with a larger sample of 120 GRBs (see the upper right panel of Fig.~\ref{fig:6}) finding it to be consistent with the following relation:
\begin{equation}
\log E_{\rm peak} = 2 + 0.5\log E_{\rm iso}.
\end{equation}
with units the same as in Eqs.~(\ref{equ2}) and (\ref{equ3}). Afterwards, \citet{qin13} analysed a sample of 153 GRBs with measured $z$, $E_{\rm peak}$, $E_{\rm iso}$ and $T_{90}$, observed by various instruments up to 2012 May. The distribution of the logarithmic deviation of $E_{\rm peak}$ from the Amati relation displayed a clear bimodality which was well represented by a mixture of two Gaussian distributions. Moreover, it was suggested to use the logarithmic deviation of the $E_{\rm peak}$ value for distinguishing GRBs in the $E_{\rm peak}$ versus $E_{\rm iso}$ plane. This procedure separated GRBs into two classes: the Amati type bursts, which follow the Amati relation, and the non-Amati type bursts, which do not follow it. For the Amati type bursts it was found that
\begin{equation}
\log E_{\rm peak} = (2.06 \pm 0.16) + (0.51 \pm 0.12) \log E_{\rm iso}
\end{equation}
with $r=0.83$ and $P<10^{-36}$, while for non-Amati bursts:
\begin{equation}
\log E_{\rm peak} = (3.16 \pm 0.65) + (0.39 \pm 0.33) \log E_{\rm iso}
\end{equation}
with $r=0.91$ and $P<10^{-7}$. In both relations $E_{\rm peak}$ is in keV, and $E_{\rm iso}$ is in units of $10^{52}\,{\rm erg}$.

In addition, it was pointed out that almost all Amati type bursts are LGRBs at higher energies, as opposed to non-Amati type bursts which are mostly SGRBs. An improvement to this classification procedure is that the two types of GRBs are clearly separated, hence different GRBs can be easily classified.

\citet{heussaff13}, applying particular selection criteria for the duration and the spectral indices, obtained a set of {\it Fermi} GRBs and analysed their locations in the $E_{\rm peak}-E_{\rm iso}$ plane. The sample, composed of 43 GRBs with known redshifts, yielded the following relation:
\begin{equation}
\log E_{\rm peak} = 2.07 + 0.49 \log E_{\rm iso},
\end{equation}
with $\rho=0.70$, $P=1.7\times10^{-7}$, and the same units as in previous relations of this type.

\citet{amati13} pointed out that an enlarged sample of 156 LGRBs with known $z$ and $E_{\rm peak}$ also follows the Amati relation with a slope $\simeq 0.5$ (see the bottom right panel of Fig.~\ref{fig:6}). Additionally, \citet{basak12} showed that a time-resolved Amati relation also holds within each single GRB with normalization and slope consistent with those obtained with time-averaged spectra 
and energetics/luminosity, and is even better than the time-integrated relation \citep{basak13}. Time-resolved $E_{\rm peak}$ and $E_{\rm iso}$ are obtained at different times during the prompt phase (see also \citealt{ghirlanda10,lu12,frontera12} and Sect.~\ref{sect36}).

\subsubsection{The \texorpdfstring{$E_{\rm peak}-E_{\gamma}$}{Lg} correlation}\label{ghirlanda}

The $E_{\rm peak}-E_{\gamma}$ relation (also called the Ghirlanda relation) was first discovered by \citet{Ghirlanda2004},
who used 40 GRBs with $z$ and $E_{\rm peak}$ known at their time of writing. Considering the time $T_{\rm break}$, its value can be used to deduce 
$E_{\gamma}$ from $E_{\rm iso}$. Indeed, even if only a little less than half of the bursts have observed jet breaks (47\%), 
from \citep{SPH99} we know that
\begin{equation}
\theta_{\rm jet} = 0.161\left(\frac{T_{\rm break}}{1+z}\right)^{3/8}\left(n\eta_{\gamma}E_{\rm iso}\right)^{1/8}, 
\end{equation} 
where $T_{\rm break}$ is measured in days, $n$ is the density of the circumburst medium in particles per ${\rm cm}^3$, $\eta_{\gamma}$ is the radiative efficiency, and $E_{\rm iso}$ is in units of $10^{52}\,{\rm erg}$. Here, $\theta_{\rm jet}$ is in degrees and it is the angular radius (the half opening angle) subtended by the jet. For GRBs with no measured $n$, the median value $n=3\,{\rm cm}^{-3}$ of the distribution of the computed densities, extending between 1 and $10\,{\rm cm}^{-3}$, was considered \citep{frail2000,yost02,panaitescu02,schaefer03a}.

Later, \citet{liang05} using a sample of 15 GRBs with measured $z$, $E_{\rm peak}$ and $T_{\rm break}$, considered a purely phenomenological $T^{*}_{\rm break}$ of the optical afterglow light curves, thus avoiding the assumption of any theoretical model, contrary to what was done by \citet{Ghirlanda2004}. The functional form of this correlation is given by:
\begin{equation}
\log E_{\gamma} = (0.85 \pm 0.21)+ (1.94 \pm 0.17)\log E^{*}_{\rm peak}-(1.24 \pm 0.23)\log T^{*}_{\rm break}, 
\end{equation} 
where $E_{\gamma}$ is in units of $10^{52}\,{\rm erg}$, $E^{*}_{\rm peak}$ in units of $100\,{\rm keV}$, $T^{*}_{\rm break}$ is measured in days, and $\rho=0.96$ and $P < 10^{-4}$.

\citet{nava06} found that the Ghirlanda relation, assuming a wind-like circumburst medium, is as strong as the one considering a homogeneous medium. They analysed the discrepancy between the correlations in the observed and in the comoving frame (with Lorentz factor identical to the fireball's one). Since both $E_{\rm peak}$ and $E_{\gamma}$ transform in the same way, the wind-like Ghirlanda relation remains linear also in the comoving frame, no matter what the Lorentz factor's distribution is. The wind-like relation corresponds to bursts with the same number of photons emitted. Instead, for the homogeneous density medium scenario, it is common to consider a tight relation between the Lorentz factor and the total energy, thus limiting the emission models of the prompt radiation. Using 18 GRBs with firm $z$, $E_{\rm peak}$ and $T_{\rm break}$, \citet{nava06} found for the homogeneous density case
\begin{equation}
\log \frac{E^*_{\rm peak}}{100\,{\rm keV}}=0.45^{+0.02}_{-0.03}+(0.69\pm 0.04) \log \frac{E_{\gamma}}{2.72\times 10^{52}\,{\rm erg}},
\end{equation}
with $\rho=0.93$ and $P=2.3\times10^{-8}$. Instead, for the wind case
\begin{equation}
\log \frac{E^*_{\rm peak}}{100\,{\rm keV}}=0.48^{+0.02}_{-0.03}+(1.03\pm 0.06) \log \frac{E_{\gamma}}{2.2\times 10^{50}\,{\rm erg}},
\end{equation}
with $\rho=0.92$ and $P=6.9\times10^{-8}$.

\citet{ghirlanda07} tested the $E_{\rm peak}-E_{\gamma}$ correlation using 33 GRBs (16 new bursts detected 
by {\it Swift} with firm $z$ and $E_{\rm peak}$ up to December 2006, and 17 pre-{\it Swift} GRBs). They claimed that for computing the $T_{\rm break}$ it is required that:
\begin{enumerate}
\item the detection of the jet break should be in the optical,
\item the optical light curve should continue up to a time longer than the $T_{\rm break}$, 
\item the host galaxy flux and the flux from a probable SN should be removed,
\item the break should not depend on the frequency in the optical, and a coincident break in the X-ray 
light curve is not necessary, because the flux in X-rays could be controlled by another feature,
\item the considered $T_{\rm break}$ should be different from the one at the end of the plateau emission 
(the time $T_a$ in \citealt{Willingale2007}), otherwise the feature affecting the X-ray flux is also influencing the optical one.
\end{enumerate}
Therefore, considering all these restrictions, the sample was reduced to 16 GRBs, all compatible with the following $E_{\rm peak}-E_{\gamma}$ relation:
\begin{equation}
\log \frac{E_{\rm peak}}{100\,{\rm keV}}=(0.48\pm 0.02)+(0.70\pm 0.04) \log \frac{E_{\gamma}}{4.4\times 10^{50}\,{\rm erg}}.
\end{equation}
No outliers were detected. Therefore, the reduced scatter of the $E_{\rm peak}-E_{\gamma}$ relation corroborates the use of GRBs as standardizable candles.

\subsubsection{Physical interpretation of the energetics vs. peak energy relations}

\citet{lloydpetr2000} investigated the physical explanation of the $E_{\rm peak}-S_{\rm tot}$ correlation assuming the emission process to be a synchrotron radiation from internal and external shocks. Indeed, they claimed that this correlation is easily obtained considering a thin synchrotron radiation by a power law distribution of electrons with $\Gamma$ larger than some minimum threshold value, $\Gamma_m$. Moreover, the internal shock model illustrates the tight $E_{\rm peak}-S_{\rm tot}$ relation and the emitted energy better than the external shock model.

\citet{lloydpetrosian02} pointed out that the GRB particle acceleration is not a well analysed issue. Generally, the main hyphotesis is that the emitted particles are accelerated via recurrent scatterings through the (internal) shocks. They found that the recurrent crossings of the shock come from a power law distribution of the particles with a precise index, providing a large energy synchrotron photon index. Moreover, the connection between $E_{\rm peak}$ and the photon flux can be justified by the variation of the magnetic field or electron energy in the emission events. Finally, they claimed that in the majority of GRBs, the acceleration of particles is not an isotropic mechanism, but occurs along the magnetic field lines.

\citet{AmatiEtal02} confirmed the findings of \citet{lloyd2000} that the $\log E_{\rm peak}\sim 0.5\log E_{\rm iso}$ relation is obtained assuming an optically thin synchrotron shock model. This model considers electrons following the $N(\Gamma) = N_0 \Gamma^{-p}$ distribution for $\Gamma > \Gamma_m$ with $\Gamma_m$, GRB duration, and $N_0$ constant in each GRB. However, the above assumptions are not fully justifiable. In fact the duration is different in each GRB and $E_{\rm iso}$ might be smaller in the case of beamed emission.

\citet{Amati2006} pointed out the impact that the correlation has on the modeling of the prompt emission and on the possible unification of the two classes of GRBs and XRFs. In addition, this correlation is often applied for checking
GRB synthesis models (e.g. \citealt{Zhang02,ghirlanda13}).

In every model, $E_{\rm peak}$ and $E_{\rm iso}$ depend on $\Gamma$, and the $E_{\rm peak}-E_{\rm iso}$ relation can help to relate the parameters of the synchrotron shock model and inverse Compton model \citep{Zhang02,schaefer03a}. Specifically, \citet{Zhang02} and \citet{rees05} found that, for an electron distribution given by a power law and produced by an internal shock in a fireball with velocity $\Gamma$, the peak energy is given as
\begin{equation}
\log E^*_{\rm peak} \sim -2\log \Gamma+0.5\log L-\log t_{\nu},
\end{equation}
where $L$ is the total fireball luminosity and $t_{\nu}$ the variability timescale. However, to recover the $E_{\rm peak}-E_{\rm iso}$ relation from this relation, $\Gamma$ and $t_{\nu}$ should be similar for each GRB, a condition that cannot be easily supported. A further issue arises when one considers that $L\propto\Gamma^{N}$, with $N$ between 2 and 3 in different models \citep{Zhang02,schaefer03a,ramirez05}. An explanation could be that direct or Comptonized thermal radiation from the fireball photosphere \citep{Zhang02,ramirez05,ryde05,rees05,beloborodov2010,guiriec2011,hascoet2013,guiriec2013,vurm2015,guiriec2015a,
guiriec2015b} can affect significantly the GRB prompt emission. This can be a good interpretation of the really energetic spectra presented for many events \citep{frontera2000,preece2000,ghirlanda03} and the flat shape in GRB average spectra. In such cases, $E_{\rm peak}$ depends on the peak temperature $T_{bb,{\rm peak}}$ of photons distributed as by a blackbody, and therefore it is associated to the luminosity or emitted energy. For Comptonized radiation from the photosphere the relations are
\begin{equation}
\log E_{\rm peak} \sim \log \Gamma +\log T_{bb,{\rm peak}} \sim 2 \log\Gamma-0.25 \log L 
\end{equation}
or
\begin{equation}
\log E_{\rm peak} \sim \log \Gamma + \log T_{bb,{\rm peak}} \sim -0.5\log r_0+0.25\log L, 
\end{equation}
where $r_0$ is a particular distance between the central engine and the energy radiating area, such that the Lorentz factor evolves as $\Gamma\simeq r/r_0$ up to some saturation radius $r_s$ \citep{rees05}. As suggested by \citet{rees05}, in this scenario the $E_{\rm peak}-E_{\rm iso}$ relation could be recovered for particular physical cases just underneath the photosphere, though it would rely on an undefined number of unknown parameters.

Also for high-energetic GRBs (i.e., $E_{\rm iso}\approx 10^{55}\,{\rm erg}$) the nonthermal synchrotron emission model can explain the $E_{\rm peak}-E_{\rm iso}$ correlation. This can be possible either considering the minimum Lorentz factor and the normalization of the power law distribution of the emitting electrons constant in each GRB, or by constraints on the slope of the relation between $\Gamma$ and the luminosity \citep{lloyd2000,Zhang02}.

\citet{panaitescu09} used 76 GRBs with measured redshifts to analyse the case in which the $E_{\rm peak}-E_{\rm iso}$ relation for LGRBs is due to the external shock generated by a relativistic outflow interacting with the ambient medium. He considered the effect of each parameter defining the $E_{\rm peak}-E_{\rm iso}$ relation on the radial distribution of the external medium density and pointed out that the $\log E_{\rm peak}\sim 0.5\log E_{\rm iso}$ relation is recovered if the external medium is radially stratified. For some combinations of radiative (synchrotron or inverse-Compton) and dissipation (such as RS or FS) mechanisms, it is concluded that the external medium requires a particle 
density distributed distinctly from $R^{-2}$, with $R$ the distance at which the GRB radiation is generated. This tendency should be commonly associated to uniform mass-loss rate and final velocity.

\citet{mochkovitch14} checked whether the $E_{\rm peak}-E_{\rm iso}$ relation can be recovered in a case when the prompt emission is due to internal shocks, or alternatively if the correlation can give some limits for the internal shock scenario defined through the impact of only two shells. Simulated GRB samples were obtained considering different model parameter distributions, such as the emitted power in the relativistic emission and $\Gamma$. Simulated $E_{\rm peak}-E_{\rm iso}$ distributions were plotted for each case and analysed together with the observed relation (based on 58 GRBs). The sample contained only luminous {\it Swift} GRBs with $F_{\rm peak}> 2.6\,{\rm ph}\,{\rm cm}^{-2}\,{\rm s}^{-1}$ in the $15-150\,{\rm keV}$ energy band. In conclusion, a correspondence between the model and data was found, but exclusively if the following restrictions for the dynamics of the emission and for the dispersion of the energy are assumed:
\begin{enumerate}
 \item the majority of the dispersed energy should be radiated in few electrons;
 \item the spread between the highest and the lowest Lorentz factor should be small;
 \item if the mean Lorentz factor grows as $\bar{\Gamma} \propto \dot{E}^{1/2}$ (where $\dot{E}$ is the rate of injected energy, or mean emitted power, in the relativistic outflow), the $E_{\rm peak}-E_{\rm iso}$ relation is not retrieved and $E_{\rm peak}$ is diminishing with larger $E_{\rm iso}$. However, the $E_{\rm peak}-E_{\rm iso}$ relation can be regained if $\bar{\Gamma} \propto \dot{E}^{1/2}$  is a lower constraint for a particular $\dot{E}$;
 \item when the timescale or the width of the variability of the Lorentz factor is associated with $\bar{\Gamma}$, $E_{\rm peak}-E_{\rm iso}$ relation is recovered.
\end{enumerate}

\begin{figure}[htbp]
\centering
\includegraphics[width=8.3cm,height=7cm,angle=0,clip]{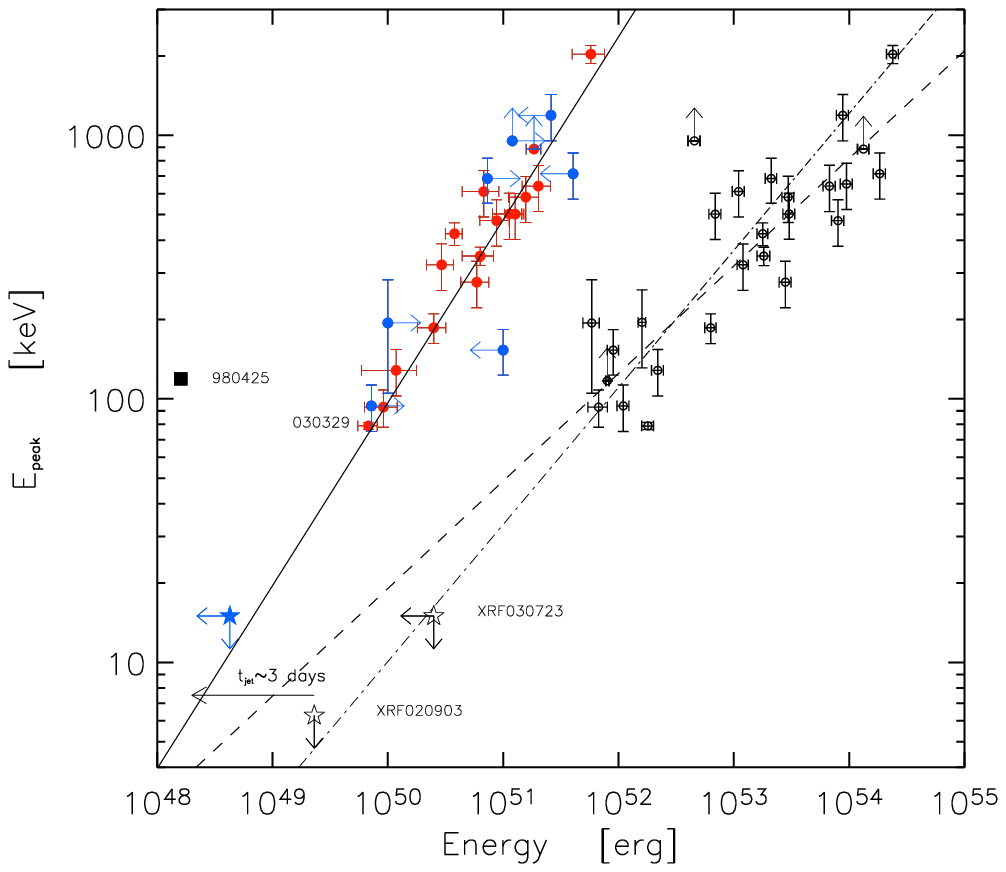}
\includegraphics[width=7.9cm,height=7.3cm,angle=0,clip]{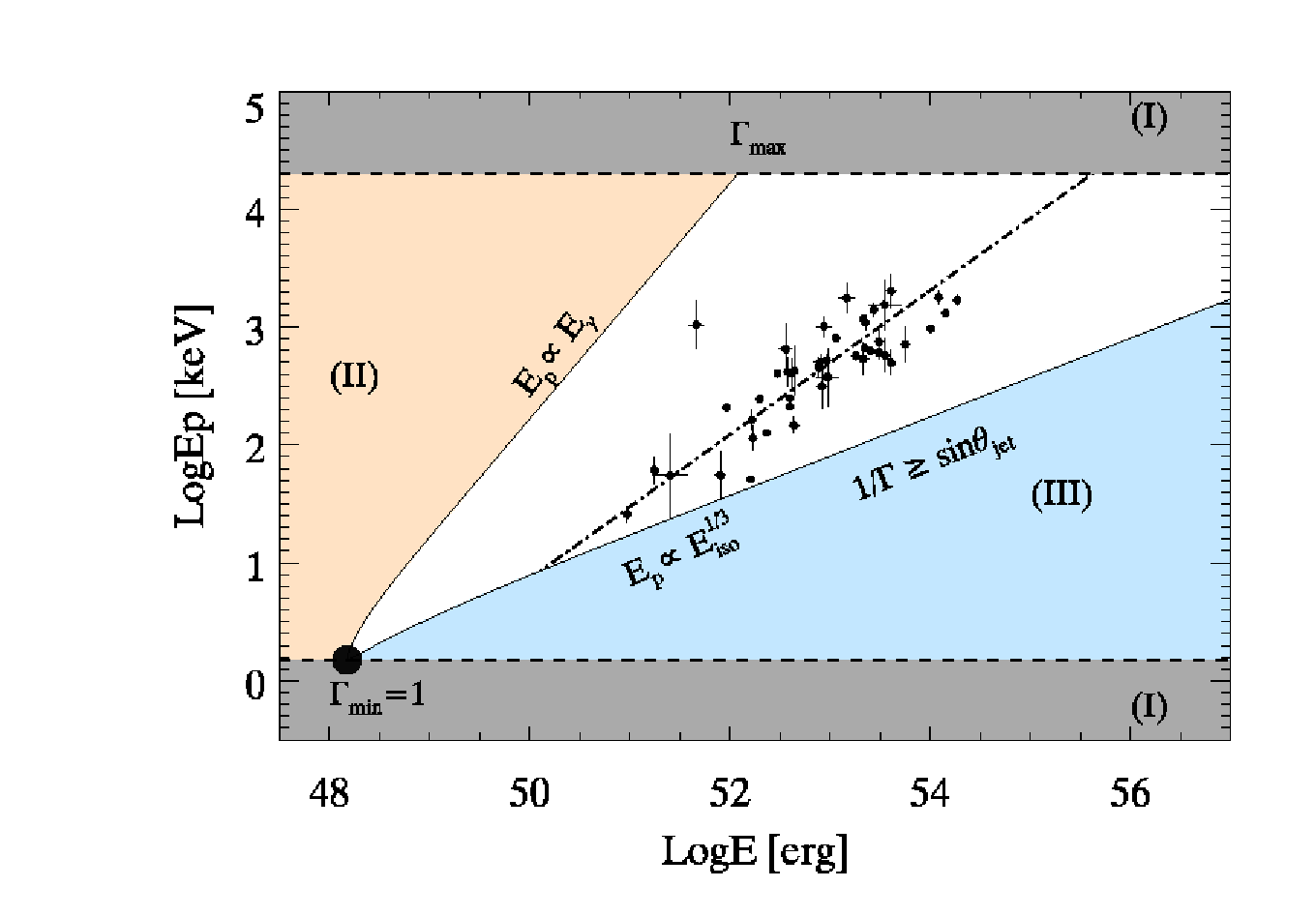}   
\caption{\footnotesize {\bf Left panel:} $E^*_{\rm peak}-E_{\gamma}$ relation for GRBs with known redshift. 
The filled circles represent $E_{\gamma}$ for the events where a jet break was detected. Grey 
symbols indicate lower/upper limits. The solid line represents the best fit, i.e. 
$\log E_{\rm peak} \sim 53.68 + 0.7\log E_{\gamma}$. Open circles denote $E_{\rm iso}$ for the GRBs. The dashed line 
represents the best fit to these points and the dash-dotted line is the relation shown by \citet{AmatiEtal02}. 
(Figure after \cite{Ghirlanda2004}; see Fig. 1 therein. @ AAS. Reproduced with permission.) 
{\bf Right panel:} rest frame plane of GRB energy. The large black dot indicates that all 
simulated GRBs were assigned $E^*_{\rm peak}=1.5\,{\rm keV}$ and $E^*_{\gamma}=1.5\times 10^{48}\,{\rm erg}$. Since 
$\Gamma>1$ but less then 8000, regions (I) are forbidden. Since for all the simulated GRBs 
$\theta_{\rm jet}\leq90^\circ$, they cannot stay in region (II). When $\Gamma$ is small, the beaming cone turns out to 
be larger than the jet. Therefore, the isotropic equivalent energy is given by 
$\log E_{\rm iso}=\log E_{\gamma}+\log(1+\beta_0)+2\log \Gamma$, lower than the energy computed by 
$\log E_{\rm iso}=\log E_{\gamma}-\log(1-\cos \theta_{\rm jet})$. This brings in a constraint, 
$\log E_{\rm peak}\sim 1/3\times \log E_{\rm iso}$, and GRBs cannot lie to the right of this constraint. Hence, 
region (III) is not allowed. The black dots indicate the actual GRBs of the {\it Swift} sample. The fit to the 
Swift sample is displayed as the dot-dashed line. (Figure after \cite{ghirlanda13}; see Fig. 1 therein.)}
\label{fig:9}
\end{figure}

For the Ghirlanda relation \citep{Ghirlanda2004}, with the assumption that the line of sight is within the jet angle, the $E_{\rm peak}-E_{\gamma}$ relation indicates its invariance when moving from the rest frame to the comoving frame. As a result, the number of radiated photons in each GRBs is comparable and should be about $10^{57}$. The last characteristic could be important for understanding the dynamics of GRBs and the radiative mechanisms (see also right panel of Fig.~\ref{fig:9}).

\citet{collazzi11} found that the mean $E^*_{\rm peak}$ is near to $511\,{\rm keV}$, the electron rest-mass energy $m_ec^2$. Therefore, it is claimed that the tight shape of the $E_{\rm peak}$ distribution does not stem only from selection effects. No studied mechanism can drive this effect, however with the $E^*_{\rm peak}$ compatible with the effective temperature of the $\gamma$-ray radiating area, the almost constant temperature needs some mechanism similar to a thermostat, keeping the temperature at a steady value. It was suggested that such a mechanism could be an electron-positron annihilation.

\citet{ghirlanda13}, using a simulated sample, analysed if different intrinsic distributions of $\Gamma$ and $\theta_{\rm jet}$ can replicate a grid of observational constraints. With the assumption that in the comoving frame each GRB has similar $E_{\rm peak}$ and $E_{\gamma}$, it was found that the distributions of $\Gamma$ and $\theta_{\rm jet}$ cannot be power laws. Instead, the highest concordance between simulation and data is given by log-normal distributions, and a connection between their maxima, like $\theta_{\rm jet,max}^{2.5}\Gamma_{\rm max}={\rm const}$. In this work $\theta_{\rm jet}$ and $\Gamma$ are important quantities for the calculation of the GRB energetics. Indeed, from a sample of $\approx 30$ GRBs with known $\theta_{\rm jet}$ or $\Gamma$ it was found that the $E_{\gamma}$ distribution is centered at $10^{50}-10^{51}\,{\rm erg}$ and it is tightly related to $E_{\rm peak}$. It was obtained that
\begin{equation}
\log E_{\rm peak}\sim\log\frac{E_{\gamma}}{5-2\beta_0}.
\label{Equ:EpeakEgamma}
\end{equation}
Present values of $\Gamma$ and $\theta_{\rm jet}$ rely on incomplete data sets and their distributions could be affected by biases. Neverthless, \citet{ghirlanda13} claimed that greater values of $\Gamma$ are related to smaller $\theta_{\rm jet}$ values, i.e. the faster a GRB, the narrower its jet.

Furthermore, GRBs fulfilling the condition $\sin \theta_{\rm jet} < 1/\Gamma$ might not display any jet break in the afterglow light curve, and \citet{ghirlanda13} predicted that this group should comprise $\approx 6\%$ of the on-axis GRBs. Finally, their work is crucial as it allowed to find that the local rate of GRBs is $\approx 0.3\%$ of the local SNe Ib/c rate, and $\approx 4.3$\% of the local hypernovae (i.e., SNe Ib/c with wide-lines) rate.

\subsection{Correlations between the luminosity and the peak energy}\label{sect36}

\subsubsection{The \texorpdfstring{$L_{\rm iso}-E_{\rm peak}$}{Lg} correlation}\label{LisoEpeak}

\begin{figure}[htbp]
\centering
\includegraphics[width=12cm, height=8cm,angle=0]{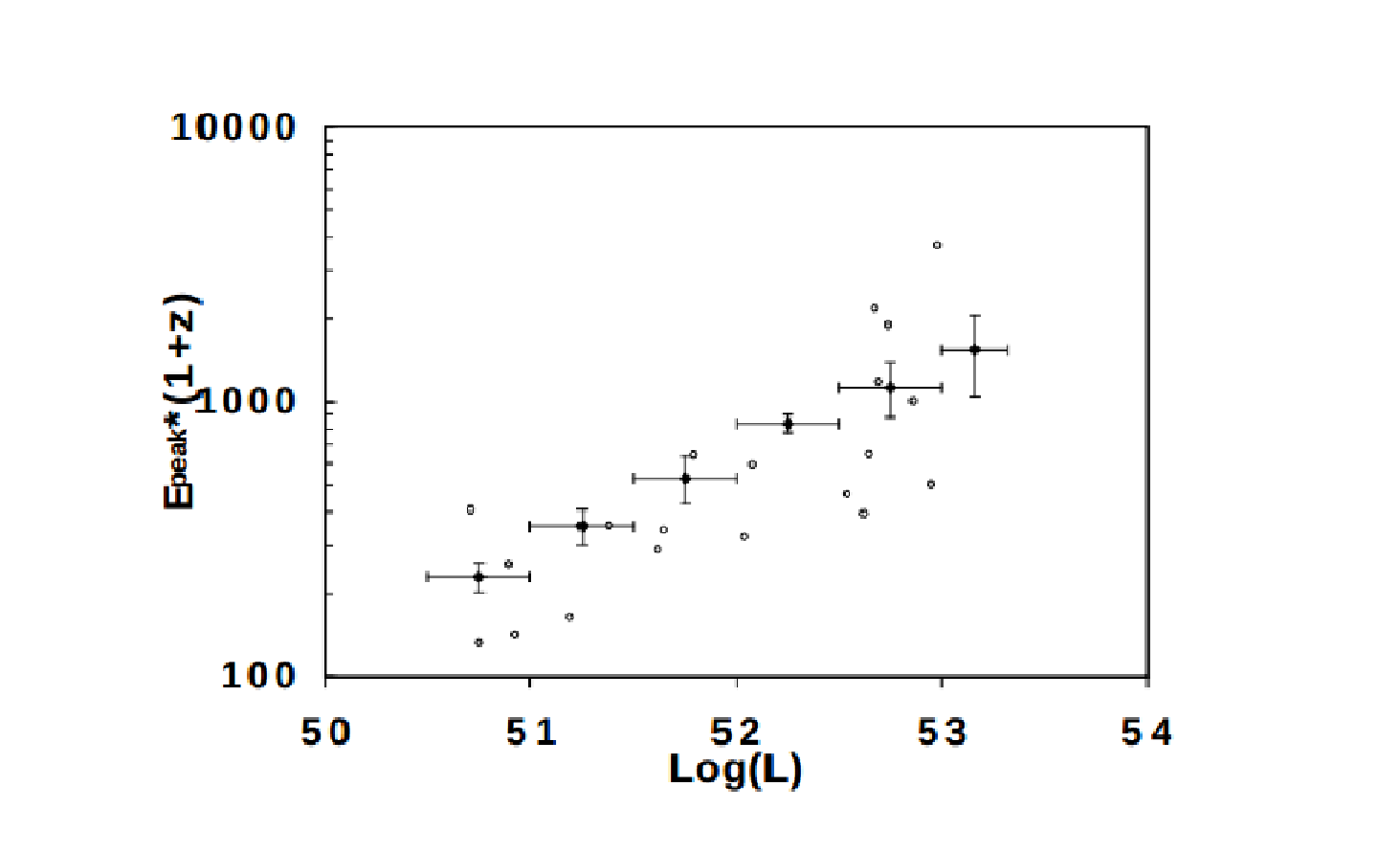}
\caption{\footnotesize Direct fit of $\log L_{\rm iso}-\log E_{\rm peak}$ data. 
This is shown here for two independent data sets for which the luminosities are derived by two independent methods. 
The first data set consists of 20 bursts with spectroscopically measured redshifts (the open circles). The second one 
is for 84 bursts (whose binned values are shown as filled diamonds, and the horizontal bars are the bin widths) whose 
luminosity (and then redshift) were determined with the spectral lag and variability light curve parameters. Both data 
sets show a highly significant and similar power law relations. (Figure after \cite{schaefer03a}; see Fig. 3 therein. @ AAS. Reproduced with permission.)}
\label{fig:4}
\end{figure}

The $L_{\rm iso}-E_{\rm peak}$ relation was discovered by \citet{schaefer03a} who used 84 GRBs with known $E_{\rm peak}$ from the BATSE catalogue \citep{schaefer01}, and 20 GRBs with luminosities based on optically measured redshift \citep{AmatiEtal02,schaefer03b}. It was found that (see Fig.~\ref{fig:4}) for the 20 GRBs
\begin{equation}
\log E_{\rm peak} \sim (0.38 \pm 0.11)\log L_{\rm iso},
\end{equation}
with $r=0.90$ and $P=3\times 10^{-8}$, and among the 84 GRBs the relation was
\begin{equation} \label{eq: prediction}
\log E_{\rm peak} \sim (0.36 \pm 0.03)\log L_{\rm iso}.
\end{equation}
The underlying idea is that the $L_{\rm iso}$ varies as a power of $\Gamma$, as we have already discussed in Sect.~\ref{PhysicalofLpeak-taulag}, and the $E_{\rm peak}$ also varies as some other power of $\Gamma$, so that $E_{\rm peak}$ and $L_{\rm iso}$ will be correlated to each other through their dependence on $\Gamma$. For the general case where the luminosity varies as $\Gamma^{N}$ and $E_{\rm peak}$ varies as $\Gamma^{M}$, and therefore $\log E_{\rm peak}$ will vary as $(M+1)/N\times \log L_{\rm iso}$.

\begin{figure}[htbp]
\centering
\includegraphics[width=12cm, height=8cm,angle=0]{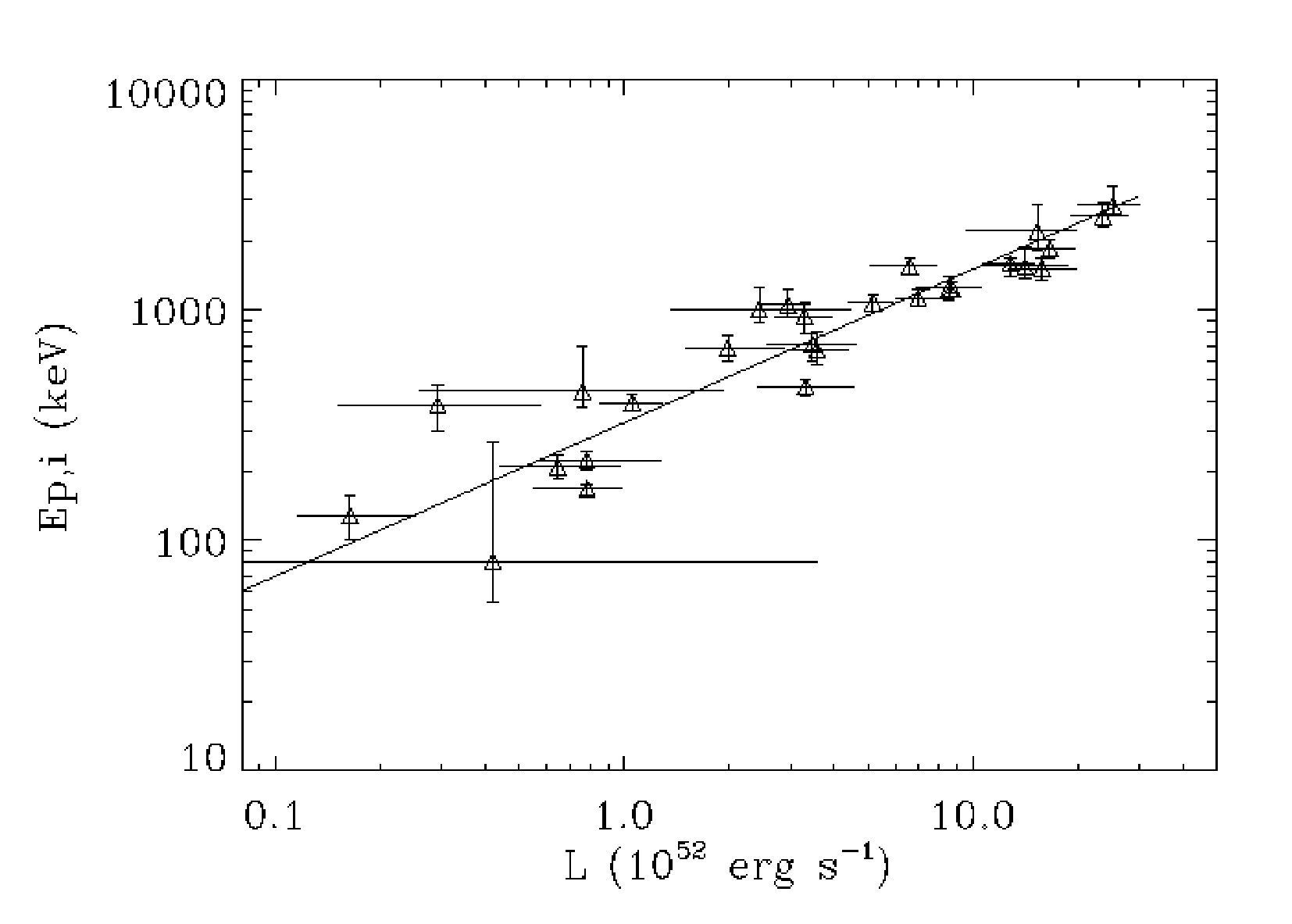}
\caption{\footnotesize $E^*_{\rm peak}$ vs. $L_{\rm iso}$, obtained from data for 
GRBs 990123 and 990510. The solid line is the best-fit power law. (Figure after \cite{frontera12}; see Fig. 6 therein. @ AAS. Reproduced with permission.)}
\label{figfrontera}
\end{figure}

\citet{frontera12}, using a sample of 9 GRBs detected simultaneously with the Wide Field Camera (WFC) on board the {\it BeppoSAX} satellite, and by the BATSE instrument, reported the results of a systematic study of the broadband ($2-2000\,{\rm keV}$) time-resolved prompt emission spectra. However, only 4 of those GRBs (970111, 980329, 990123, 990510) were bright enough to allow a fine time-resolved spectral analysis, resulting in a total of 40 spectra. Finally, the study of the time-resolved dependence (see also the end of Sect.~\ref{Amati}) of $E_{\rm peak}$ on the corresponding $L_{\rm iso}$ was possible for two bursts with known redshift (i.e., 990123 and 990510), and found using the least squares method (see Fig.~\ref{figfrontera}):
\begin{equation}
\log E^*_{\rm peak} \sim (0.66 \pm 0.03)\log L_{\rm iso},
\end{equation}
with $\rho = 0.94$ and $P=1.57\times 10^{-13}$.

Afterwards, \citet{nava12}, using a sample of 46 {\it Swift} GRBs with measured $z$ and $E_{\rm peak}$, found a strong $L_{\rm iso}-E_{\rm peak}$ correlation, with a functional form of
\begin{equation}
\log E^*_{\rm peak} = -(25.33 \pm 3.26)+(0.53 \pm 0.06)\log L_{\rm iso},
\end{equation}
with $\rho=0.65$ and $P=10^{-6}$; $E_{\rm peak}$ is in keV, and $L_{\rm iso}$ is in units of $10^{51}\,{\rm erg}\,{\rm s}^{-1}$. Furthermore, using 12 GRBs with only an upper limit on $z$ (3 events) or no redshift at all (3 events), or with a lower limit on $E_{\rm peak}$ (3 events) or no estimate at all (3 events), they found that these bursts also obey the obtained $L_{\rm iso}-E_{\rm peak}$ relation.

\subsubsection{The \texorpdfstring{$L_{\rm peak}-E_{\rm peak}$}{Lg} correlation}\label{Yonetoku}

\begin{figure}[htbp]
\centering
\includegraphics[width=8cm, height=6cm,angle=0,clip]{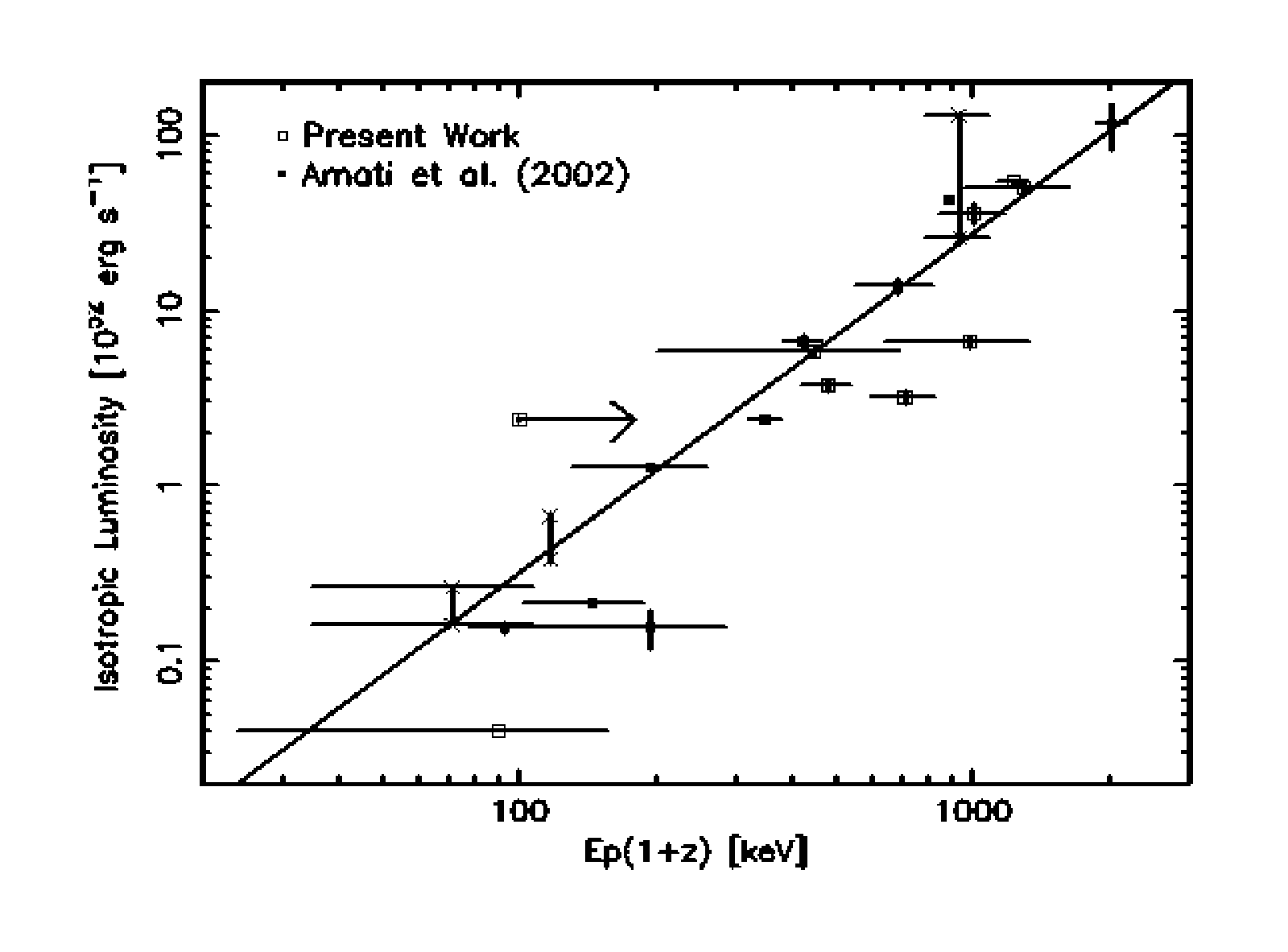}
\includegraphics[width=8cm, height=5.8cm,angle=0,clip]{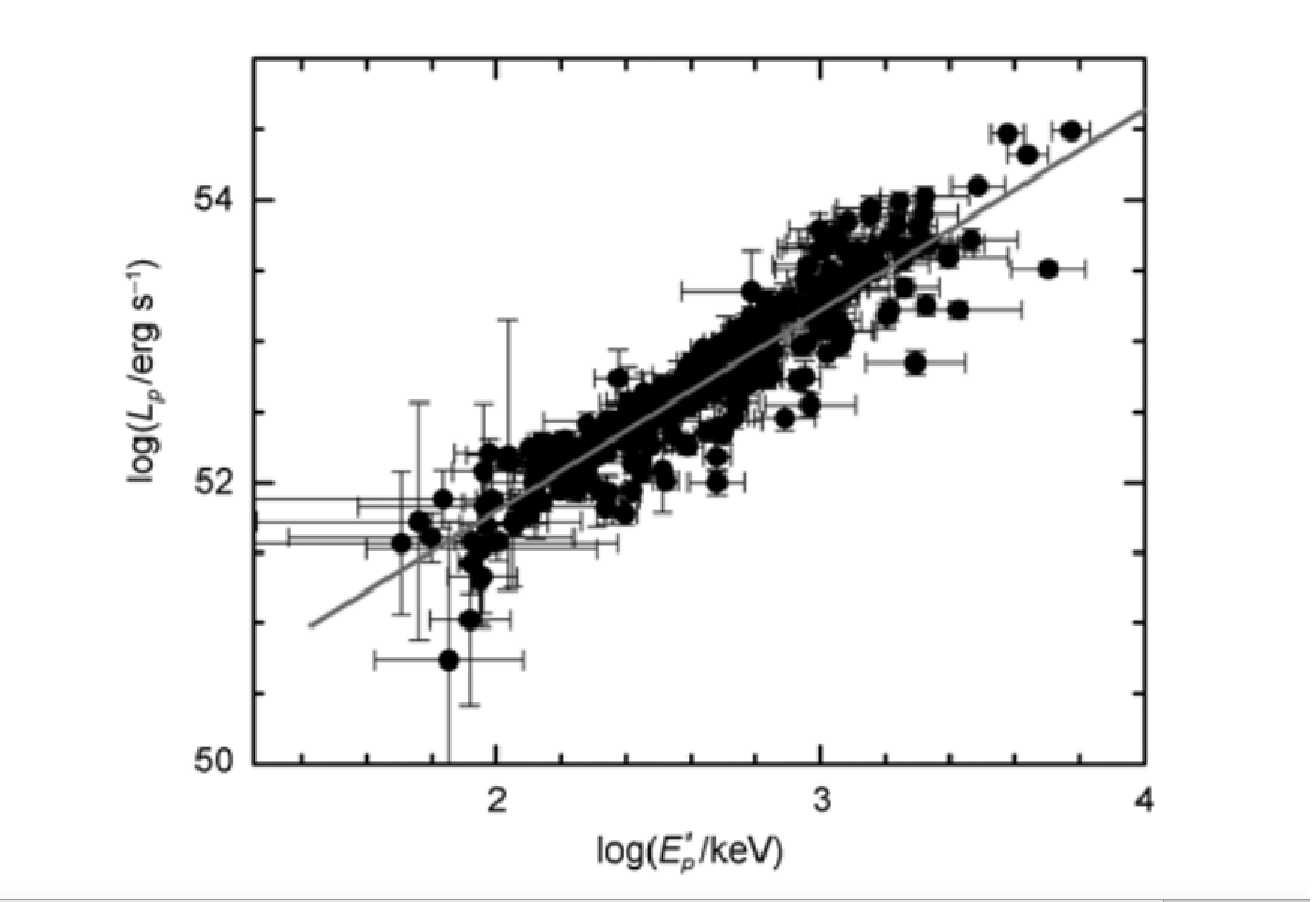}
\caption{\footnotesize {\bf Left panel:} the $\log L_{\rm peak}-\log E_{\rm peak}$ relation. The 
open squares mark the BATSE data. {\it BeppoSAX} events, which are converted into the energy range of 
$30-10000\,{\rm keV}$, are shown as filled squares and the cross points. The solid line indicates the best-fit line. 
(Figure after \cite{yonetoku04}; see Fig. 1 therein. @ AAS. Reproduced with permission.) 
{\bf Right panel:} $\log L_{\rm peak }$ vs. $\log E_{\rm peak}$ for 276 time-resolved spectra within the decay pulses 
for the sample. The solid line stands for the best fit to the data. (Figure after \cite{Lu2010}; see Fig. 4 therein.
Copyright @ 2010 Springer.)}
\label{fig:8}
\end{figure}

It was also found that the Amati relation holds even if $E_{\rm iso}$ is substituted with $L_{\rm iso}$ and $L_{\rm peak}$, which is not surprising given that these ``energy indicators'' are strongly correlated. To this end, the Yonetoku correlation (\citealt{yonetoku04}, see the left panel of Fig.~\ref{fig:8}), relates $E_{\rm peak}$ with $L_{\rm peak}$. The relation was obtained employing 11 GRBs with known redshifts detected by BATSE, together with {\it BeppoSAX} GRBs from \citep{AmatiEtal02}. This relation uses the $L_{\rm peak}$ of the burst instead of $L_{\rm iso}$, and it is tighter than previous prompt correlations. The best-fit line is given by
\begin{equation}
\log L_{\rm peak} \sim (2.0 \pm 0.2)\log E^*_{\rm peak},
\end{equation}
with $r=0.958$, $P=5.31 \times 10^{-9}$, and the uncertainties are $1\sigma$ error. This relation agrees well with the standard synchrotron model \citep{Zhang02,lloyd2000}. Finally, it has been used to estimate pseudo-redshifts of 689 BATSE LGRBs with unknown distances and to derive their formation rate as a function of $z$.

\citet{ghirlanda04a} selected 36 bright SGRBs detected by BATSE, with an $F_{\rm peak}$ on the $64\,{\rm ms}$ timescale in the $50-300\,{\rm keV}$ energy range exceeding $10\,{\rm ph}\,{\rm cm}^{-2}\,{\rm s}^{-1}$. In 7 cases, the signal-to-noise-ratio was too low to reliably constrain the spectral best-fit parameters. One case yielded missing data. Hence, the sample consisted of 28 events. Due to unknown redshifts, $E^*_{\rm peak}$, $E_{\rm iso}$ and $L_{\rm peak}$ were expressed as functions of the redshift in the range $z\in[0.001,10]$. It was found that SGRBs are unlikely to obey the Amati relation, $E_{\rm iso}-E^*_{\rm peak}$, but the results were consistent with the $L_{\rm peak}-E^*_{\rm peak}$ relation of \citet{yonetoku04}. Hence, assuming that this relation indeed holds for SGRBs, their pseudo-redshifts were estimated and found to have a similar distribution as LGRBs, with a slightly smaller average redshift.

Afterwards, \citet{Yonetoku2010} investigated the prompt emission of 101 GRBs with measured redshifts and a reported $F_{\rm peak}$ detected until the end of 2009. The sample comes from events detected in a number of independent missions: the satellites used for this purpose are {\it KONUS}, {\it Swift}, {\it HXD-WAM} and {\it RHESSI}. Using this data set, the $E_{\rm peak}-L_{\rm peak}$ correlation was revised, and its functional form could be written as
\begin{equation}
\log L_{\rm peak}=(52.43 \pm 0.037)+ (1.60 \pm 0.082)\log E^*_{\rm peak},
\end{equation}
with $r=0.889$ for $99$ degrees of freedom and an associated $P=2.18 \times 10^{-35}$; $L_{\rm peak}$ is expressed in ${\rm erg}\,{\rm s}^{-1}$, and $E^*_{\rm peak}$ in units of $355\,{\rm keV}$. To provide reference to previous works, the $1-10^4\,{\rm keV}$ energy band in the GRB rest frame was used to calculate the bolometric energy and $L_{\rm peak}$. Finally, it was demonstrated that this relation is intrinsic to GRBs and affected by the truncation effects imposed by the detector threshold.

\citet{Lu2010}, using time-resolved spectral data for a sample of 30 pulses in 27 bright GRBs detected by BATSE, investigated the $L_{\rm peak}-E_{\rm peak}$ relation in the decay phases of these pulses (see right panel of Fig.~\ref{fig:8}). Quite all of the pulses followed a narrow $L_{\rm peak}-E_{\rm peak}$ relation given by
\begin{equation}
\log L_{\rm peak} \sim (1.42 \pm 0.03)\log E^*_{\rm peak}, 
\end{equation}
with $r = 0.91$ and $P < 10^{-4}$, but the power law index varied. The statistical or observational effects could not account for the large scatter of the power law index, and it was suggested to be an intrinsic feature, indicating that no relation common for all GRB pulses $L_{\rm peak}-E_{\rm peak}$ would be expected. However, in the light of {\it Fermi} observations that revealed deviations from the Band function (\citealt{abdo2009,guiriec2010,ackermann2010,ackermann2011,ackermann2013}; see also \citealt{lin2016}), it was proposed recently that the GRB spectra should be modelled not with the Band function itself (constituting a non-thermal component), but with additional black-body (BB, thermal) and power law (PL, non-thermal) components \citep{guiriec2013,guiriec2015a,guiriec2015b,guiriec2016}. The non-thermal component was well described within the context of synchrotron radiation from particles in the jet, while the thermal component was interpreted by the emission from the jet photosphere. The PL component was claimed to originate most likely from the inverse Compton process. The results point toward a universal relation between $L_{\rm peak}$ and $E^*_{\rm peak}$ related to the non-thermal components.

\citet{tsutsui13} analysed 13 SGRB candidates (i.e., an SGRB with $T^*_{90}<2\,{\rm s}$), from among which they selected 8 events considering them as secure ones. An SGRB candidate is regarded as a misguided SGRB if it is located within the $3\sigma_{\rm int}$ dispersion region from the best-fit $E^*_{\rm peak}-E_{\rm iso}$ function of the correlation for LGRBs, while the others are regarded as secure SGRBs. The relation obtained with secure GRBs is the following:
\begin{equation}
\log L_{\rm peak}=(52.29 \pm 0.066) +(1.59 \pm 0.11)\log E^*_{\rm peak},
\label{Tsutsuishort}
\end{equation}
with $r=0.98$ and $P= 1.5 \times 10^{-5}$, where $E^*_{\rm peak}$ (in units of $774.5\,{\rm keV}$) is from the time-integrated spectrum, while $L_{\rm peak}$ (in ${\rm erg}\,{\rm s}^{-1}$) was taken as the luminosity integrated for $64\,{\rm ms}$ at the peak considering the shorter duration of SGRBs. Application of this relation to 71 bright BATSE SGRBs resulted in pseudo-redshifts distributed in the range $z\in[0.097,2.258]$, with $\langle z\rangle=1.05$, which is apparently lower than $\langle z\rangle=2.2$ for LGRBs. Finally, \citet{yonetoku14}, using 72 SGRBs with well determined spectral features as observed by BATSE, determined their pseudo-redshifts and luminosities by employing the $L_{\rm peak}-E_{\rm peak}$ correlation for SGRBs found by \cite{tsutsui13}. It was found that the obtained redshift distribution for $z\leq 1$ was in agreement with that of 22 {\it Swift} SGRBs, indicating the reliability of the redshift determination via the $E^*_{\rm peak}-L_{\rm peak}$ relation.

\subsubsection{Physical interpretation of the luminosity vs. peak energy relations}
As pointed out by \cite{schaefer01} and \cite{schaefer03a}, $E_{\rm peak}$ and $L_{\rm iso}$ are correlated because of their dependence on $\Gamma$. The $L_{\rm iso}-E_{\rm peak}$ relation could shed light on the structure of the ultra relativistic outflow, the shock acceleration and the magnetic field generation \citep{Lloyd2002}. However, since only few SGRBs are included in the samples used, the correlations and interpretations are currently only applicable to LGRBs.

\cite{schaefer01} and \cite{schaefer03a} claimed that the values of $E_{\rm peak}$ are approximately constant for all the 
bursts with $z \ge 5$. However, with the launch of the {\it Swift} satellite in the end of 2004 the hunt for ``standard candles" via a number of GRB correlations is still ongoing. Thus, the great challenge is to find universal constancy in some GRB parameters, despite the substantial diversity exhibited by their light curves. If this goal is achieved, GRBs might prove to be a useful cosmological tool \citep{wang15}.

\citet{liang04} defined a parameter $\omega=(L_{\rm iso}/10^{52}\,{\rm erg}\,{\rm s}^{-1})^{0.5}/(E_{\rm peak}/200\,{\rm keV})$ and discussed possible implications of the $E_{\rm peak}-L_{\rm iso}$ relation for the fireball models. They found that $\omega$ is limited to the range $\simeq 0.1-1$. They constrained some parameters, such as the combined internal shock parameter, $\zeta_i$, for the internal as well as external shock models, with an assumption of uncorrelated model parameters. Their distributions suggest that the production of prompt $\gamma$-rays within internal shocks dominated by kinetic energy is in agreement with the standard internal shock model. Similarly in case when the $\gamma$-rays come from external shocks dominated by magnetic dissipation. These results imply that both models can provide a physical interpretation of the $L_{\rm iso}\propto E^2_{\rm peak}$ relation as well as the parameter $\omega$.

To explain the origin of this correlation, \citet{mendoza2009} considered simple laws of mass and linear momentum conservation on the emission surface to give a full description of the working surface flow parameterized by the initial velocity and mass injection rate. They assumed a source-ejecting matter in a preferred direction $x$ with a velocity $v(t)$ and a mass ejection rate $\dot{m}(t)$, both dependent on time $t$ as measured from the jet's source; i.e., they studied the case of a uniform release of mass and the luminosity was measured considering simple periodic oscillations of the particle velocity, a common assumption in the internal shock model scenario.

Due to the presence of a velocity shear with a considerable variation in $\Gamma$ at the boundary of the spine and sheath region, a fraction of the injected photons are accelerated via a Fermi-like acceleration mechanism such that a high energy power law tail is formed in the resultant spectrum. \citet{Ito2013} showed in particular that if a velocity shear with a considerable variance in $\Gamma$ is present, the high energy part of the observed GRB photon spectrum can be explained by this photon acceleration mechanism. The accelerated photons may also account for the origin of the extra hard power law component above the bump of the thermal-like peak seen in some peculiar GRBs (090510, 090902B, 090926A). It was demonstrated that time-integrated spectra can also reproduce the low energy spectra of GRBs consistently due to a multi-temperature effect when time evolution of the outflow is considered.

Regarding the Yonetoku relation, its implications are related to the GRB formation rate and the luminosity function of GRBs. In fact, the analysis of \citet{yonetoku04} showed that the existence of the luminosity evolution of GRBs, assuming as a function a simple power law dependence on the redshift, such as $g(z) = (1+z)^{1.85}$, may indicate the evolution of GRB progenitor itself (mass) or the jet evolution. To study the evolution of jet opening angle they considered two assumptions: either the maximum jet opening angle decreases or the total jet energy increases. In the former case, the GRB formation rate obtained may be an underestimation since the chance probability to observe the high redshift GRBs will decrease. If so, the evolution of the ratio of the GRB formation rate to the star formation rate becomes more rapid. On the other hand, in the latter case, GRB formation rate provides a reasonable estimate.

Recently \citet{frontera2016}, building on the spectral model of the prompt emission of \citet{titarchuk2012}, gave a physical interpretation of the origin of the time resolved $L_{\rm iso}-E_{\rm peak}$ relation. The model consists of an expanding plasma shell, result of the star explosion, and a thermal bath of soft photons. \citet{frontera2016} showed analytically that in the asymtotic case of the optical depth $\tau\gg 1$ the relation $\log L_{\rm iso}-\log E_{\rm peak}$ indeed has a slope of $1/2$. This, in turn, is evidence for the physical origin of the Amati relation (see Sect.~\ref{Amati}).

\subsection{Comparisons between \texorpdfstring{$E_{\rm peak}-E_{\rm iso}$}{Lg} and \texorpdfstring{$E_{\rm peak}-L_{\rm peak}$}{Lg} correlation}

For a more complete dissertation we compare the $E_{\rm peak}-E_{\rm iso}$ correlation with the $E_{\rm peak}-L_{\rm peak}$ correlation. To this end, \citet{ghirlanda2005} derived the $E_{\rm peak}-L_{\rm peak}$ relation with a sample of 22 GRBs with known $z$ and well determined spectral properties. This relation has a slope of $0.51$, similar to the one proposed by \citet{yonetoku04} with 12 GRBs, although its scatter is much larger than the one originally found.

\citet{Tsutsui2009} investigated these two relations using only data from the prompt phase of 33 low-redshift GRBs with $z\leq 1.6$. In both cases the correlation coefficient was high, but a significant scatter was also present. Next, a partial linear correlation degree, which is the degree of association between two random variables, was found to be $\rho_{L_{\rm peak},E_{\rm iso},E_{\rm peak}} = 0.38$. Here, $\rho_{1,2,3}$ means the correlation coefficient between the first and the second parameter after fixing the third parameter. This fact indicates that two distance indicators may be independent from each other even if they are characterized by the same physical quantity, $E_{\rm peak}$, and similar quantities, $L_{\rm peak}$ and $E_{\rm iso}$. To correct the large dispersion of the Yonetoku correlation, \cite{Tsutsui2009} introduced a luminosity time constant $T_L$ defined by $T_L = E_{\rm iso}/L_{\rm peak}$ as a third parameter and a new correlation was established in the form
\begin{equation}
\log L_{\rm peak}=(-3.87\pm 0.19)+(1.82\pm 0.08)\log E_{\rm peak}-(0.34\pm 0.09)\log T_L,
\end{equation}
with $r=0.94$ and $P=10^{-10}$. Here, $L_{\rm peak}$ is in units of $10^{52}\,{\rm erg}\,{\rm s}^{-1}$, $E_{\rm peak}$ is in keV, and $T_L$ in seconds. In this way the systematic errors were reduced by about 40\%, and the plane represented by this correlation might be regarded as a ``fundamental plane" of GRBs.

Later, \citet{Tsutsui2010} reconsidered the correlations among $E_{\rm peak}$, $L_{\rm peak}$ and $E_{\rm iso}$, using the
database constructed by \cite{Yonetoku2010}, which consisted of 109 GRBs with known redshifts, and $E_{\rm peak}$, 
$L_{\rm peak}$ and $E_{\rm iso}$ well determined. The events are divided into two groups by their data quality. One (gold data set) consisted of GRBs with $E_{\rm peak}$ determined by the Band function with four free parameters. GRBs in the other group (bronze data set) had relatively poor energy spectra so that their $E_{\rm peak}$ were determined by the Band function with three free parameters (i.e., one spectral index was fixed) or by the cut-off power law (CPL) model with three free parameters. Using only the gold data set, the intrinsic dispersion, $\sigma_{\rm int}$, in $\log L_{\rm peak}$ is 0.13 for the $E_{\rm peak}-T_L-L_{\rm peak}$ correlation, and 0.22 for the $E_{\rm peak}-L_{\rm peak}$ correlation. In addition, GRBs in the bronze data set had systematically larger $E_{\rm peak}$ than expected by the correlations constructed with the gold data set. This indicates that the quality of the sample is an important issue for the scatter of correlations among $E_{\rm peak}$, $L_{\rm peak}$, and $E_{\rm iso}$.

The difference between the $E_{\rm peak}-L_{\rm peak}$ correlation for LGRBs from \citep{ghirlanda10} and the one from \citep{Yonetoku2010} is due to the presence of GRB060218. In the former, it was considered an ordinary LGRB, while in the latter, an outlier by a statistical argument. Because GRB060218 is located far from the $L_{\rm peak}-E_{\rm peak}$ correlation in \citep{Yonetoku2010} (more than $8\sigma$), it makes the best-fit line much steeper.

Regarding the high-energetic GRBs, \cite{ghirlanda10} considered 13 GRBs detected by {\it Fermi} up to the end of July 2009, and with known redshift. They found a tight relation: 
\begin{equation}
\log E^*_{\rm peak} \sim 0.4\log L_{\rm iso},
\end{equation}
with a scatter of $\sigma=0.26$. A similarly tight relation exists between $E^*_{\rm peak}$ and $E_{\rm iso}$:
\begin{equation}
\log E^*_{\rm peak} \sim 0.5\log E_{\rm iso}.
\end{equation}
The time integrated spectra of 8 {\it Fermi} GRBs with measured redshift were consistent with both the $E_{\rm peak}-E_{\rm iso}$ and the $E_{\rm peak}-L_{\rm iso}$ correlations defined by 100 pre-{\it Fermi} bursts.

Regarding the study of SGRBs within the context of these two correlations, \citet{tsutsui13} used 8 SGRBs out of 13 SGRB candidates to check whether the $E_{\rm peak}-E_{\rm iso}$ and $E_{\rm peak}-L_{\rm peak}$ correlations exist for SGRBs as well. It was found that the $E_{\rm peak}-E_{\rm iso}$ correlation seemed to hold in the form
\begin{equation}
\log E_{\rm iso} =(51.42\pm 0.15)+(1.58\pm 0.28)\log E^*_{\rm peak},
\end{equation}
with $r = 0.91$, $P = 1.5\times10^{-3}$, $E_{\rm iso}$ in ${\rm erg}\,{\rm s}^{-1}$ and $E^*_{\rm peak}$ in units of $774.5\,{\rm keV}$. They also found that the $E_{\rm peak}-L_{\rm peak}$ correlation with a functional form as in Eq.~(\ref{Tsutsuishort}) is tighter than the $E_{\rm peak}-E_{\rm iso}$ one. Both correlations for SGRBs indicate that they are less luminous than LGRBs, for the same $E_{\rm peak}$, by factors $\simeq 100$ (for $E_{\rm peak}-E_{\rm iso}$), and $\simeq 5$ (for $E_{\rm peak}-L_{\rm peak}$). It was the first time that the existence of distinct $E_{\rm peak}-E_{\rm iso}$ and $E_{\rm peak}-L_{\rm peak}$ correlations for SGRBs was argued.

\subsection{The \texorpdfstring{$L_{X,p}-T^*_p$}{Lg} correlation and its physical interpretation}

Using data gathered by {\it Swift}, \citet{Willingale2007} proposed a unique phenomenological function to estimate some relevant parameters of both the prompt and afterglow emission. Both components are well fitted by the same functional form:
\begin{equation}
f_i(t)=
\begin{cases} F_i e^{\alpha_i(1-\frac{t}{T_i})}e^{-\frac{t_i}{t}}, & \ t
< T_i \,, \\
F_i(\frac{t}{T_i})^{-\alpha_i} e^{-\frac{t_i}{t}}, & t\ge T_i \,.
\end{cases}
\end{equation}
\begin{figure}[htbp]
\centering
\includegraphics[width=13cm,height=13cm,angle=0]{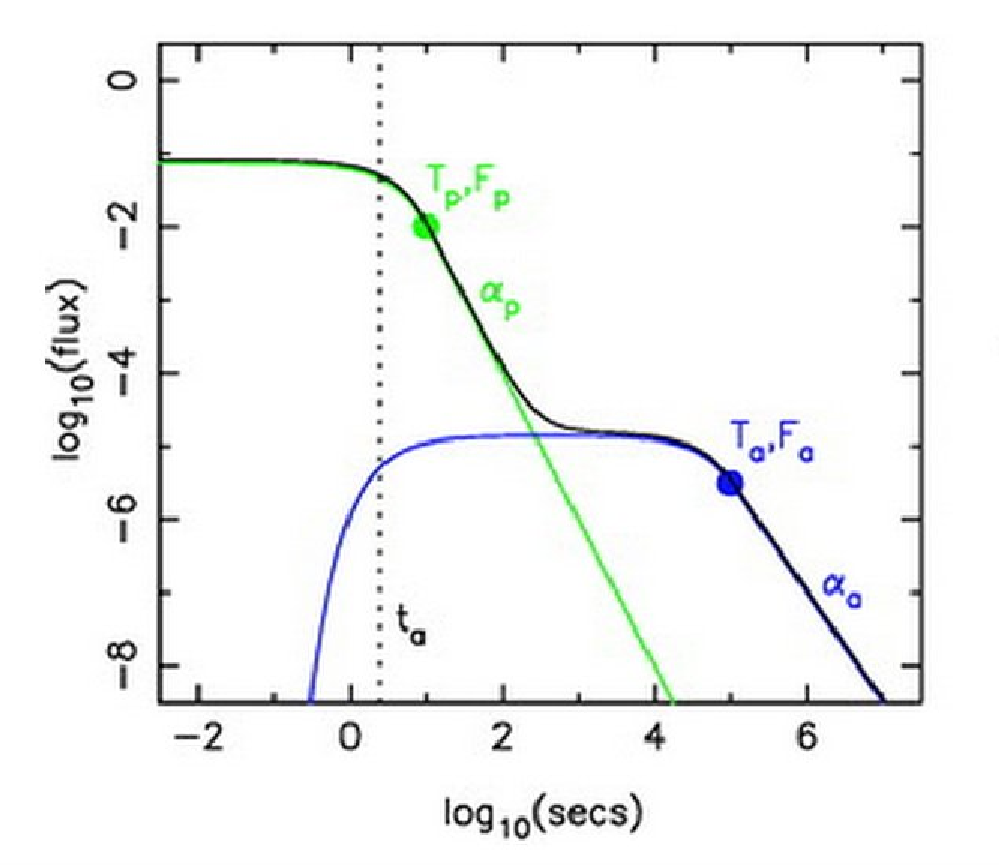}
\caption{\footnotesize Functional form of the decay and the fitted parameters. 
The prompt component (green curve) has no rise because time zero is set at the peak. The afterglow component 
(blue curve) rises at time $T_a$ as shown. (Figure after \cite{Willingale2007}; see Fig. 1 therein. @ AAS. Reproduced with permission.)}
\label{fig:10}
\end{figure}
The index $i$ can take the values $p$ or $a$ to indicate the prompt and afterglow, respectively. The complete light curve, $f_{\rm tot}(t) = f_p(t) + f_a(t)$, is described by two sets of four parameters each: \{$T_i, F_i, \alpha_i, t_i$\}, where $\alpha_i$ is the temporal power law decay index, the time $t_i$ is the initial rise timescale, $F_i$ is the flux and $T_i$ is the break time. Fig.~\ref{fig:10} schematically illustrates this function.

Following the same approach as adopted in \citep{Dainotti2008}, \citet{Qi2010} investigated the prompt emission properties of 107 GRB light curves detected by the XRT instrument onboard the {\it Swift} satellite in the X-ray energy band ($0.3-10\,{\rm keV}$). They found that there is a correlation between $L_{X,p}$ and $T^{*}_p$. Among the 107 GRBs, they used only 47, because some of the events did not have a firm redshift and some did not present reliable spectral parameters in the prompt decay phase. Among the 47 GRBs, only 37 had $T^{*}_p>2\,{\rm s}$, and 3 of them had $T^{*}_p>100\,{\rm s}$.

\begin{figure}[htbp]
\includegraphics[width=5.2cm,height=4.3cm,angle=0]{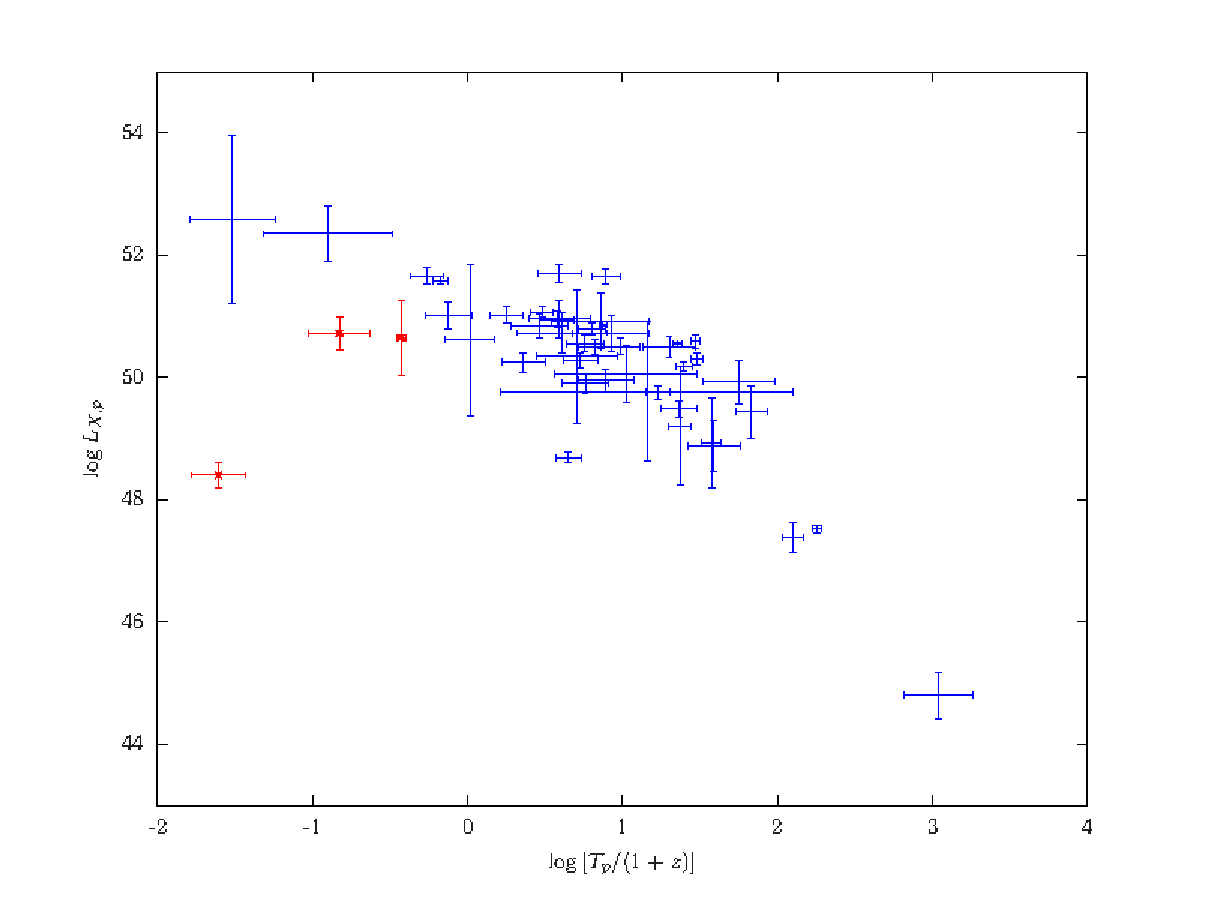}
\includegraphics[width=5.2cm,height=4.3cm,angle=0]{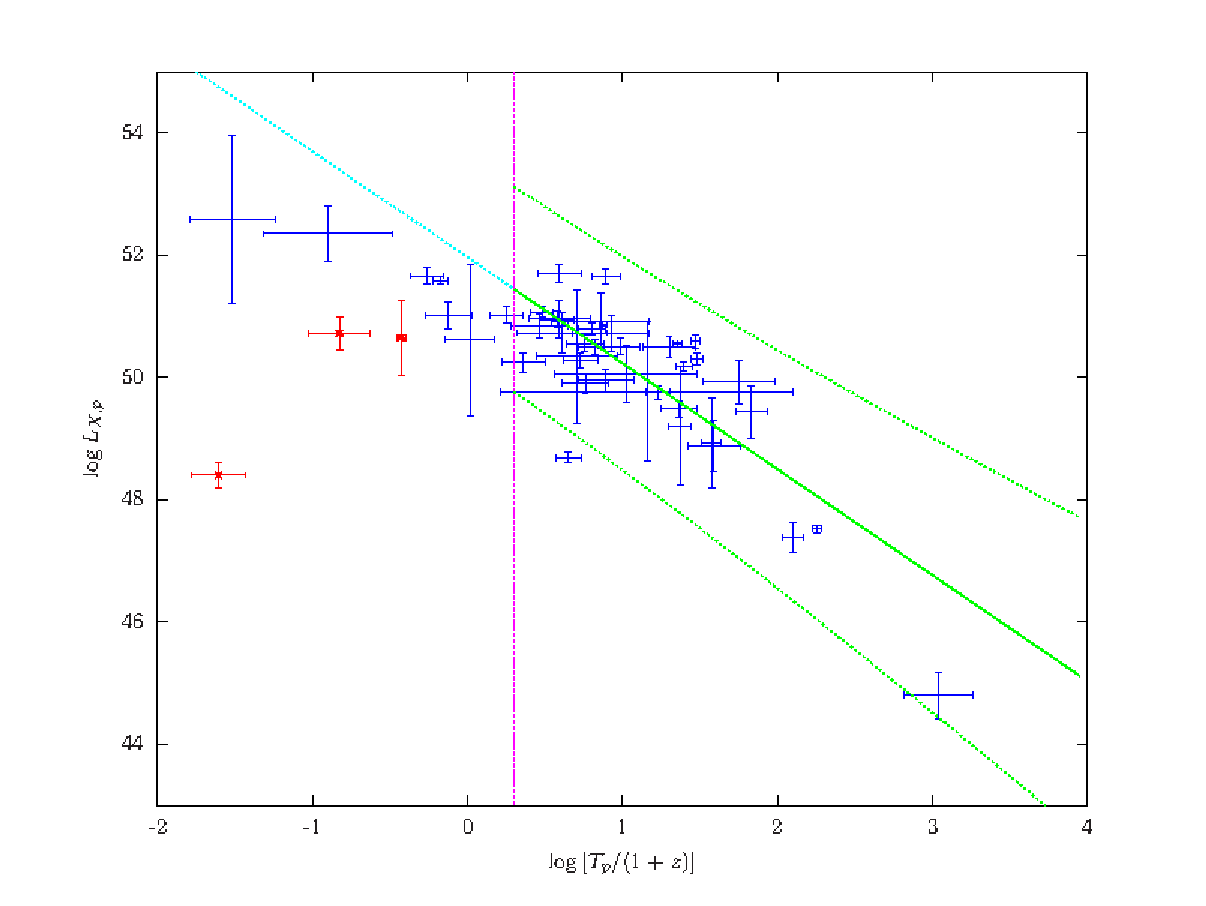}
\includegraphics[width=5.2cm,height=4.3cm,angle=0]{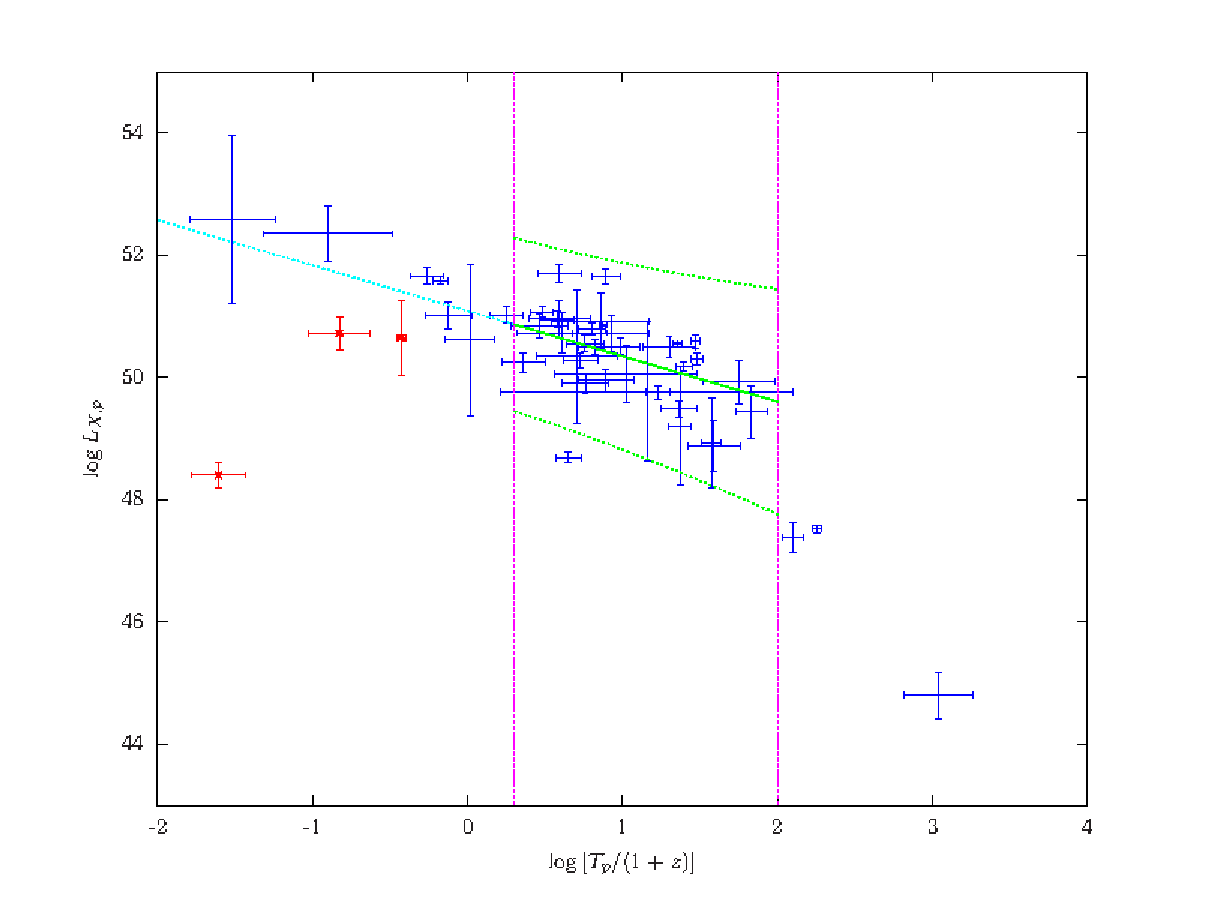}
\caption{\footnotesize {\bf Left panel:} $\log L_{X,p}$ (in ${\rm erg}\,{\rm s}^{-1}$) vs. $\log T^{*}_p$ (in s) 
for the whole sample of 47 GRBs. The red dots represent SGRBs (i.e., $T_{90} < 2\,{\rm s}$). (Figure after \cite{Qi2010}; see Fig. 1 therein. @ AAS. Reproduced with permission.) 
{\bf Middle panel:} best fit of the $\log L_{X,p}$ (in ${\rm erg}\,{\rm s}^{-1}$) vs. $\log T^{*}_p$ (in s) relation 
following Eq.~(\ref{Equ:Qi}) and the corresponding $2\sigma$ confidence region. Only GRBs with 
$T^{*}_p>2\,{\rm s}$ are included in the fit. (Figure after \cite{Qi2010}; see Fig. 2 therein. @ AAS. Reproduced with permission.)
{\bf Right panel:} Best fit of the $\log L_{X,p}$ (in ${\rm erg}\,{\rm s}^{-1}$) vs $\log T^*_p$ (in s) relation 
following Eq.~(\ref{Equ:Qi}) and the corresponding $2\sigma$ confidence region. In this case only the 34 GRBs with 
$2\,{\rm s}< T^{*}_p< 100\,{\rm s}$ are included in the fit. (Figure after \cite{Qi2010}; see Fig. 3 therein. @ AAS. Reproduced with permission.)}
\label{fig:Qi}
\end{figure}

The functional form for this correlation could be written in the following way:
\begin{equation}
\log L_{X,p} = a + b\log T^{*}_p,
\label{Equ:Qi}
\end{equation}
where $L_{X,p}$ is in ${\rm erg}\,{\rm s}^{-1}$, and $T^{*}_p$ is in seconds. The fits were performed via the \citet{Dagostini2005} fitting method applied to the following data sets:
\begin{enumerate}
\item the total sample of 47 GRBs (see the left panel of Fig.~\ref{fig:Qi}),
\item 37 GRBs with $T^{*}_p > 2\,{\rm s}$ (see the middle panel of Fig.~\ref{fig:Qi}),
\item 34 GRBs with $2\,{\rm s} < T^{*}_p < 100\,{\rm s}$ (see the right panel of Fig.~\ref{fig:Qi}).
\end{enumerate}
The results of these fittings turned out to give different forms of Eq.~(\ref{Equ:Qi}). In case 1., $a=50.91 \pm 0.23$ and $b=-0.89 \pm 0.19$ were obtained. The slope $b$ is different in cases 2. and 3., $b=-1.73$ and $b=-0.74$, respectively. The best fit with the smallest $\sigma_{\rm int}$ comes from case 3. Remarkably, in this case the slope $b$ is close to the slope ($-0.74^{+0.20}_{-0.19}$) of a similar $\log L_X-\log T^*_a$ relation \citep{Dainotti2008}. 

\cite{Qi2010} noticed a broken linear relation of the $L_{X,p}-T^*_p$ correlation. More specifically, an evidence of curvature appears in the middle panel of Fig.~\ref{fig:Qi}. One can see, from the left panel of Fig.~\ref{fig:Qi}, that if the best-fit line is extended to the range of $T^{*}_p < 2\,{\rm s}$, all the GRBs with $T^{*}_p < 2\,{\rm s}$ are located below this line. However, the small sample of GRBs used in their analysis is still not sufficient to conclude whether the change in the slope is real or just a selection bias caused by outliers. If there is a change in the slope this may suggest that GRBs could be classified into two groups, long and short, based on their values of $T^{*}_p$ instead of $T_{90}$, since $T^{*}_p$ is an estimate of the GRB duration based on temporal features of the light curves, and $T_{90}$ is a measure based on the energy. This idea has actually been proposed for the first time by \cite{OBrien2007}. It is worth noting that while $T_{90}$ and $T_p$ are both estimates of the GRB duration, the correlation does not hold if $T_p$ is replaced with $T_{90}$. For an analysis of an extended sample and comparison of $T_{45}$ versus $T_p$ also, see \citep{Dainotti11b}.

Regarding the physical interpretation, the change of the slope in the $L_{X,p}-T^*_p$ relation at different values of $T^{*}_p$ in \citep{Qi2010} can be due to the presence of few GRBs with a large $T^{*}_p$, but it might also be due to different emission mechanisms. Unfortunately, the paucity of the sample prevents putting forward any conclusion due to the presence of (potential) outliers in the data set. A more detailed analysis is necessary to further validate this correlation and better understand its physical interpretation.

\subsection{The \texorpdfstring{$L_f-T_f$}{Lg} correlation and its physical interpretation}\label{Willingale2010}

In most GRBs a rapid decay phase (RDP) soon after the prompt emission is observed \citep{Nousek2006}, and this RDP appears continue smoothly after the prompt, both in terms of temporal and spectral variations \citep{Obrien06}. This indicates that the RDP could be the prompt emission's tail and a number of models have been proposed to take it into account (see \citealt{zhang07c}), in particular the high latitude emission (HLE). This model states that once the prompt emission from a spherical shell turns off at some radius, then the photons reach the observer from angles apparently larger (relative to the line of sight) due to the added path length caused by the curvature of the emitting region. The Doppler factor of these late-arriving photons is smaller.

A successful attempt to individually fit all the distinct pulses in the prompt phase and in the late X-ray flares observed by the complete {\it Swift}/BAT+XRT light curves has been performed by \citet{willingale2010} using a physically motivated pulse profile. This fitting is an improved procedure compared to the \citet{Willingale2007} one. The pulse profile has the following functional form:
\begin{equation}
P= \left\{\left[\min\left(\frac{T-T_{\rm{ej}}}{T_f},1 \right)^{\alpha+2}-\left(\frac{T_f-T_{\rm{rise}}}{T_f}\right)^{\alpha + 2}\right]\left[ 1- \left( \frac{T_f-T_{\rm{rise}}}{T_f} \right)^{\alpha + 2} \right]^{-1} \right\} \left( \frac{T-T_{\rm{ej}}}{T_f} \right)^{-1},
\end{equation}
where $T_0=T_f-T_{\rm{rise}}$ (with $T_{\rm rise}$ the rise time of the pulse) is the arrival time of the first photon 
emitted from the shell. It is assumed here that the emission comes from an ultra-relativistic thin shell spreading over a finite range of radii along the line of sight, in the observer frame measured with respect to the ejection time, $T_{\rm{ej}}$. From these assumptions it is possible to model the rise of the pulse through $\alpha$, $T_{\rm{rise}}$ and $T_f$ (see also Fig.~\ref{fig1}).

The combination of the pulse profile function $P(t,T_{f},T_{\rm rise})$ and the blue-shift of the spectral profile $B(x)$ produces the rise and fall of the pulse. $B(x)$ is approximated with the Band function in the form
\begin{equation}
B(x)=B_{\rm norm}\times \left\{
\begin{array}{ll}
x^{(\alpha-1)} e^{-x}, & x \le \alpha-\beta \\
x^{(\beta-1)}(\alpha-\beta)^{(\alpha-\beta)} e^{-(\alpha-\beta)}, & x > \alpha-\beta \\
\end{array}
\right.
\label{eqA}
\end{equation}
where $x=\left(E/E_{f}\right)\left[\left(T-T_{\rm ej}\right)/T_{f}\right]^{-1}$, with $E_f$ the energy at the spectral break, and $B_{\rm norm}$ is the normalization.

\begin{figure}[htbp]
\centering
\includegraphics[width=16.2cm,angle=0,clip]{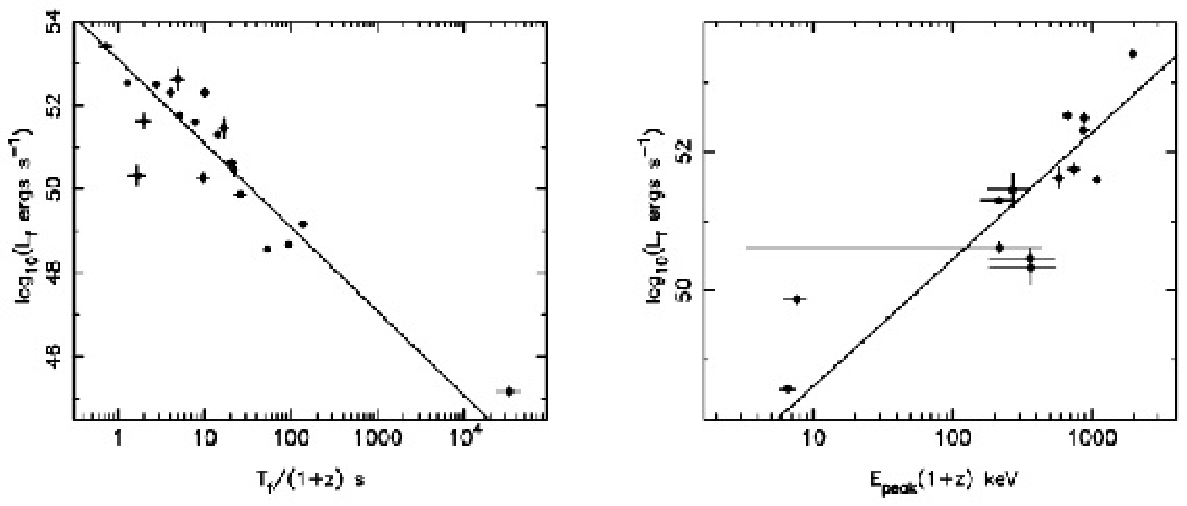}

\vspace{-6.5cm}
\caption{\footnotesize {\bf Left panel:} $L_f$ (in ${\rm erg}\,{\rm s}^{-1}$) vs. $T^{*}_f$.
{\bf Right panel:} $L_f$ (in ${\rm erg}\,{\rm s}^{-1}$) vs. $E_{\rm peak}$. 
(Figures after \cite{willingale2010}; see Fig. 16 therein.)}
\label{fig:12}
\end{figure}

Using this motivated pulse profile, \citet{willingale2010} found that, within a sample of 12 GRBs observed by {\it Swift} in the BAT and XRT energy bands, $L_f$ is anti-correlated with $T^{*}_f$ in the following way:
\begin{equation}
\log L_f \sim -(2.0 \pm 0.2)\log T^{*}_f.
\end{equation}
Therefore, high luminosity pulses occur shortly after ejection, while low luminosity pulses appear at later time (see the left panel of Fig.~\ref{fig:12}). Moreover, \citet{willingale2010} also found a correlation between $L_f$ and $E_{\rm peak}$ as shown in the right panel of Fig. \ref{fig:12}. This is in agreement with the known correlation between the $L_{\rm peak}$ for the whole burst and the $E_{\rm peak}$ of the spectrum during the time $T_{90}$ \citep{yonetoku04,tsutsui13}; for comparison with the $L_{\rm peak}-E_{\rm peak}$ correlation, see also Sect.~\ref{Yonetoku}.

In the 12 light curves considered by \citet{willingale2010}, 49 pulses were analysed. Although several pulses with a hard peak could not be correctly fitted, the overall fitting to the RDP was satisfactory and the HLE model was shown to be able to take into account phase of the GRB emission. However, it is worth to mention the hard pulse in GRB061121 which requires a spectral index $\beta_S = 2.4$, larger than the value expected for synchrotron emission, i.e. $\beta_S = 1$.

\citet{Lee2000} and \citet{Quilligan2002} discussed analogous correlations, although these authors considered the width of a pulse rather than $T_f$, which is in fact closely correlated with pulse width. Many authors afterwards \citep{Littlejohns2013,bosnjak14b,evans2014,hakkila2014,laskar2014,Littlejohns2014,Roychoudhury2014,ceccobello2015,kazanas15,laskar2015,peng2015} have used the motivated pulse profile of \citet{willingale2010} for various studies on the prompt emission properties of the pulses.

Regarding the physical interpretation, in \citep{willingale2010} the flux density of each prompt emission pulse is depicted by an analytical expression derived under the assumption that the radiation comes from a thin shell, as we have already described. The decay after the peak involves the HLE \citep{Genet2009} along the considered shell which is delayed and modified with a different Doppler factor due to the curvature of the surface \citep{Ryde2002,Dermer2007}. 

\section{Summary}\label{summary}
In this work we have reviewed the bivariate correlations among a number of GRB prompt phase parameters and their characteristics. It is important to mention that several of these correlations have the problem of double truncation which affects the parameters. Some relations have also been tested to prove their intrinsic nature like the $E_{\rm peak}-S_{\rm tot}$, $E_{\rm peak}-E_{\rm iso}$ and $L_{\rm peak}-E_{\rm peak}$ relations. For the others we are not aware of their intrinsic forms and consequently how far the use of the observed relations can influence the evaluation of the theoretical models and the ``best'' cosmological settings. Therefore, the evaluation of the intrinsic correlations is crucial for the determination of the most plausible model to explain the prompt emission. In fact, though there are several theoretical interpretations describing each correlation, in many cases more than one is viable, thus showing that the emission processes that rule GRBs still need to be further investigated. These correlations might also serve as discriminating factors among different GRB classes, as several of them hold different forms for SGRBs and LGRBs, hence providing insight into the generating mechanisms. Hopefully those correlations could lead to new standard candles allowing to explore the high-redshift universe. 

\acknowledgments
MT acknowledges support in a form of a special scholarship of Marian Smoluchowski Scientific 
Consortium Matter-Energy-Future from KNOW funding, grant number KNOW/48/SS/PC/2015. 
The work of R.D.V. was supported by the Polish National Science 
Centre through the grant DEC-2012/04/A/ST9/00083.

\addcontentsline{toc}{section}{References}
\bibliography{biblioReview4}

\end{document}